\documentclass[11pt]{article}

\usepackage{arxiv}
\usepackage[utf8]{inputenc}

\usepackage{appendix}

\usepackage{amsthm}
\newtheorem{definition}{Definition}
\usepackage[normalem]{ulem}
\usepackage{xcolor}
\usepackage{bm}
\usepackage{amsmath}
\usepackage{mathtools}
\usepackage{algorithm}
\usepackage{algcompatible}
\usepackage{algpseudocode}
\usepackage{scalerel}
\usepackage{float}
\usepackage{framed}
\usepackage{footnote} %enable footnores
\usepackage{xpatch}
\usepackage{subcaption}
\usepackage{threeparttable}
\usepackage{booktabs}
\usepackage{bbm}
\usepackage{multirow}
\usepackage{cancel}
\usepackage{natbib}
\usepackage{amsfonts}
\usepackage{setspace} % increase space bettwen algorithm states

\newcommand{\R}{\mathbb{R}} % real numbers
\newcommand{\E}{\mathbb{E}}
\newcommand{\V}{\mathbb{V}}
\newcommand{\by}{\bm{y}}
\newcommand{\bz}{\bm{z}}
\newcommand{\bu}{\bm{u}}
\newcommand{\btz}{\bm{\tilde{z}}}
\newcommand{\btu}{\bm{\tilde{u}}}

\newcommand{\bx}{\bm{x}}
\newcommand{\btx}{\bm{\tilde{x}}}
\newcommand{\Norm}{\text{N}}

\newcommand{\Unif}{\text{Unif}}

\newcommand{\Pois}{\text{Pois}}

\newcommand{\GP}{\text{GP}}
\newcommand{\KL}{\text{KL}}

\newcommand{\CS}{C^{\text{\tiny{S}}}}
\newcommand{\CNS}{C^{\text{\tiny{NS}}}}
\newcommand{\bCS}{\bm{C}^{\text{\tiny{S}}}}
\newcommand{\bCNS}{\bm{C}^{\text{\tiny{NS}}}}

\newcommand{\iidsim}{\overset{\text{iid}}{\sim}}

\newcommand{\tell}{\tilde{\ell}}
\newcommand{\btX}{\tilde{\bm{X}}}
\newcommand{\bX}{\bm{X}}
\newcommand{\bC}{\bm{C}}
\newcommand{\KLD}{\text{KL}\infdivx}
\newcommand{\btheta}{\bm{\theta}}
\newcommand{\btzeta}{\tilde{\bm{\zeta}}}

\newcommand{\btxi}{\tilde{\bm{\xi}}}

\newcommand{\brho}{\bm{\rho}}
\newcommand{\bphi}{\bm{\phi}}
\newcommand{\bmu}{\bm{\mu}}

\newcommand{\bvarphi}{\bm{\varphi}}

\newcommand{\sign}{\text{sign}}

\newcommand\mystarb{{\hspace{-1.6pt}{\scaleto{*}{3.4pt}}}}

\DeclarePairedDelimiter\abs{\lvert}{\rvert} 
\DeclarePairedDelimiterX{\norm}[1]{\lVert}{\rVert}{#1} % norm
\let\oldnorm\norm
\def\norm{\@ifstar{\oldnorm}{\oldnorm*}}
\DeclarePairedDelimiterX{\infdivx}[2]{(}{)}{#1\;\delimsize\|\;#2}
\let\Algorithm\algorithm
\renewcommand\algorithm[1][]{\Algorithm[#1]\setstretch{1.35}}  % increase space bettwen algorithm states
\algnewcommand{\IfThenElse}[3]{% \IfThenElse{<if>}{<then>}{<else>}
	\State \algorithmicif\ #1\ \algorithmicthen\ #2\ \algorithmicelse\ #3 \unskip\ \algorithmicend\ \algorithmicif}  %Single line if
\algnewcommand{\IfElse}[2]{% \IfThenElse{<if>}{<then>}{<else>}
	\State \algorithmicif\ #1\ \algorithmicthen\ #2\ \unskip\ \algorithmicend\ \algorithmicif}  %Single line if
\algnewcommand{\IfThen}[2]{% \IfThenElse{<if>}{<then>}{<else>}
	\State \algorithmicif\ #1\ \algorithmicthen\ #2\ }  %Single line if
\algnewcommand{\ElseThen}[1]{% \IfThenElse{<if>}{<then>}{<else>}
	\State \algorithmicelse\ #1 \unskip\ \algorithmicend\ \algorithmicif }  %Single line if
\algnewcommand{\IfElsenonumber}[2]{% \IfThenElse{<if>}{<then>}{<else>}
	\Statex  \hskip 5.5em g. \algorithmicif\ #1\ \algorithmicthen\ #2\ \unskip\ \algorithmicend\ \algorithmicif}  %Single line if
\algnewcommand{\IfElsenonumberh}[2]{% \IfThenElse{<if>}{<then>}{<else>}
	\Statex  \hskip 5.5em h. \algorithmicif\ #1\ \algorithmicthen\ #2\ \unskip\ \algorithmicend\ \algorithmicif}  %Single line if
\algnewcommand\algorithmicforeach{\textbf{for each}}
\algdef{S}[FOR]{ForEach}[1]{\algorithmicforeach\ #1\ \algorithmicdo}
\DeclareMathSymbol{\widehatsym}{\mathord}{largesymbols}{"62}

\renewcommand{\hat}{\widehat}
\newcommand\newexp{{\hspace{1.75pt}{\scaleto{h,k}{5.5pt}}}}

\algnewcommand{\IfElsenonumberi}[2]{% \IfThenElse{<if>}{<then>}{<else>}
	\Statex  \hskip 5.5em i. \algorithmicif\ #1\ \algorithmicthen\ #2\ \unskip\ \algorithmicend\ \algorithmicif}  %Single line if

\title{On MCMC for variationally sparse Gaussian processes: A pseudo-marginal approach}

\author{
  Karla Monterrubio-G\'omez  \\
  MRC Human Genetics Unit\\
  University of Edinburgh \\
  \texttt{kmonterr@ed.ac.uk} \\
   \And
 Sara Wade \\
  School of Mathematics\\
   University of Edinburgh\\
  \texttt{sara.wade@ed.ac.uk} \\
 
}

\begin{document}
\maketitle
\begin{abstract}
Gaussian processes (GPs) are frequently used in machine learning and statistics to construct powerful models. However, when employing GPs in practice, important considerations must be made, regarding the high computational burden, approximation of the posterior, choice of the covariance function and inference of its hyperparmeters. To address these issues, \citet{hensman2015mcmc} combine variationally sparse GPs with Markov chain Monte Carlo (MCMC) to derive a scalable, flexible and general framework for GP models. Nevertheless, the resulting approach requires intractable likelihood evaluations for many observation models. To bypass this problem, we propose a pseudo-marginal (PM) scheme that offers asymptotically exact inference as well as computational gains through doubly stochastic estimators for the intractable likelihood and large datasets. In complex models, the advantages of the PM scheme are particularly evident, and we demonstrate this on a two-level GP regression model with a nonparametric covariance function to capture non-stationarity.  
\end{abstract}


\section{Introduction}
Gaussian processes (GPs) are frequently used in machine learning and statistics to construct powerful models. 
%, which are appealing due to their analytical properties and probabilistic formulation.  
In a Bayesian setting, GPs provide a probabilistic approach to model unknown functions; specifically, the GP prior assumes that the function evaluated at any finite set of inputs has a Gaussian distribution with consistent parameters, specified by the mean function and symmetric positive definite covariance (or kernel) function.  
The flexible, probabilistic and nonparametric nature of GP models makes them appropriate and useful in a wide range of applications, including geostatistics \citep{matheron}, atmospheric sciences \citep{berrocal2010}, biology \citep{bat}, inverse problems \citep{kaipio2006statistical}, and more.
%. In geostatistics, GPs have been under the name of Kriging \citep{matheron}. They are also common in other applications; for instance, in atmospheric sciences \citep{berrocal2010}, biology \citep{bat}, and inverse problems \citep{kaipio2006statistical}.
However, when employing GPs in practice, important considerations must be made, specifically, to address the computational burden, approximation of the posterior, form of the covariance function and inference of its hyperparmeters. 

First, GPs suffer from a high computational burden, due to the need to store and invert large and dense covariance matrices.  To overcome this, various schemes have been proposed, including local approximations \citep{rasmussen2002infinite, trespBCM}, predictive processes \citep{banerjee2008}, basis function approximations \citep{cressie2008}, sparse formulations of the precision matrix \citep{lindgren2011, grigorievskiy2017, durrande2019}, and others (see \cite{heaton2019} and \citet[Chap. 8]{rasmussen2006gaussian} for reviews of approaches in spatial statistics and machine learning, respectively). In machine learning, the sparse GP approximation based on a set of inducing points or pseudo inputs is one of the most popular approaches, due to its general applicability, with no requirements or assumptions on the data structure or covariance function.  Early work in this direction includes \cite{seeger2003fast} and \cite{snelson2006sparse}, and in this paper, we focus on the variational inducing point framework introduced in \citet{titsias2009variational}.

Second, while GP priors result in tractable posterior inference in a normal regression setting, many statistical and machine learning tasks require other likelihoods, necessitating approximation of the posterior. Markov chain Monte Carlo (MCMC) provides a general way to simulate from the posterior and is often considered a gold standard in Bayesian inference, due to its asymptotic guarantees. However, the computational complexity and high-dimensionality of GP models prohibits the use of MCMC algorithms for large datasets.  Variational methods are a popular alternative to MCMC schemes and commonly employed in machine learning for faster, approximate inference \citep{blei2017variational}. 
%commonly employed in the machine learning literature as a faster, approximate alternative to MCMC schemes \citep{blei2017variational}. 
More specifically, for GP models, variational inference is an active area of research \citep[e.g.][]{titsias2009variational, matthews2016sparse, Hensmanbig, cutajar2019deep, damianou2013}, but models are typically restricted to a fixed family of likelihoods.

Lastly, the form of the covariance function and its hyperparmeters  crucially determine properties of the unknown function, such as the spatial correlation, smoothness, and periodicity.  Typically, parametric forms are specified for the covariance function, and the hyperparameters are inferred via a hierarchical or an empirical Bayes approach, which is critical to allow the model to adapt to the true smoothness of the function and achieve desirable posterior consistency and coverage \citep{sniekers2015}. In a hierarchical approach, MCMC algorithms have been proposed \citep{yu2011,filippone2014pseudo} but require costly operations on large, dense covariance matrices at every iteration. Variational schemes employ (approximate) empirical Bayes, or maximum marginal likelihood estimation, and while such operations are still required, the number of iterations is typically much reduced. Furthermore, the parametric assumption of the covariance function limits the model's ability to recover changing behavior of the function, e.g. different smoothness levels, across the input space. Thus, more complex model structures have been proposed to combine multiple GPs for increased model flexibility \citep{dunlop2017deep,gadd20}, which pose further challenges for inference.

To address these issues, we build on the work of \citet{hensman2015mcmc}, which 
combines the variational inducing point framework of \citet{titsias2009variational} with  MCMC  to derive a scalable yet flexible and general framework for GP models.
Their sparse variational method results in a low-dimensional approximate posterior, where MCMC is employed to draw samples.
Thus, the variationally sparse MCMC framework benefits from (i) the sparse variational method to alleviate the computational burden, (ii) a general scheme for any factorized likelihood, and (iii) full posterior inference of the hyperparameters. 
The  flexibility, generality and theoretical guarantees of MCMC allow them to apply the framework to general tasks, removing restrictions on the likelihood, and importantly, avoid any further distributional assumptions on the the low-dimensional posterior, e.g. independence, that are typically required in full variational schemes. 

%Their sparse variational method alleviates the computational burden and subsequently employs MCMC to draw samples from the  low-dimensional approximate posterior. 
%Thus, the variationally sparse MCMC framework benefits from (i) the sparse variational method to alleviate the computational burden, (ii) a general scheme for any factorized likelihood, and (iii) full posterior inference of hyperparameters. 
%The  flexibility and theoretical guarantees of MCMC allows them to apply the framework to general tasks, removing restrictions on the likelihood, and importantly, avoid any further distributional assumptions on the the low-dimensional posterior, e.g. independence, that are typically required in full variational schemes. Moreover, the framework allows for full posterior inference of hyperparameters

%Their sparse variational method results in a low-dimensional approximate posterior, where MCMC is employed to draw samples. Thus, the variationally sparse MCMC framework benefits from (i) the sparse variational method to alleviate the computational burden, (ii) a general scheme for any factorized likelihood, and (iii) full posterior inference of hyperparameters
%The flexibility and theoretical guarantees of MCMC to sample from a complicated low-dimensional variational posterior. I
%Importantly, this last step avoids placing any further distributional assumptions on the low-dimensional posterior, e.g. independence, that are typically required in full variational schemes.

Nevertheless,  the resulting optimal approximate posterior requires  exponentiated expected log-likelihood evaluations that are intractable for many observation models, including classification problems, robust GP regression with the student $t$ likelihood, stochastic volatility models, and various others \citep{nickisch2008, hernandez2011, Neal1997, wu2014gaussian}.  In this case, \cite{hensman2015mcmc} suggest approximating the required expectations with Gauss-Hermite quadrature. This clearly introduces an additional level of approximation, which can have adverse effects on the accuracy and computational cost of the method. In complex models, such effects may be particularly severe, and we demonstrate this on the two-level GP regression model with a nonparametric covariance function to capture non-stationarity \citep{monterrubio2020}. 
%; however, we demonstrate that such approximation can be inaccurate and computationally expensive for certain type of functions.

We propose an alternative approach to bypass this problem by replacing the intractable likelihood evaluation with a computationally cheap unbiased estimate based on the block-Poisson estimator introduced by \citet{quiroz2016block}. Our proposed framework offers asymptotically exact inference for the low-dimensional variational posterior through a pseudo-marginal scheme as well as computational gains through doubly stochastic estimators for the intractable likelihood evaluations and large datasets. 

This paper is organised as follows. We start, in Section~\ref{section:hensman}, with a review of variationally sparse GP models, where we focus on the work of \cite{hensman2015mcmc}. Section~\ref{section:PM} introduces the signed block-Poisson pseudo-marginal (S-BP-PM) scheme for variationally sparse GPs. %highlighting the differences with a classical variationally sparse GP framework. 
Section~\ref{section:motivation} showcases the limitations of the Gauss-Hermite quadrature approximation employing, as an example, a non-stationary 2-level GP model as well as details of the S-BP-PM scheme for this model. Later, Section~\ref{section:Simstudy} demonstrates our proposed scheme on a 1-dimensional 2-level GP regression model. 
%and  Section~\ref{section:Realdata} provide results on simulated and real datasets, respectively.
Finally, Section~\ref{section:Conslusions} concludes summarising the main findings.
%and possible extensions and directions for future research.

\section{Variationally sparse GPs } \label{section:hensman}

The observed data are assumed to consist of outputs $y_n$, which may be real-valued or more generally binary, counts, etc., with corresponding input locations $\bx_n \in \R^D$ for $n=1,\ldots,N$. The likelihood is assumed to factorise across data points, dependent on an latent function $z:{\R}^D\rightarrow \R$ that maps the input locations to the real line:
$$ p(\by \mid \bm{z}, \brho)=\prod_{n=1}^{N} p(y_n\mid z(\bx_n), \brho ),$$
where $\by=(y_1,\ldots,y_N)^T$, $\bm{z}=(z_1,\ldots,z_N)^T$ with $z_n \equiv z(\bx_n)$, and $\brho$ contains any additional likelihood parameters. The unknown function $z$ has a Gaussian process prior with zero mean and covariance function $C_{{\bphi}} (\cdot,\cdot)$ parametrised by ${{\bphi}}$, namely, $z(\cdot) \sim \GP ( 0,C_{{\bphi}} (\cdot,\cdot) )$. 

%A huge burden to employ GPs in practice is the high computational complexity, which scales cubically with the number of data points. 
To overcome the computational complexity of GPs, 
\citet{snelson2006sparse} proposed the sparse pseudo-input framework. The key idea of this approach is to augment the data with a set of $M \ll N$ inducing or pseudo-points $\btX= (\tilde{\bx}_1, \ldots\tilde{\bx}_M )^T$  and collect the values of the latent functions at the inducing points into the vectors $\btz=(\tilde{z}_1,\ldots,\tilde{z}_M)^T,$ where $\tilde{z}_m \equiv z(\btx_m)$, and we refer to the $\tilde{z}_m$ as the \textit{inducing variables}. By properties of GPs, the augmented prior is
\begin{align*}
\begin{split}
\pi(\bz,\btz \mid {\bphi}) &= \pi( \bz \mid  \btz, {\bphi}) \pi( \btz \mid {\bphi}) \\
&= \Norm \left(\bz \mid \bC_{\bz,\btz} \bC_{\btz,\btz}^{-1} \btz \, , \bC_{\bz,\bz}- \bC_{\bz,\btz}\bC_{\btz,\btz}^{-1} \bC_{\btz,\bz} \right) \Norm \left(\btz \mid 0, \bC_{\btz,\btz}\right),
\end{split}
\end{align*}
where we make use of the short notation $\bC_{\bz,\bz}$ and $\bC_{\btz,\btz}$ to denote the covariance matrix constructed by evaluating the kernel at the inputs $\bX$ and  $\btX$, respectively, and  $\bC_{\bz,\btz}$ to denote the cross-covariance matrix between the the function evaluated at the inputs $\bX$ and inducing points $\btX$. Under the augmented model, the posterior of the parameters and latent variables is given by:
\begin{align*}
\pi(\bz,\btz, \brho, {\bphi}\mid \by,X) \propto \prod_{n=1}^N p(y_n \mid z_n, \brho) \pi(\bz\mid \btz,  {\bphi}) \pi(\btz\mid {\bphi}) \pi({\bphi}) \pi(\brho).
\end{align*}
The variationally sparse approach of \cite{hensman2015mcmc} restricts the approximate variational posterior to take the form:
\begin{equation}
q(\bz,\btz, \brho, {\bphi}) \propto \pi(\bz|\btz,  {\bphi})  q(\btz, \brho, {\bphi}). \label{eq:app_v}
\end{equation}
Note that in the right-hand side of Eq.~\eqref{eq:app_v}, the first term corresponds to the prior predictive distribution of $\bz$ given $\btz$, while the second is the joint approximate posterior of the inducing variables, parameters, and hyperparameters. Thus, the variationally sparse approach assumes that conditioned on the inducing variables and hyperparameters, the latent function at the observed input locations does not depend on the data. This assumption is crucial to achieve the desired scalability, but the accuracy of this approximation clearly depends on the number and locations of the inducing points.
Under this assumption, \citet{hensman2015mcmc} showed that the optimal low-dimensional variational posterior (the second term in Eq.~\eqref{eq:app_v}), which is obtained by minimizing the  Kullback-Leibler (KL) divergence between the approximate and true posterior,
$\KLD{ q(\bz,\btz, \brho, {\bphi})}{\pi(\bz,\btz, \brho, {\bphi}\mid \by,X) }$, %,  corresponds to Eq.~\eqref{eq:app_v}, where the  low-dimensional variational posterior (the second term in Eq.~\eqref{eq:app_v}) 
takes the form
\begin{align}
q(\btz,\brho, {\bphi}) \propto \exp\left( \sum_{n=1}^N \E_{\pi(z_n \mid \btz,{\bphi})}[ \log(p(y_n \mid z_n, \brho)) ] \right) \pi(\btz \mid
{\bphi})  \pi({\bphi}) \pi(\brho). \label{eq:varposterior}
\end{align}
Computation of Eq.~\eqref{eq:varposterior} involves expectations over univariate Gaussian random variables, which are available analytically only for certain tasks, specifically, for Gaussian or Poisson likelihoods. In all other settings, \citet{hensman2015mcmc} suggest to approximate with Gauss-Hermite quadrature. In addition, to avoid placing any further restrictions on the form of the optimal variational posterior, \citet{hensman2015mcmc} propose to use MCMC methods to sample from a whitened version of Eq.~\eqref{eq:varposterior}.  Whitening is employed because  Eq.~\eqref{eq:varposterior}  exhibits high correlations between $\btz$ and $ \bphi$,  which can result in poor mixing. Thus, the target distribution is:
\begin{align}
q(\btxi,\brho, {\bphi}) \propto \exp\left( \sum_{n=1}^N \E_{\pi(z_n \mid \btxi,{\bphi})}[ \log(p(y_n \mid z_n, \brho)) ] \right) \Norm(\btxi \mid 0, \bm{I}_M)  \pi({\bphi}) \pi(\brho), \label{eq:whitvarposterior}
\end{align}
with $\btz=\bm{L}(\bphi)\btxi$,  where $\btxi \sim \Norm (0, \bm{I}_M)$ and ${\bm{L}(\bphi)}{\bm{L}(\bphi)}^T=\bC_{\btz,\btz}$.  

While this results in a scalable and flexible approach for general GP-based models, the Gauss-Hermite quadrature clearly introduces an extra level of approximation, which can adversely affect the accuracy and cost. 
The error of the quadrature approximation has been carefully studied in literature \citep{gautschi1981survey,mastroianni1994error,arasaratnam2007discrete}, and in general, Gauss-Hermite quadrature of order $J$ provides a good approximation if the integrand, denoted by $g(\cdot)$, is a polynomial of order $2J-1$ or less.  When this is not the case, the error of the approximation corresponds to $J! g^{(2J)}(\epsilon) /{2J!} ,$ where $g^{(2J)}(\cdot)$ is the $2J$-th derivative of $g(\cdot)$. In simple settings, such as binary classification, the computational complexity for a likelihood evaluation is $\mathcal{O}(NM^2+NJ)$ and sufficient approximation may be achieved with relatively small $J$. However, the computational cost can increase drastically for more complex models, e.g. $\mathcal{O}(JNM^2)$ in the two-level GP model of Section \ref{section:motivation}, which also requires a large $J$ for good approximation.  Moreover, for some tasks, such as multinomial classification, multivariate quadrature is needed, which further exacerbates costs and accuracy issues.

When quadrature approximation is required, the scheme of \citet{hensman2015mcmc} belongs to the class of approximate MCMC methods \citep{marjoram2003}. Recently, \citet{vihola2020} proposed an importance sampling correction of approximate MCMC to yield exact inference. While this could be applied to the scheme of \citet{hensman2015mcmc}, instead, in the next section, we construct a  pseudo-marginal sampler  based on the block-Poisson estimator of \cite{quiroz2016block}. Importantly, our scheme bypasses the problems of the approximate MCMC scheme, by providing both asymptotically exact inference for the optimal varitional posterior and reduced computational complexity.

\section{PM for variationally sparse GPs } \label{section:PM}

The pseudo-marginal (PM) approach introduced by \citet{beaumont2003estimation} and \citet{andrieu2009pseudo} provides a route to do exact Bayesian inference in models with intractable  or expensive likelihoods.
PM samplers employ a non-negative unbiased estimator of the likelihood in place of the intractable or expensive function in a Metropolis-Hasting (MH) algorithm to produce asymptotically exact samples from the posterior distribution. In particular, PM schemes have been previously employed to do inference in GP models \citep[see, e.g.][]{PM, murray2016pseudo, xiong2017adaptive}.

The key ingredient of PM schemes is the unbiased estimator of the likelihood, and as such, different approaches exist to produce the unbiased estimator in different scenarios. For expensive likelihoods due to tall data, subsampling-based strategies include the Rhee-Glynn estimator \citep{rhee2015,bardenet2017} or PM Firefly \citep{maclaurin2015,bardenet2017}. For intractable likelihoods, unbiased estimators have been proposed using, for example, importance sampling \citep{PM}, annealed importance sampling \citep{filippone2014bayesian}, particle filters \citep{andrieu2010particle}, and  generalized Poisson estimation \citep{fearnhead2008}. However, the variance of estimator needs to be carefully controlled; if too high, the likelihood may be overestimated, making it difficult for the chain to leave the current state. Indeed, \citet{doucet2015efficient} recommend keeping the variance of the log-likelihood estimator to 1.5, in order to balance computation time with low variance of the MCMC estimates.       

Recently, \citet{quiroz2016block} proposed combining an importance sampling sign correction \citep{lyne2015russian} with a product of Poisson estimators \citep{poissonP} to derive a signed block-Poisson pseudo-marginal scheme for fast, exact inference in tall datasets. 
Their block-Poisson estimator is appealing for several reasons. First, through the product form, correlation is introduced between the log estimated likelihood at current and proposed states, resulting in an efficient dependent PM scheme \citep{deligiannidis2018}, that can accommodate noisier likelihood estimates. In addition, it has lower variance than the Rhee-Glynn estimator \citep{rhee2015,bardenet2017} and makes use of control variates for variance reduction. Lastly, a sign correction permits employing a soft-lower bound (as opposed to the strict lower bound in \citet{bardenet2017}), which is more computationally efficient.

In this work, we build on  \citet{quiroz2016block} to construct a signed block-Poisson PM scheme for variationally sparse GPs, that is both computationally efficient for large datasets and offers asymptotically exact inference for the low-dimensional variational posterior. This is accomplished through a variant of the PM scheme that employs a doubly stochastic estimator for the intractable exponentiated expected log-likelihood, by also data subsampling.

\subsection{Doubly stochastic block-Poisson estimator}\label{subsec:blockPoisson}

Our goal is find an unbiased estimator, $	\widehat{E}$, of the intractable  exponentiated expected log-likelihood term in Eq.~\eqref{eq:whitvarposterior}:
$$E=  \exp\left( \sum_{n=1}^N \E_{\pi(z_n \mid \btxi,{\bphi})}[ \log(p(y_n \mid z_n, \brho)) ] \right),$$ 
%is intractable, and our goal is to find an unbiased estimator, $	\widehat{E}$, of $E$, 
that is also computationally efficient for large sample sizes. To do so, we follow \citet{quiroz2016block} by employing subsampling for computational efficiency and control variates to reduce the variance of our estimator. In addition, we employ a further layer of stochasticity in our estimator to deal with the intractable expectation.

Using the simplifying notation $l(y_n \mid z_n, \brho)=\log(p(y_n \mid z_n, \brho)) $, the first step in this direction is to
define the difference $d=\sum_{n=1}^N d_n,$ with%
\begin{equation}%
d_n=\E_{\pi(z_n \mid \btxi,{\bphi})}[ l(y_n \mid z_n, \brho) ]  - \bar {\nu}_n, \label{eq:dn_v2}
\end{equation}%
where the control variate $ \bar{\nu}_n$ is an approximation to $\E_{\pi(z_n \mid \btxi,{\bphi})}[ l(y_n \mid z_n, \brho) ] $. Thus, we have $$\log(E)= d+ \sum_{n=1}^{N}  \bar{\nu}_n.$$ Specifically, assuming differentiability of $l(y \mid z, \brho)$ with respect to $z$,  we define $\bar{\nu}_n$ through a first-order Taylor-expansion around $\E[z_n]$ (with the expectation taken with respect to $\pi(z_n \mid \btxi,{\bphi})$): 
\begin{equation}%
\nu_n(z_n)=   l(y_n \mid \E[z_n], \brho)+(z_n-\E[z_n])  l^{\prime}(y_n \mid \E[z_n], \brho)
\label{eq:vn_v2},\end{equation}%
such that
\begin{equation}%
\bar{\nu}_n= \E_{ \pi(z_n \mid \btxi,{\bphi})} [\nu_n(z_n)] = l (y_n \mid \E[z_n], \brho).  \label{eq:controlv_v2}
\end{equation} %
We note that more generally, a multivariate Taylor expansion may be used in models involving multiple Gaussian processes. 

By defining $\nu_n(z_n)$ through Eq.~\eqref{eq:vn_v2}, the difference in Eq.~\eqref{eq:dn_v2} can be equivalently written as 
$$d_n=\E_{ \pi(z_n \mid \btxi,{\bphi})} \left[ l (y_n \mid z_n, \brho) -\nu_n(z_n)  \right],$$ and an unbiased difference estimator for $d_n$ is
\begin{equation}%
\widehat{d}_n =l (y_n \mid z_n, \brho) - \nu_n(z_n), \label{eq:difest_v2}%
\end{equation} with $z_n \sim \pi(z_n \mid \btxi,{\bphi})$.
We then use subsampling techniques to obtain
\begin{equation}\widehat{d}_B=\frac{N}{B} \sum_{b=1}^{B}\widehat{d}_{\alpha_{b}}, \label{eq:dhat_v2}\end{equation}
where $\alpha_{b}\iidsim\Unif(1,\ldots, N)$ indexes a subsample of size $B$ that is taken with replacement. The variance of the estimator in Eq.~\eqref{eq:dhat_v2} is $\V[\widehat{d}_B]=\gamma/B$ with $\gamma=N^2\V[\widehat{d}_{\alpha_{b}}]$ denoting the intrinsic variance of the estimator.

Thus, $\widehat{d}_B +\sum_{n=1}^N \bar{\nu}_n$ provides an unbiased estimator of the expected log-likelihood. Clearly, simply exponentiating this does not provide an unbiased estimator of $E$, and in order to do so, we employ block-Poisson estimation \citep{quiroz2016block}. 
In the doubly stochastic block-Poisson estimator in Definition \ref{def:BP},  we re-write the difference estimator in Eq.~\eqref{eq:difest_v2} in terms of random variables that do not depend on $(\btxi,\bphi)$. Thus, we introduce uniform random variables, $\chi_1, \ldots, \chi_N$, and apply the inverse CDF 
transformation to produce samples $z_n$ needed to evaluate Eq.~\eqref{eq:difest_v2}.

\begin{definition} \label{def:BP}
	The doubly stochastic block-Poisson estimator is
	\begin{align}
	\widehat{E}  = \exp\left( \sum_{n=1}^N  l (y_n \mid \E[z_n], \brho) \right ) \prod_{k=1}^\kappa \exp \left(  \frac{a+\kappa}{\kappa}\right) \prod_{h=1}^{\mathcal{H}_k} \left(  \frac{ \widehat{d}_B^\newexp -a }{\kappa}\right) \label{eq:blockP_v2}
	\end{align}
	with $\kappa \in \mathbb{Z}^{+},$ ${\mathcal{H}_1},\ldots,{\mathcal{H}_\kappa}  \sim \Pois(1)$, $a \in \mathbb{R}$ a lower bound for  
	$$\widehat{d}_B^\newexp= \frac{N}{B} \sum_{b=1}^{B} \widehat{d}_{\alpha_b^\newexp}, \text{ with} \quad \widehat{d}_{\alpha_b^\newexp} = l(y_{\alpha_b^\newexp} \mid z_{\alpha_{b}^\newexp}, \brho)-\nu_{\alpha_{b}^\newexp}(z_{\alpha_{b}^\newexp}),$$ where $\alpha_b^\newexp$ is uniformly sampled from $(1,\ldots, N);$
	$\chi_{b}^\newexp \sim \Unif(0,1)$, and the the log-likelihood and Taylor expansion in $\widehat{d}_{\alpha_b^\newexp}$ are evaluated with $z_{\alpha_{b}^\newexp}=F_{z_{\alpha_{b}^\newexp}}^{-1}(\chi_{b}^\newexp)$, where $F_{z_n}^{-1}$ denotes the inverse CDF of Gaussian prior predictive of $z_n$ given $(\btxi,\bphi)$. 
\end{definition}

%As efficient performance in PM algorithms is closely related to low variability of the estimator, \citet{quiroz2016block} assert that one of the key advantages of the estimator in Eq.~\eqref{eq:blockP} is that the blocking structure induces a  controllable correlation between the estimates, which in turn reduces the variance of the estimators.	

Evaluation of the likelihood estimator in Eq.~\eqref{eq:blockP_v2} requires computation of both the control variates across every data point and, on average, $\kappa$ difference estimators. Note that the control variates only depend on the predictive means $\E[ z_n | \btz]=\bC_{z_n,\btz}\bC_{\btz,\btz}^{-1}\btz$, with complexity $\mathcal{O}(NM+M^3)$. The difference estimator, instead, requires also the predictive variance $\V( z_n | \btz)=\bC_{z_n,z_n}-\bC_{z_n,\btz}\bC_{\btz,\btz}^{-1}\bC_{\btz, z_n}$, but only across all subsets, with complexity $\mathcal{O}(\kappa B M^2+M^3)$. 
Thus, evaluation of the block-Poisson estimator is $\mathcal{O}(NM+ \kappa B M^2+M^3)$, in contrast to the $\mathcal{O}(NM^2 +M^3)$ required by the approximate scheme of \citet{hensman2015mcmc}. 

%\SW{note that computing the predictive mean $\E[ z_n | \btz]=\bC_{z_n,\btz}\bC_{\btz,\btz}^{-1}\btz$ is $\mathcal{O}(NM+M^3)$, while computing the predictive variance  $\V( z_n | \btz)=\bC_{z_n,z_n}-\bC_{z_n,\btz}\bC_{\btz,\btz}^{-1}\bC_{\btz, z_n}$ is $\mathcal{O}(NM^2)$. However, the variance is not needed to compute the control variates $\bar{\nu}_n$, which only depend on the predictive mean. Thus, evaluation of the block-Poisson estimator is $\mathcal{O}(NM+ \sum_{k=1}^{\kappa} \mathcal{H}_k )B M^2+M^3)$. Check with Karla if this is computed efficiently.}

We highlight that \citet{quiroz2016block}
focus on tall data with tractable likelihoods, and thus, provide an alternate construction of the difference estimator, considering two types of control variates, parameter expanded and data expanded, that reduce the computational complexity to a small number of evaluations that does not depend on $N$. \citet{quiroz2019} examine properties of such control variates in balancing computation time with variance reduction. Variants of these control variates could be developed here to further reduce the computational complexity and remove dependency on $N$; however, we do not pursue this, in order to maintain low variance of the doubly stochastic estimator.

\subsection{Signed block PM MCMC}

\begin{algorithm}[!t]
	\small 
	\caption{Signed block-Poisson pseudo-marginal sampler (S-BP-PM) }\label{algo:SBPM_v2}
	\begin{algorithmic}[1]
		\Require Target distribution: $\breve{q}(\btxi,\bphi, \brho),$ initial states: $\btxi^{ \scaleto{(0)}{5pt}},$ ${\bphi}^{ \scaleto{(0)}{5pt}},$ $\brho^{ \scaleto{(0)}{5pt}}$, iterations: $T$, number of Poisson estimators: $\kappa$,  batch size: $B$; lower bound: $a$, initial means: $c_1,\ldots,c_n$, initial random variates $(\mathcal{V}_1,\ldots, \mathcal{V}_G)$, initial estimator ${\widehat{E}}$ \vspace{1mm}.
		\For{$t=1,\ldots,T$}  \vspace{1mm}
				\State Draw ${ \brho}^{ *}$ from $q(\brho| \brho^{(t-1)}) $ 
				\State Compute control variates: $l_n=l(y_{n} | c_n, \brho^*)$ for $n=1, \ldots, N$ 
				\State  Sample a block $g$
				%$g\sim \Unif(1,\ldots, \kappa)$ 
				and random variates $\mathcal{V}_g^*$.  
		\State  Compute $ \log(|\widehat{E}^*|)$ using Eq.~\eqref{eq:blockP_v2} \vspace{1mm}
			\State Set $\brho^{\scaleto{(t)}{5pt}} \gets \brho^*$; $ \mathcal{V}_g \gets \mathcal{V}^{*}_g$;
					  $\log(|\widehat{E}|) \gets \log(|\widehat{E}^*|) $; and $ \mathcal{S}^{\scaleto{(t)}{5pt}}_{\brho} \gets \sign(\widehat{E}^*) $ with 
				\State \hskip2.5em	  probability:
				\({\text{min}} \left\{ 1, \frac{  \abs*{\widehat{E}^*} \pi(\brho^*) q( \brho^{(t-1)}|\brho^*)}{   \abs*{\widehat{E}}\pi( \brho^{(t-1)})q( \brho^* |\brho^{(t-1)}) }\right\} \) 
				\State Else  $\brho^{\scaleto{(t)}{5pt}} \gets \brho^{\scaleto{(t-1)}{5pt}}$  and $ \mathcal{S}^{\scaleto{(t)}{5pt}}_{\brho} \gets \mathcal{S}^{\scaleto{(t-1)}{5pt}}_{\brho} $
			\State Draw ${ \btxi}^{ *}$ from $q(\btxi| \btxi^{(t-1)}) $ 
				\State Compute GP predictive means $c_n^*$ and control variates: $l_n=l(y_{n} | c_n^{*}, \brho^{\scaleto{(t)}{5pt}})$ 
				\State \hskip2.5em for $n=1, \ldots, N$ 
				\State  Sample a block $g$
				%$g\sim \Unif(1,\ldots, \kappa)$ 
				and random variates $\mathcal{V}_g^*$. 
					\State  Compute $ \log(|\widehat{E}^*|)$ using Eq.~\eqref{eq:blockP_v2} \vspace{1mm}
					\State Set $\btxi^{\scaleto{(t)}{5pt}} \gets \btxi^*$; $ \mathcal{V}_g \gets \mathcal{V}^{*}_g$;
					  $\log(|\widehat{E}|) \gets \log(|\widehat{E}^*|) $; $ \mathcal{S}^{\scaleto{(t)}{5pt}}_{\btxi} \gets \sign(\widehat{E}^*) $; and 
					  \State \hskip2.5em $ c_{1:N} \gets c_{1:N}^* $ with 
					  probability:
				\({\text{min}} \left\{ 1, \frac{  \abs*{\widehat{E}^*} \pi(\btxi^*) q( \btxi^{(t-1)}|\btxi^*)}{   \abs*{\widehat{E}}\pi( \btxi^{(t-1)})q( \btxi^* |\btxi^{(t-1)}) }\right\} \) 
				\State Else  $\btxi^{\scaleto{(t)}{5pt}} \gets \btxi^{\scaleto{(t-1)}{5pt}}$  and $ \mathcal{S}^{\scaleto{(t)}{5pt}}_{\btxi} \gets \mathcal{S}^{\scaleto{(t-1)}{5pt}}_{\btxi} $
				\State Draw ${ \bphi}^{ *}$ from $q(\bphi| \bphi^{(t-1)}) $ 
				\State Compute GP predictive means $c_n^*$ and control variates: $l_n=l(y_{n} | c_n^{*}, \brho^{\scaleto{(t)}{5pt}})$ 
				\State \hskip2.5em for $n=1, \ldots, N$ 
				\State  Sample a block $g$
				%$g\sim \Unif(1,\ldots, \kappa)$ 
				and random variates $\mathcal{V}_g^*$. 
					\State  Compute $ \log(|\widehat{E}^*|)$ using Eq.~\eqref{eq:blockP_v2} \vspace{1mm}
					\State Set $\bphi^{\scaleto{(t)}{5pt}} \gets \bphi^*$; $ \mathcal{V}_g \gets \mathcal{V}^{*}_g$;
					  $\log(|\widehat{E}|) \gets \log(|\widehat{E}^*|) $; $ \mathcal{S}^{\scaleto{(t)}{5pt}}_{\bphi} \gets \sign(\widehat{E}^*) $; and 
					  \State \hskip2.5em $ c_{1:N} \gets c_{1:N}^* $ with 
					  probability:
				\({\text{min}} \left\{ 1, \frac{  \abs*{\widehat{E}^*} \pi(\bphi^*) q( \bphi^{(t-1)}|\bphi^*)}{   \abs*{\widehat{E}}\pi( \bphi^{(t-1)})q( \bphi^* |\bphi^{(t-1)}) }\right\} \) 
				\State Else  $\bphi^{\scaleto{(t)}{5pt}} \gets \bphi^{\scaleto{(t-1)}{5pt}}$  and $ \mathcal{S}^{\scaleto{(t)}{5pt}}_{\bphi} \gets \mathcal{S}^{\scaleto{(t-1)}{5pt}}_{\bphi} $	
		 
		\EndFor
	\end{algorithmic}
	%\caption{Let $U_k=(\mathcal{H}_k,\alpha_{1:B}^{1:\mathcal{H}_k,k}, \chi_{1:B}^{1:\mathcal{H}_k,k})$ denote the set of random variates for block $k$ of the block-Poission estimator.}
\end{algorithm}

The algorithm is a simple extension of the signed dependent PM scheme developed in \citet{quiroz2016block}, using a Metropolis-within-Gibbs (MwG) sampler to iterate over the components $ (\btxi,\brho,\bphi)$.  

\paragraph{Sign correction.} In order to employ the block-Poisson estimator, one must define a lower bound $a$ for $\widehat{d}_B$. \citet[Section 3.3]{quiroz2016block} explain that employing a soft lower bound, i.e. a lower bound that makes $ \tau=\Pr(\widehat{E}\geq 0) \approx 1$, is computationally more efficient than employing a strict lower bound. One can then use the absolute value of $\widehat{E}$ to avoid possible negative likelihood estimates and apply a sign correction to obtain asymptotically exact estimates from the MCMC samples  \citep{lyne2015russian}.

%Something about block correlation here: 
\paragraph{Block correlation.} 
%As efficient performance in PM algorithms is closely related to low variability of the estimator, \citet{quiroz2016block} assert that one of the key advantages of the estimator in Eq.~\eqref{eq:blockP} is that the blocking structure induces a  controllable correlation between the estimates, which in turn reduces the variance of the estimators.
A key  advantage of the estimator in Eq.~\eqref{eq:blockP_v2} is that through blocking structure, an efficient correlated PM scheme can be constructed. %correlation can be introduced between the estimates at the current and proposed states of the MH algorithm.
Specifically, at each step, we only update the random variables 
%$\mathcal{V}_{k}= \lbrace \mathcal{H}_k, \alpha_{1:B}^{1,k},\ldots,  \alpha^{ \mathcal{H}_k,k}_{1:B}, \chi_{1:B}^{1,k},\ldots, \chi_{1:B}^{ \mathcal{H}_k,k} \rbrace$ for $k=1, \ldots, \kappa$, where ${\alpha}^{h,k}_{1:B } \coloneqq( {\alpha}^{h,k}_{1},\ldots \alpha_{B}^{h,k} )$ and $\chi^{h,k}_{1:B}\coloneqq( \chi_{1}^{h,k},\ldots \chi_{B}^{h,k} )$. 
$\mathcal{V}_g = (\mathcal{H}_g,(\alpha_1^{h,g},\ldots,\alpha_B^{h,g},\xi_1^{h,g},\ldots,\xi_B^{h,g})_{h=1}^{\mathcal{H}_g})$ of a single block $g$, drawn uniformly from 1 to $\kappa$.  The remaining random variables of all other blocks are fixed.  As shown in \citet{quiroz2016block}, this induces the correlation $\rho = \text{Cor}( \log |\widehat{E}^*|,\log |\widehat{E}|) \approx 1 - 1/\kappa $ between the log of the estimated likelihood at the proposed and current states. Moreover, one can target a prespecified correlation $\rho \approx  1 - 1/G$ by grouping the $\kappa$ Poisson estimators into a corresponding number of groups, $G$.

%Algorithm~\ref{algo:SBPM_v2} details the proposed MCMC sampler that employs the signed block-Poisson estimator. The scheme uses a Metropolis-within-Gibbs (MwG) sampler to iterate over the components of $ \breve{q}(\btxi,\brho,\bphi)$. %The noise variance $\sigma_\varepsilon^2$ and second level length-scale $\bm{\lambda}$, are sampled with adaptive rwMH steps. 
%For the whitened spatially varying length-scale, we employ ELLSS, and for the non-stationary function $\btz$, we use a MH step with a Gaussian proposal that approximates the true variational conditional posterior; more precisely, the proposal is:
%\begin{equation}
%\Norm \left(  \sigma^{-2}_\varepsilon\CNS_{\btz,\btz} \left(\CNS_{\btz,\btz}+\sigma^{-2}_\varepsilon \hat{\bm{P}} \right)^{-1} \hat{B}^T\by,\CNS_{\btz, \btz}\left( \CNS_{\btz,\btz} + \sigma^{-2}_\varepsilon \hat{\bm{P}} \right)^{-1} \CNS_{\btz,\btz}\right), \label{eq:proposalz}
%\end{equation}
%with $\hat{B}$ an $N \times M$ matrix with rows $\hat{\bm{\beta}}_n=\left[ \CNS_{z_n, \btz}\right]_{\mu_{\ell_n}}$ and  $\hat{\bm{P}}= \sum_{n=1}^N \hat{P}_n,$ with $\hat{P}_n=\left[ \CNS_{ \btz,z_n} \CNS_{z_n, \btz} \right]_{\mu_{\ell_n}}, $ where we use $\mu_{\ell_n}=\exp(c_n+w_n^2/2)$ to evaluate the expressions in square brackets with $c_n$ and $w_n^2$ as in Eq.~\eqref{eq:candw}.
%%\begin{equation*}%
%%\hat{\bm{\beta}}_n =\left[ \CNS_{z_n, \btz}\right]_{\mu_{\ell_n}=\exp(c_n,w_n^2/2)}, \quad 
%%\hat{P}_n = \left[ \CNS_{ \btz,z_n} \CNS_{z_n, \btz} \right]_{\mu_{\ell_n}=\exp(c_n,w_n^2/2)},
%%\end{equation*}
\paragraph{Metropolis-within-Gibbs} We employ an MwG framework to sample the parameters $(\btxi,\brho,\bphi)$. The whitened parameters $\btxi$ may be sampled via elliptical slice sampling \citep[ELLSS,][]{murray2010elliptical}, preconditioned Crank-Nicolson \citep{cotter2013}, or recent adaptive extensions \citep{wallin2018infinite}. For example, for the two-level GP model in Section \ref{SEC:SBP}, we employ independent Metropolis-Hastings (MH) at the first level and ELLSS at the second level. An adaptive random walk \citep{roberts2009examples} may be used for the likelihood parameters $\brho$ and GP hyperparameters $\bphi$. Alternatively, Hamiltonian Monte Carlo samplers may be employed, following  \citet{dang2019}.

%The computational complexity of each iteration of Algorithm~\ref{algo:SBPM} is $\mathcal{O}( (\sum_{k=1}^{\kappa} \mathcal{H}_k )B M^2+M^3+NM))$. 
%We emphasise that for all the parameter updates, we can compute the difference estimator $\widehat{d}_{\alpha_{b}^\newexpg}$, for all $h$ and $k$, in parallel.
%While our current implementation does not make use of parallel computing, we adapt Algorithm~\ref{algo:SBPM} to vectorise some of the operations; for instance, $\E[z_n]$ can be easily vectorised.

Algorithm~\ref{algo:SBPM_v2} summarizes the PM scheme. It requires prespecification of the algorithmic parameters $(\kappa, a, B)$ as well as the inducing locations $\btX$, which are discussed in the following subsections. 
%Additionally, we note that Algorithm~\ref{algo:SBPM_v2} belongs to the so-called grouped independence MH methods \citep{beaumont2003estimation}. While such algorithms are exact in the sense that the limiting distribution corresponds to the posterior of interest \citep{andrieu2009pseudo}, they can suffer from slow mixing and convergence \citep{drovandi2018accelerating}. In contrast,  Monte Carlo within Metropolis (MCWM) schemes, which recompute the estimator at each iteration, tend to exhibit better mixing at the price of draws from an approximate posterior \citep{andrieu2009pseudo}. Motivated by this, we employ MCWM during the burnin phase.
%\KM{We actually dont need the MCWM phase, specially if we initialise with the quadrature pre-run}
%\SW{a lot of acronyms - check that there are defined}

\subsection{Algorithmic parameters} \label{subsec:Alg_parameters}
We follow the guidelines of \citet{quiroz2016block} to optimally select the key algorithmic parameters, namely the lower bound $a$, the number of Poisson estimators $\kappa$, and the number of subsamples $B$. 
First, by minimising the variance of the estimator, the authors suggest the lower bound $a= \bar{d} - \kappa$, where $\bar{d}$ is an approximation of the difference $d$. 
Next, they set $\kappa$ and $B$ by minimising a measure of computational time (CT). 
This quantity describes the cost required to produce an estimator of equivalent precision as that based on a single Monte Carlo draw from the target %\citep[see Section 4.3]{quiroz2016block} 
and is derived under a normality assumption for $\hat{d}_B$. The measure balances the inefficiency of the MCMC (IF), the computational cost of the likelihood evaluation, and the probability of a negative likelihood estimate and is defined as:
\begin{equation}
\text{CT}= (NM+\kappa BM^2 +M^3)\frac{\text{IF}}{ (2\tau -1)^2}.
\label{eq:CT_v2}
\end{equation}
%with $\tau=\Pr(\widehat{E}\geq 0)$, and
%$$\text{IF} = 1+2\E_f\left( \frac{1- \vartheta}{\vartheta}\right),  \quad \vartheta=\exp(-\beta + \omega^2/2)\Phi \left(\frac{\beta}{\omega}- \omega \right) + \Phi \left(\frac{-\beta}{\omega} \right),$$
%with $\beta:= f+ \sigma^2_{\log \abs*{\widehat{E}}}$, $\omega:=\sigma_{\log \abs*{\widehat{E}}} (1-\rho^2)^{1/2}$ and $f \sim \Norm( \tfrac{1}{2}\sigma^2_{\log \abs*{\widehat{E}}},  \sigma^2_{\log \abs*{\widehat{E}}})$, where $\sigma^2_{\log \abs*{\widehat{E}}}$ denotes the variance of the logarithm of $\abs*{\widehat{E}}$ and $\rho$ the induced correlation, which is controlled by the number of blocks. 
Expressions for IF and $\tau$ are provided in \citet{quiroz2016block}, and approximations can be obtained given the values of $\kappa$, $B$, $\gamma$, and $\rho$. 
We note that $\text{IF}$ and $\tau$, and in general quantities such as  $\gamma$ and $E$, are conditioned on the values of the state, but to simplify notation we suppress this dependency.

%\SW{revise this. describe what CT is. We follow the guidelines of \citet{quiroz2016block} to optimally select the number of subsamples by minimizing a measure of the computational time which balances 1) inefficiency of the MCMC (IF), 2) computational cost of the likelihood evaluation (for us this is $NM + M^3+ \kappa B M^2$, and 3) the probability of a negative estimate. }

%\subsubsection{Tuning algorithmic parameters}

%Following the guidelines of \citet{quiroz2016block}, we present in
Algorithm~\ref{algo:tuningpar_v2} details the steps required to obtain the optimal tuning parameters $(\bar{d},\kappa,B)$. %needed for the S-BP-PM sampler. 
At the first step, in order to obtain an approximation of the difference $\bar{d}$ and an estimate of $\gamma$,
%the maximum variance of the estimator $\gamma_{max}$,, 
we run a pilot MCMC  to generate a small number of approximate samples, $S$, from the posterior of interest based on a subsample. To do so, one can employ %an MCMC  sampler that uses $\bar{\nu}_n$ 
a Taylor expansion or Gauss-Hermite quadrature to approximate the intractable likelihood. %expectation in Eq.~\eqref{eq:target}. %In our case, we employ the already implemented MCMCM  algorithm based on Gauss-Hermite quadrature (see Chapter~\ref{Chap6}).
%Note also that the pilot MCMC  scheme can be run on a small subset of the data to speed up computations. 
We employ a conservative estimate of $\gamma$ that is the maximum of the estimated $\hat{\gamma}^{(s)}$ across the pilot MCMC  draws. Next,  the optimal $\kappa$ is obtained based on a grid search to minimise the CT in Eq~\eqref{eq:CT_v2}, computed based on our estimate of $\gamma$ and for a fixed value of subsamples $B$. In general, the number of subsamples $B$ can also be optimized, but \citet{quiroz2016block} instead recommend setting $B=30$, thus dedicating more computational resources to more batches over larger batch sizes. 

\begin{algorithm}[!tp]
	\small
	\caption{Optimal tuning parameters for S-BP-PM}\label{algo:tuningpar_v2}
	\begin{algorithmic}[1]
		\Require Target distribution: $q(\btxi,\brho,\bphi)$, small number of samples: $S$, subsample size: $B^\prime$,  batch size $B$, a grid of $\kappa$ values,  assumption: $\widehat{d}_B^\newexp \iidsim \Norm (d, \gamma/B)$. \vspace{1mm}
		\State Run pilot MCMC with approximated likelihood (e.g. Gauss-Hermite or Taylor) to produce $S$ samples from $q(\btxi,\brho,\bphi)$ based on a subsample. \vspace{1mm}
		\For{ $s=1, \ldots S$} 
		\For{ $b=1, \ldots B^\prime$}			
		\State Sample $\alpha_b \sim \Unif(1,\dots N)$ 
		\State Sample  $z_{\alpha_b} \sim  \Norm( c_{\alpha_b},  w_{\alpha_b}^2)$ where \vspace{-3mm}
		\begin{equation*} 
		\begin{split}%
		c_{\alpha_b} &= \bC_{z_{\alpha_b}, \btz}(\bC_{\btz,\btz})^{-1}  L(\bphi^{(s)})\btxi^{(s)}, \\
		w^2_{\alpha_b}  &=  \bC_{z_{\alpha_b}, z_{\alpha_b}}-   \bC_{z_{\alpha_b}, \btz}(\bC_{\btz,\btz})^{-1} \bC_{\btz,z_{\alpha_b}}
		\end{split}%
		\end{equation*} \vspace{-5.5mm} 
		\State Compute %\vspace{-2.5mm} 
		$ \nu_{\alpha_b}^{(s)}(z_{\alpha_b})= l (y_{\alpha_b} \mid c_{\alpha_b}, \brho^{(s)})+(z_{\alpha_b}-c_{\alpha_b})  l^{\prime}(y_{\alpha_b} \mid c_{\alpha_b}, \brho^{(s)})$ %\vspace{-5.4mm}
		\State Evaluate: 	
		${\widehat{d}_{\alpha_b}}^{(s)}= l(y_{\alpha_b} \mid z_{\alpha_b},  \brho^{(s)})-\nu_{\alpha_b}^{(s)}(z_{\alpha_b})$ \vspace{-.7mm}	
		\EndFor	 	\vspace{1mm}
		\State Compute: $\widehat{\sigma^2}^{(s)} =\frac{1}{B^\prime-1} \sum_{b=1}^{B^\prime} \left(  {\widehat{d}_{\alpha_b}}^{(s)} - \frac{1}{B^\prime}  \sum_{b=1}^{B^\prime}  {\widehat{d}_{\alpha_b}}^{(s)} \right)^2$	\vspace{1mm}
		\State Compute: $\widehat{\gamma}^{(s)} = N^2\widehat{\sigma^2}^{(s)}$ \vspace{1mm}
		\State Compute $ {\widehat{d}}_{{B^\prime}}^{(s)}=\frac{N}{B^\prime} \sum_{b=1}^{B^\prime} {\widehat{d}_{\alpha_b}}^{(s)}$
		\EndFor		
		\State Set $\gamma_{max} = \max \widehat{\gamma}^{(s)}$	\vspace{1mm}
		\State Compute $\bar{d} = \frac{1}{S} \sum_{s=1}^{S}{\widehat{d}}_{{B^\prime}}^{(s)}$ \vspace{1mm}
		%	\State \textbf{return} $\gamma_{max}$ and $\bar{d}$
		\ForEach{$\kappa$}
		\State Compute: $\sigma^2_{\log \abs*{\widehat{E}}}= \kappa( v^2 +\eta^2)$ with \hspace{-1.07mm}\Comment{Approximate expectations by truncation}\vspace{-3mm} 
		\begin{equation*} 
		\begin{split}%
		v &=\log \left( \sqrt{\frac{\gamma_{max}}{B \kappa^2}} \right) +\frac{1}{2} \left( \log 2 + \E_p [ \psi^{(0)} (1/2+p)] \right),\\
		\eta^2&= \frac{1}{4} \left(  \E_p [ \psi^{(1)} (1/2+p)] + \V_p[ \psi^{(1)} (1/2+p)]\right),
		\end{split}
		\end{equation*}\vspace{-1.6mm} \hspace{2mm} where $p \sim \Pois(B\kappa^2/2\gamma_{max})$, and $\psi^{(i)}$ the polygamma function of order $i$. \vspace{1.4mm} 		
		\State Compute the probability of a positive $\hat{E}$: \Comment{$\Phi$ denotes CDF of a standard Gaussian } \vspace{-.3mm}
		$$\tau =\frac{1}{2} \left(  1+ \exp \left[ 2 \kappa \left( \Phi \left( \frac{\kappa \sqrt{B}}{ \sqrt{\gamma_{max}}} \right)-1 \right) \right]\right)$$ 
	   \IfThenElse{$\kappa<100$}{$\rho= 1- \frac{1}{\kappa}$}{$\rho= 1- \frac{1}{100}$}
		%	\State Set $\rho= 1- \frac{1}{\kappa}$
		\State Employ GH quadrature to approximate: {\Comment{ $f \sim \Norm \left(\tfrac{1}{2} \sigma^2_{\log \abs*{\widehat{E}}},  \sigma^2_{\log \abs*{\widehat{E}}} \right)$}} \vspace{-3mm}%
		\begin{equation*}%
		\hat{\E}_f \approx \E_f\left( \frac{1- \vartheta}{\vartheta}\right), \quad \vartheta=\exp(-\beta + \omega^2/2)\Phi \left(\frac{\beta}{\omega}- \omega \right) + \Phi \left(\frac{-\beta}{\omega} \right),
		\end{equation*}
		\vspace{-3mm}  \hspace*{5mm} with $\beta:= f+ \sigma^2_{\log \abs*{\widehat{E}}}$, $\omega=\sigma_{\log \abs*{\widehat{E}}} (1-\rho^2)^{1/2}$. 	\vspace{5mm}		   
		\State Compute the inefficiency: $\text{IF} = 1+ 2  \hat{\E}_f$ \vspace{1mm}
		\State Compute the computational time: \vspace{-3mm}
		$$\text{CT}^\star=(\kappa B M^2 + N M + M^3) \frac{ \text{IF}}{(2\tau -1)^2}$$
		\EndFor	
		\State \textbf{return} $\kappa$ with the minimum $\text{CT}^\star$ and $\bar{d}$.
	\end{algorithmic}
\end{algorithm}

\subsection{Inducing points}
A poor selection of the inducing points can lead to unsatisfactory posterior and predictive estimates. While a strategy based on K-means clustering of the inputs is fast and computationally cheap, \citet{hensman2015mcmc} demonstrate that significant improvements can be achieved by optimizing the inducing points. Specifically, following  \citet{hensman2015mcmc}, the inducing points can be optimized based on an initial Gaussian approximation to the posterior, prior to running the MCMC. Recently, \citet{rossi2020sparse} advocate for a fully Bayesian treatment of the inducing points in approximations based on fully independent training conditionals \citep[FITC,][]{quinonero2005}. Beyond this, \citet{uhrenholt2020probabilistic} place a point process prior on the inducing points, to learn not only the locations but also the number of inducing points. Extending and incorporating these advancements within our scheme is a promising direction of research.

To select the number of inducing points, a common approach is to increase $M$ until there is no improvement in an approximation to the marginal likelihood. Indeed, this is typically the case in variational inference when one aims to optimize a lower bound on the marginal likelihood. \cite{burt2019rates} study convergence rates of variationally sparse GP regression to provide asymptotic guidelines for the choice of $M$. For example, they show that with the squared exponential kernel, it suffices to take $M = \mathcal{O}(\log^D(N))$.
Alternatively, we discuss in Appendix~\ref{app:inducing} an heuristic approach to select the number of inducing points select the number of inducing points by optimizing a measure that combines both accuracy and computational cost.

%In this case, this measure can also provide a guide to choosing the number of inducing points \SW{add this to inducing point discussion? To choose the number of inducing points, we could consider an efficiency measure that combines the accuracy and computational cost (measure by CT). To approximate the accuracy, we consider the cheap first-order Taylor approximate of the negative predictive log density (NLPD), defined as $\text{NLPD}=\sum_{n=1}^N l(y_n \mid \E[z_n|\widehat{\btz}],\widehat{\rho})$, where we simply plug-in the posterior estimates of $(\btz, \brho,\bphi)$ based on the pilot run. The efficiency measure is then defined as NLPD/CT.}

\subsection{Posterior and predictive estimates}

Note that the block-Poisson estimator in Eq.~\eqref{eq:blockP_v2} is only unbiased without the absolute value. Therefore, MCMC draws $(\btxi,\brho, \bphi)$ do not follow $q(\btxi,\brho, \bphi)$, but rather
%\begin{equation*}
%\breve{q}(\btxi,\btzeta,\btheta, \bvarphi)=  \frac{\int \abs*{\widehat{E}} \pi(\btxi,\btzeta,\btheta, \bvarphi)  \pi(\mathcal{V}_{1}, \ldots, \mathcal{V}_G)  \,d\mathcal{V}_{1}, \ldots, \mathcal{V}_G }{\int  \abs*{\widehat{E}} \pi(\btxi,\btzeta,\btheta, \bvarphi) \pi(\mathcal{V}_{1}, \ldots, \mathcal{V}_\kappa) \,d\mathcal{V}_{1}, \ldots, \mathcal{V}_\kappa  \,d\btxi,\btzeta,\btheta, \bvarphi},
%\end{equation*}
\begin{equation*}
\breve{q}(\btxi,\brho, \bphi) \propto \int \abs*{\widehat{E}} \pi(\btxi,\brho, \bphi))  \pi(\mathcal{V}_{1}, \ldots, \mathcal{V}_G)  \,d\mathcal{V}_{1}, \ldots, \mathcal{V}_G,
\end{equation*}
where $\pi(\btxi,\brho, \bphi)$ is the prior over $(\btxi,\brho, \bphi)$ and $\pi(\mathcal{V}_{1}, \ldots, \mathcal{V}_G)$ is the  density of the auxiliary random variables involved in computation of the block-Poisson estimator. %with $\mathcal{V}_{k}= \lbrace \mathcal{H}_k, \alpha_{1:B}^{1,k},\ldots,  \alpha^{ \mathcal{H}_k,k}_{1:B}, \chi_{1:B}^{1,k},\ldots, \chi_{1:B}^{ \mathcal{H}_k,k} \rbrace$ for $k=1, \ldots, \kappa$, where ${\alpha}^{h,k}_{1:B } \coloneqq( {\alpha}^{h,k}_{1},\ldots \alpha_{B}^{h,k} )$ and $\chi^{h,k}_{1:B}\coloneqq( \chi_{1}^{h,k},\ldots \chi_{B}^{h,k} )$. 
Nevertheless, one can estimate expectations with respect to  $q(\btxi,\brho, \bphi)$ by taking the MCMC draws from $\breve{q}(\btxi,\brho, \bphi)$ and applying an importance sampling step that corrects for the sign of the estimator \citep{lyne2015russian}.  For instance, assume one wants to compute the expectation of some function $g(\Psi)$ depending on a subset $\Psi$ of the parameters $(\brho,\btxi,\bphi)$. Then,
$$\E_{q}[ g(\Psi)] = \frac{\E_{\breve{q}}[ g(\Psi)  \mathcal{S}]}{ \E_{\breve{q} }[ \mathcal{S}] },$$
with $\mathcal{S}= \sign(\widehat{E}).$ Thus, one can estimate $\E_{q}[ g(\Psi)]$ through
$$\E_{q}[ g(\Psi)] \approx \frac{\sum_{t=1}^{T} \sum_{j=1}^J  \mathcal{S}_{j}^{(t)} g(\Psi_j^{(t)})}{\sum_{t=1}^{T} \sum_{j=1}^J \mathcal{S}_{j}^{(t)} } ,$$
%g (\bphi^{(t)},\btxi^{(t-1 +\mathbbm{1}(j>1))},\brho^{((t-1 +\mathbbm{1}(j>2)))})
where $ \Psi_j^{(t)}$ %= (\bphi^{(t)},\btxi^{(t-1 +\mathbbm{1}(j>1))},\brho^{((t-1 +\mathbbm{1}(j>2)))})$ 
and $\mathcal{S}_{j}^{(t)}$ denote the state and sign, respectively, at iteration $t$ when updating the $j$th parameter of $\Psi$ in the corresponding Gibbs step. 
%Note that this requires to store the sign of $\widehat{E}$ at each iteration $t= 1, \ldots, T$ of the MCMC procedure.
%Moreover, it is possible to obtain uncertainty estimates by evaluating the density of over a grid. 
As an example, the prediction of the unknown function at a location $x^*$ is given by $ \mathbb{E}[ z^* | \mathcal{D}] = \E_{q}[ \mathbb{E}[ z^* |\btxi, \bphi]]$, and thus can be computed by setting  $g(\btxi, \bphi) = \mathbb{E}[ z^* |\btxi, \bphi]$ to be the GP prior predictive mean.
%\bC_{z^*, \btz}(\bC_{\btz,\btz})^{-1}  L(\bphi^{(s)})\btxi^{(s)}

In addition, while credible intervals reflecting uncertainty cannot be directly computed from the MCMC  output, one can estimate the posterior probability that the parameters belong to a specified region, which in turn can be used to compute credible intervals. For example, considering  $z^*$, let $g(\btxi, \bphi) = F_{z^*}(z \mid \btxi, \bphi)$ be the CDF of the Gaussian prior predictive of $z^*$ evaluated at any $z \in \mathbb{R}$, then
$$ \Pr(z^* \leq z \mid  \mathcal{D})  \approx \frac{\sum_{t=1}^{T}  \mathcal{S}_{\btxi}^{(t)} F_{z^*}(z | \btxi^{(t)}, \bphi^{(t-1)})+ \mathcal{S}_{\bphi}^{(t)} F_{z^*}(z | \btxi^{(t)}, \bphi^{(t)})}{\sum_{t=1}^{T}  \mathcal{S}_{\btxi}^{(t)} +\mathcal{S}_{\bphi}^{(t)} } .$$
%= \E_q[ F_{z^*}(z | \btxi, \bphi)]
%\frac{\sum_{t=1}^T \mathcal{S}^{(t)} F_{z_n}(z | \btxi^{(t)}, \bphi^{(t)}) }{ \sum_{t=1}^T \mathcal{S}^{(t)}}.$$
By evaluating this over a grid of $z$ values, one can then compute credible intervals from the CDF.
%to set such lower bound to be $a= \bar{d} - \kappa$, where $\bar{d}$ is an approximation of the difference $d$ a
%We describe a strategy to compute $\gamma_{max}$ and $\bar{d}$ in the following section.

\section{Motivating model: two-level non-stationary GP regression} \label{section:motivation}
Many statistical and machine learning tasks focus on parametric, stationary covariance functions, largely due to computational convenience. However, this assumption is rarely realistic in practice, and as a consequence, various approaches exist to account for non-stationarity \citep[e.g.][]{kim2005analyzing,montagna2016computer,volodina2020diagnostics}. We focus on the 
%A 2-level non-stationary Gaussian process prior for $z(\cdot)$ can be constructed when utilising 
the family of non-stationary covariance functions introduced by \citet{paciorek2006},
\begin{align}
C^{\text{NS}}_{{\bphi}}(\bx_n,\bx_{n'})=\tau^2_z\frac{|\Sigma(\bx_n)|^{\frac{1}{4}}|\Sigma(\bx_{n'})|^{\frac{1}{4}}}{|(\Sigma(\bx_n)+\Sigma(\bx_{n'}))/2|^{\frac{1}{2}}} R_\psi\left( Q_{n\,n'} \right),
\label{eq:non-stat-family}
\end{align}
with $Q_{n\,n'}= \sqrt{ (\bx_n-\bx_{n'})^T\left((\Sigma(\bx_n)+\Sigma(\bx_{n'}))/2\right)^{-1}(\bx_n-\bx_{n'}) }$, $R_\psi$ a stationary correlation function on $\mathbb{R}$, and $\Sigma(\cdot)$ a $D \times D$ spatially varying covariance matrix, referred to as the kernel matrix. Thus, the parameters ${\bphi}$ of this  nonparametric covariance function consist of the magnitude, $\tau^2_z$; the spatially varying covariance matrices, $\Sigma(\cdot)$; and any additional parameters $\psi$ of the stationary correlation function $R_\psi$. 
Importantly, non-stationarity is introduced %in Eq.~\eqref{eq:non-stat-family} 
through the kernel matrices, and these parameters must be inferred at every observed location. To control the number of parameters, one can place assumptions on the kernel matrix and type of non-stationary present. Here, we focus on a non-stationary isotropic covariance function, which is obtained by assuming the kernel matrices are scaled identity matrices such that $\Sigma(\cdot)= \ell^2(\cdot) I_D$. In this case, we can allow the spatially varying length-scale  to be a function of the full $D$-dimensional input, with $\mathcal{O}(D)$ hyperparameters.
Therefore, the kernel can be written as,
\begin{align}
C^{\text{NS}}_{{\bphi}}(\bx_n,\bx_{n'})=\tau^2_z \frac{\ell(\bx_n)^{\frac{D}{2}}\ell(\bx_{n'})^{\frac{D}{2}}}{\left([\ell^2(\bx_n)+\ell^2(\bx_{n'})]/2\right)^{\frac{D}{2}}} R_\psi\left( \sqrt{  \frac{   \sum_{d=1}^D   (x_{n\,d}-x_{n'd})^2}{[\ell^2(\bx_n)+\ell^2(\bx_{n'})]/2} } \right).
\label{eq:non_stat_iso}
\end{align}

In the two-level non-stationary GP model, we model the log transformed length-scale process with a stationary GP prior; 
%assume a zero-centred non-stationary Gaussian process prior on the unknown function $z(\cdot)$ with the non-stationary covariance functions in \eqref{eq:non_stat_iso}. 
%Thus, a single GP is required to parametrise the kernel matrices. Specifically, we use a constant mean stationary GP prior over the log transformed length-scale process; 
that is, $u(\cdot):= \log(\ell(\cdot))\sim \GP(\mu_u, \CS_{\bvarphi}(\cdot,\cdot))$. This provides a flexible model, where information can be borrowed across neighboring observations to learn the local length scales at each location. 
While we focus on the two-level formulation, we highlight that deeper constructions with multiple levels of GPs can provide even more flexible non-stationary behavior \citep{dunlop2017deep}.  
%In addition, the model assumes independence across all the parameters, those in the non-stationary covariance, ${\bphi}$, and the noise variance, $\sigma^2_\varepsilon$. The joint prior over the parameters and hyperparameters is $ \pi(\sigma^2_\varepsilon,\tau^2_z, \psi, u(\cdot), \bvarphi)= \pi(\sigma^2_\varepsilon)\pi(\tau^2_z)\pi(\psi)\pi(u(\cdot) | \bvarphi)\pi(\bvarphi)$. Consequently, 
The hierarchical formulation of the two-level regression model is:
\begin{equation}
\begin{split}
y_n  &\sim \Norm(z(\bx_n),\sigma^2_\varepsilon), \quad  n=1, \ldots N \\
z( \cdot )  &\sim \GP(0, C^{\text{ \tiny{NS}}}_{{\bphi}}) \\
u(\cdot) := \log(\ell(\cdot)) &\sim \GP(\mu_u, \CS_{\bvarphi}(\cdot,\cdot)) \\
(\sigma^2_\varepsilon, \tau^2_z, \psi, \bvarphi ) &\sim  \pi(\sigma^2_\varepsilon) \pi( \tau^2_z) \pi(\psi)\pi( \bvarphi).
\end{split} \label{eq:hierarchical}
\end{equation}
The sparse variational strategy is well-suited for this model %in Eq.~\eqref{eq:hierarchical} 
because the complexity, in both the number of parameters and high correlation among them, makes (i) standard mean-field variational methods, which make strong independence assumptions, unsuitable and (ii) MCMC inference over the true posterior computationally expensive and challenging \citep{monterrubio2020}.

\subsection{Variationally sparse two-level GP regression}
We now derive the optimal sparse variational posterior for the two-level GP regression model. 
%Following the model in Eq.~\eqref{eq:hierarchical}, the posterior of interest is
%\begin{equation*}
%\begin{split}
%\pi(\bz,\bu, \sigma^2_\varepsilon, \tau^2_z,\psi, \bvarphi \mid \by,X) \propto \Norm(\by\mid \bz,\sigma^2_\varepsilon I_N) \pi(\bz\mid \bu, \tau^2_z, \psi) \pi(\sigma^2_\varepsilon)\pi(\tau^2_z)\\\pi(\psi)\pi(\bu \mid \bvarphi) \pi(\bvarphi),
%\end{split}
%\end{equation*}
%where $\bu=(u_1, \ldots, u_N)^T$ with $ u_n \equiv u(\bx_n)$, and $\bz=(z_1, \ldots, z_N)^T$.
%Following the sparse framework of \citet{snelson2006sparse}, 
First, we augment the model with the inducing points $ \btX=(\btx_1,\ldots,\btx_M)^T$ %with $M \ll N$, 
and, in addition to the inducing variables $\btz =(   z(\btx_1),\ldots,z(\btx_M))^T$, we collect the values of the log length-scale process at the inducing points into the vector $\btu= (u(\btx_1),\ldots,u(\btx_M))^T$. 
%we collect the values of the latent functions at the inducing points into the vector $\btz =(   z(\btx_1),\ldots,z(\btx_M))^T$. Additionally, we define the latent vector $\btu= (u(\btx_1),\ldots,u(\btx_M))^T$ collecting the values of the log length-scale process at the inducing locations. 
Letting $\bu=(u_1, \ldots, u_N)^T$ denote the the log length-scale process at the observed locations, the joint prior is:
%\[ \pi(\bu,\btu \mid \bvarphi) =  \pi( \bu \mid \btu, \bvarphi) \pi( \btu \mid \bvarphi)  =  \Norm(\bu \mid    \bm{\mu}_u^\mystarb, \Omega_u ) \Norm(\btu \mid \bm{\mu}_{u}, \CS_{\btu,\btu}),  \]
\[ \pi(\bu,\btu \mid \bvarphi)  =  \Norm(\bu \mid    \bm{\mu}_u^\mystarb, \bm{\Omega}_u ) \Norm(\btu \mid \bm{\mu}_{u}, \bCS_{\btu,\btu}),  \]
where $\bm{\mu}_u$ denotes an $M$-dimensional vector with entries $\mu_u$, $\bm{\mu}^\mystarb_u= \bm{\mu}_u + \bCS_{\bu, \btu} {(\bCS_{\btu,\btu})}^{-1} (\btu - \bm{\mu}_u ) $, and $\bm{\Omega}_u=\bCS_{\bu, \bu}- \bCS_{\bu, \btu} (\bCS_{\btu,\btu})^{-1} \bCS_{\btu,\bu}$. Additionally, 
\begin{equation*}
\begin{split}
\pi(\bz, \btz \mid \bu,\btu, \tau^2_z, \psi) %&=\pi(\bz \mid \btz, \bu,\btu, \tau^2_z, \psi) \pi(\btz \mid \btu, \tau^2_z, \psi) \\&
= \Norm(\bz \mid  \bm{\mu}_z^*, \bm{\Omega}_z) \Norm(\btz \mid 0 , \bCNS_{\btz,\btz}),
\end{split}
\end{equation*}
with   $\bm{\mu}_z^\mystarb= \bCNS_{\bz,\btz} (\bCNS_{\btz,\btz})^{-1}\btz$ and $\bm{\Omega}_z=\bCNS_{\bz,\bz}- \bCNS_{\bz,\btz} (\bCNS_{\btz,\btz})^{-1} \bCNS_{\btz,\bz}$.
%After augmentation, the posterior of the latent variables, parameters and hyperparameters is given by:
%\begin{equation*}
%\begin{split}
%\pi(\bz,\btz, \bu,  \btu , \btheta, \bvarphi \mid \by,X, \tilde{X}) \propto \Norm(\by \mid \bz,\sigma^2_\varepsilon I_N) \pi(\bz\mid \btz, \bu, \btu, \tau^2_z, \psi) \pi(\btz \mid \btu, \tau^2_z, \psi) \\ \pi(\btheta)\pi(\bu \mid \btu, \bvarphi) \pi(\btu \mid \bvarphi) \pi(\bvarphi),
%\end{split}
%\end{equation*}
%where we define $\btheta \coloneqq \{\sigma^2_\varepsilon , \tau^2_z, \psi \}$. 
The approximate variational posterior is assumed to take the form:
\[  q(\bz,\btz, \bu, \btu, \btheta) \propto \pi(\bz \mid \btz, \bu, \btu, \btheta) \pi(\bu \mid \btu, \btheta) q(\btz, \btu, \btheta),\]
where $\btheta \coloneqq \{\sigma^2_\varepsilon , \tau^2_z, \psi, \bvarphi \}$. %\SW{note: I redefined $\btheta$ to include also $\bvarphi$ to hopefully simplify notation...}  
%We then seek the variational posterior distribution which 
By minimizing the KL divergence (see Appendix A), we obtain the optimal variational posterior:
\begin{equation}
\begin{split}
q(\btz, \btu,\btheta) \propto \exp\left( \sum_{n=1}^N \E_{(z_n,u_n)}[ \log(\Norm(y_n \mid z_n,\sigma^2_\varepsilon)) ] \right) \pi(\btz \mid \btu, \btheta)  \pi(\btu \mid \btheta)  \pi(\btheta), \label{eq:varposterior_iso}
\end{split}
\end{equation}
where the expectation is taken with respect to $\pi(z_n,u_n| \btz,\btu,\btheta)$.
%The expected log-likelihood only contains intractable univariate integrals, which are expectations with respect to univariate Gaussian random variables for any kernel employed.
The elements of the expected log-likelihood in Eq.~\eqref{eq:varposterior_iso} can be written as $$\E_{(z_n,u_n)}[ \log(\Norm(y_n \mid z_n,\sigma^2_\varepsilon)) ]  =  \E_{u_n}[l(y_n|u_n,\btz,\btheta)],$$
where we define 
\begin{align*}
&\l(y_n|u_n,\btz,\btheta) = \E_{z_n}\left[ \log(\Norm(y_n \mid z_n,\sigma^2_\varepsilon)) \mid u_n\right] \\
&= -\frac{1}{2} \log(2\pi \sigma_\varepsilon^{2}) - \frac{1}{2\sigma^2_\varepsilon} \left[(y_n-\E_{z_n}[z_n|u_n])^2 + V_{z_n}(z_n|u_n)\right]\\
&= -\frac{1}{2} \log(2\pi \sigma_\varepsilon^{2}) - \frac{1}{2\sigma^2_\varepsilon} \left[ (y_n-\bCNS_{z_n, \btz} (\bCNS_{\btz,\btz})^{-1} 	\btz)^2 + \tau^2_z- \bCNS_{z_n, \btz} (\bCNS_{\btz, \btz})^{-1} \bCNS_{ \btz,z_n} \right].
\end{align*}
%with conditional expectation given by $E_{z_n}[z_n|u_n] =\bCNS_{z_n, \btz} (\bCNS_{\btz,\btz})^{-1} 	\btz$ and the conditional variance given by
%$$\E_{u_n}[ \V_{z_n}(z_n|u_n)] = \tau^2_z- \bCNS_{z_n, \btz} (\bCNS_{\btz, \btz})^{-1} \bCNS_{ \btz,z_n}.$$
We intend to explore the variational posterior in Eq.~\eqref{eq:varposterior_iso} using MCMC, however the required expectations of $l(y_n|u_n,\btz,\btheta,)$ with respect to $\pi(u_n | \btu,\btheta)$ are intractable. To overcome this, in the following subsections, we first present and discuss the limitations of a Gauss-Hermite approach and then describe the PM scheme. 

\subsection{Gauss-Hermite approach}\label{sec:GH}
%\SW{For ease of exposition, let us suppose we are interested in computing
%$$ \mathbb{E}[f(v)]=\int f(v) \Norm(v \mid \mu_v, \sigma^2_v) dv =\frac{1}{\sqrt{\pi}} \int g(h) \exp(-h^2)dh,$$
%where $g(h)=f(\sqrt{2} \sigma_v h+\mu_v)$.
%%consider the case of $\bm{\beta}_n$, where we are interested in computing,
%%\[\E_{ \pi(u_{n} | \btu,\Theta, \bvarphi)} \left[ \CNS_{z_n, \btz}\right]= \int \E_{ \pi(u_{n} | \btu,\Theta, \bvarphi)} \left[ \CNS_{z_n, \btz}\right]  \Norm(u_n \mid  \mu_u + \CS_{u_n, u_n} {(\CS_{\btu,\btu})}^{-1} (\btu - \bm{\mu}_u )  ) du_n ,\] 
%%
%%$\bm{\mu}^*_u= \bm{\mu}_u + \CS_{\bu, \btu} {(\CS_{\btu,\btu})}^{-1} (\btu - \bm{\mu}_u ) $, and $\Omega_u=\CS_{\bu, \bu}- \CS_{\bu, \btu} (\CS_{\btu,\btu})^{-1} \CS_{\btu,\bu}$
%% $u_n \sim \Norm( \mu_{u}^*)$
%%Let $z \sim \Norm(\mu_z, \sigma^2_z)$, and suppose we are interested in computing
%Gauss-Hermite quadrature of order $J$  approximates this by 
%$$ \mathbb{E}[f(v)] \approx \frac{1}{\sqrt{\pi}} \sum_{j=1}^J  w_j g(h_j),$$
%where $h_j$ (for $j=1,\ldots,J$) are the roots of the $J^{\text{th}}$ order Hermite polynomial $H_J(h)$, and 
%$$w_j= \frac{ \sqrt{\pi}2^{J-1} J! } { (J H_{J-1}(h_j))^2}.$$
%The error of the quadrature approximation is given by
%$$ \varepsilon(f) =\frac{J!}{(2J)! 2^J}g^{(2J)}( \xi),$$ for some $\xi$. Thus, Gauss-Hermite quadrature  provides a good approximation if $g$ is close to a polynomial of order $2J-1$.
%In practice, the weight and nodes are computed with numerical algorithms \citep[e.g.][]{stroud1966gaussian, rutishauser1962modification}.}

Following \citet{hensman2015mcmc}, the intractable expectations can be approximated with Gaussian-Hermite quadrature. In this case, to improve the approximation and mixing, we rewrite the expected log-likelihood as:
\begin{align*}
    \E_{u_n}[l(y_n|u_n,\btz,\btheta)] &=  -\frac{1}{2} \log(2\pi \sigma_\varepsilon^{2}) - \frac{1}{2 \sigma_\varepsilon^2} \left[   (y_n- \boldsymbol{\beta}_n(\CNS_{\btz,\btz})^{-1} \btz)^2 +\tau^2_z - \alpha_n  \right] \\
    &\quad - \frac{1}{2 \sigma_\varepsilon^2} \left[ \btz^T (\CNS_{\btz, \btz})^{-1} (\bm{P}_n - \boldsymbol{\beta}_n^T\boldsymbol{\beta}_n)(\CNS_{\btz, \btz})^{-1}\btz \right],
\end{align*}
where we denote the required intractable expectations by
\begin{equation}%
\begin{split}%
\boldsymbol{\beta}_n &=  \E_{u_{n} } \left[ \bCNS_{z_n, \btz}\right],\\
%\alpha_n&=  \E_{u_{n}} \left[\bCNS_{z_n, \btz} (\bCNS_{\btz, \btz})^{-1} \bCNS_{ \btz,z_n}\right],\\
\bm{P}_n &= \E_{u_{n} } \left[ \bCNS_{ \btz,z_n} \bCNS_{z_n, \btz} \right],
\end{split} \label{eq:expectations_iso}
\end{equation}
and define $$\alpha_n=  \E_{u_{n}} \left[\bCNS_{z_n, \btz} (\bCNS_{\btz, \btz})^{-1} \bCNS_{ \btz,z_n}\right] = \sum_{i,j=1}^{M} \left( (\CNS_{\btz, \btz})^{-1}   \odot \bm{P}_n \right)_{ij},$$
with $\odot$ denoting the Hadamard product. 
%Note that $\sum_{n=1}^N \alpha_n$ can be rewritten as $\sum_{i,j=1}^{M} \left( (\CNS_{\btz, \btz})^{-1}   \odot \bm{P} \right)_{ij}$ to avoid the computation of $N$ expectations. 
Importantly, this allows us to marginalise
%We intend to explore the variational posterior in Eq.\eqref{eq:varposterior2_iso} using MCMC, but high correlations between $\btz$ and $(\btu, \btheta, \brho)$  will result in poor mixing in a Gibbs sampling framework. However, 
the latent variables $\btz$. Furthermore, we employ whitening to break the correlation between $\btu$ and $\bvarphi$, defining $\btu=L(\bvarphi)\btzeta+ \bm{\mu}_u$,  where $\btzeta \sim \Norm (0, I_M)$ and ${L(\bvarphi)}{L(\bvarphi)}^T=\CS_{\btu,\btu}$.  Consequently, our MCMC scheme aims to simulate from the whitened marginal variational posterior,
	\begin{align}
	\begin{split} \label{eq:Isopost}
	q(\btzeta,\btheta) &\propto  \Norm(\btzeta \mid 0,I_M)  \pi(\btheta) \sigma_\varepsilon^{-N} \exp \left( -\frac{1}{2\sigma_\varepsilon^2} \sum_{n=1}^N  [y_n^2+\tau^2_z -\alpha_n]  \right) \times \\
	& \left[ \frac{|\CNS_{\btz, \btz} |^{\frac{1}{2}}}{| \CNS_{\btz, \btz}+\sigma^{-2}_\varepsilon \bm{P}|^{\frac{1}{2}} }  \exp \left( \frac{1}{2\sigma_\varepsilon^4} \by^TB\left( \CNS_{\btz,\btz} +\sigma^{-2}_\varepsilon\bm{P}\right)^{-1} B^T \by  \right)
	\right]_{_{\btu}},
	\end{split} 
	\end{align}
where $\btu$ is replaced by $L(\bvarphi)\btzeta + \bm{\mu}_u$ in the expression in brackets; $B$ an $N \times M$ matrix with rows $\bm{\beta}_n$; and $\bm{P}= \sum_{n=1}^N \bm{P}_n$.  
Furthermore, when required,  we can sample $\btz$ from its conditional variational posterior:%
\begin{equation}
\btz \mid \btu, \btheta  \sim \Norm \left(  \sigma^{-2}_\varepsilon\CNS_{\btz,\btz} \left(\CNS_{\btz,\btz}+\sigma^{-2}_\varepsilon\bm{P} \right)^{-1} B^T\by,\CNS_{\btz, \btz}\left( \CNS_{\btz,\btz} + \sigma^{-2}_\varepsilon\bm{P} \right)^{-1} \CNS_{\btz,\btz}\right). \label{eq:con_post_iso}
\end{equation}

Full details of the algorithm employed are given in the Supplementary Material. At each step, the cost of single likelihood evaluation is dominated by approximation of the intractable integrals, namely, the elements of $\bm{\beta}_n$ and $\bm{P}_n$ for each data point $n=1,\ldots,N$. This results in a computational complexity of 
 $\mathcal{O}(JNM^2)$  per expected log-likelihood evaluation, in contrast to $\mathcal{O}(N^3)$ for evaluation of the true likelihood. %We highlight that the computations  in  \algrefto{algo:GHQ}{step:par3}{step:par4}, can be done in parallel, leading to further computational gains.

\begin{figure}[!t]
	\centering
	\subcaptionbox{\scriptsize{ $\delta=0.0006$, $J=4$}}{{\includegraphics[scale=.39]{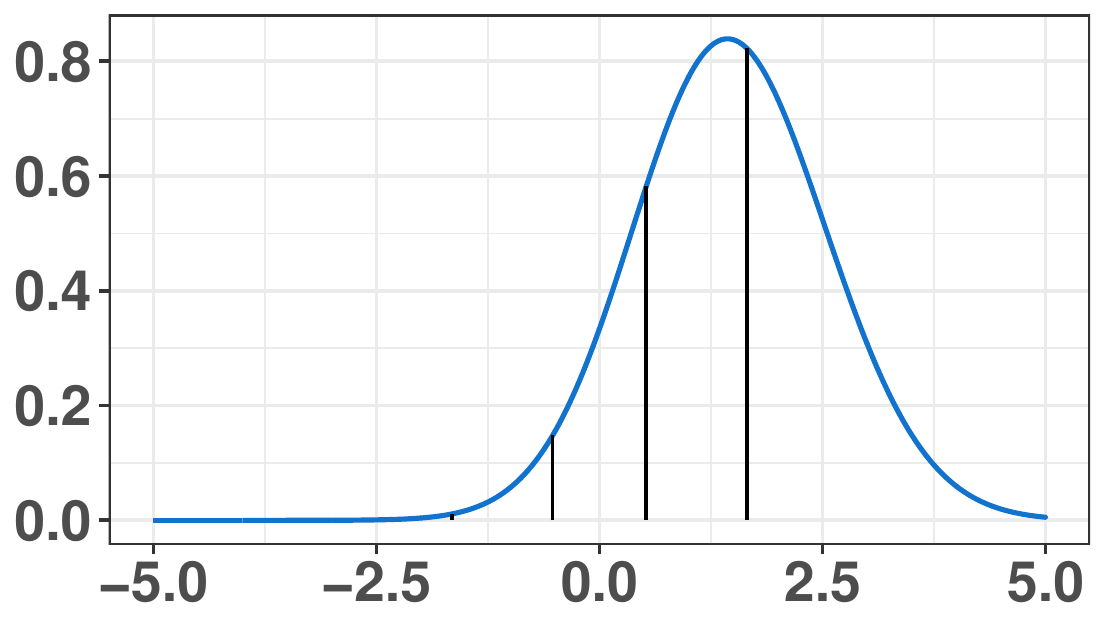}}}
	\subcaptionbox{\scriptsize{$\delta=0.0006$, $J=10$}}{{\includegraphics[scale=.39]{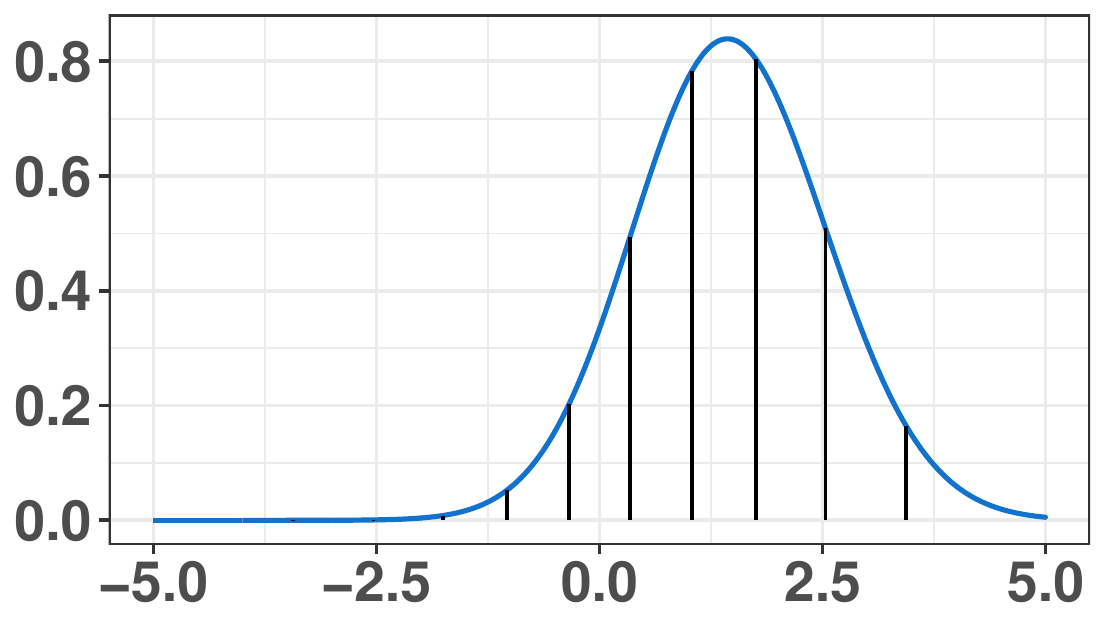}}}
	\subcaptionbox{\scriptsize{$\delta=0.0006$, $J=31$}}{{\includegraphics[scale=.39]{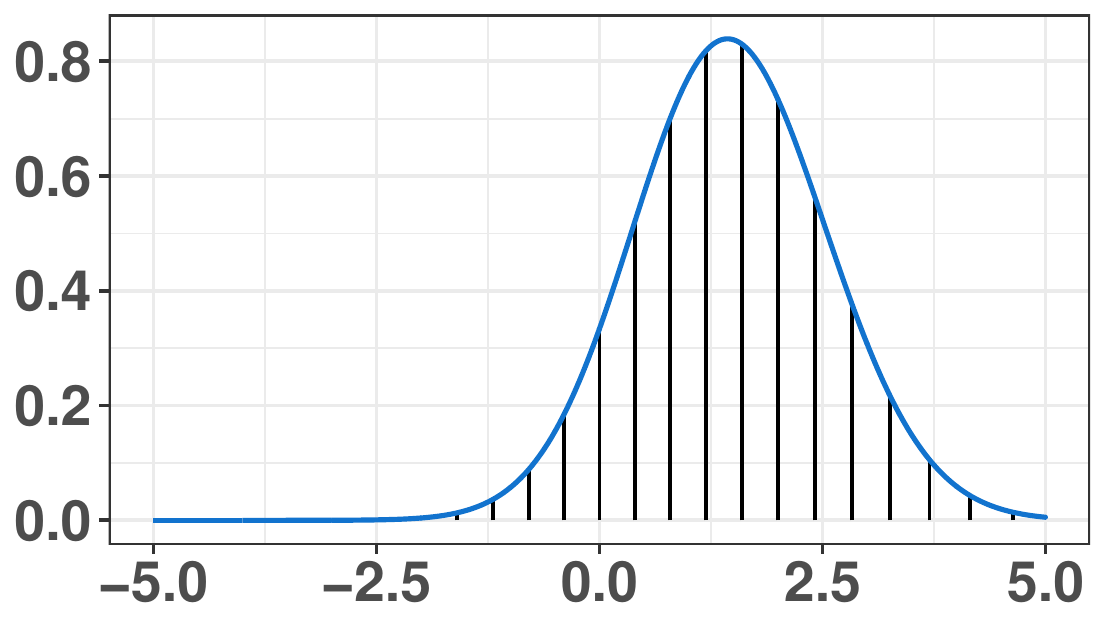}}}\\
	\subcaptionbox{\scriptsize{$\delta=0.2871$, $J=4$}}{{\includegraphics[scale=.39]{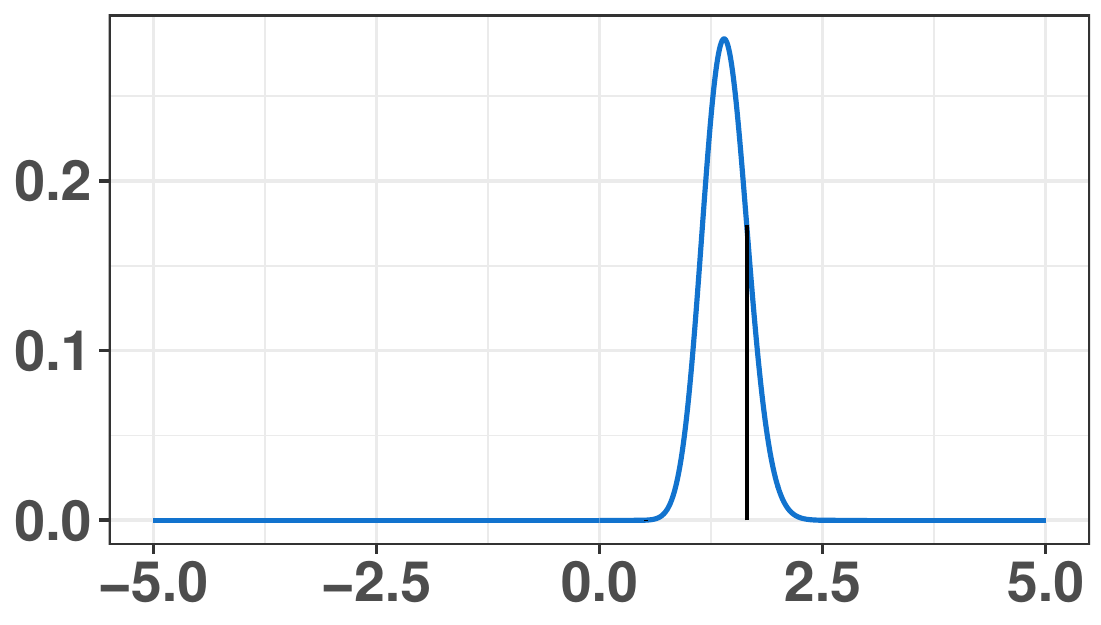}}}
	\subcaptionbox{\scriptsize{$\delta=0.2871$, $J=10$}}{{\includegraphics[scale=.39]{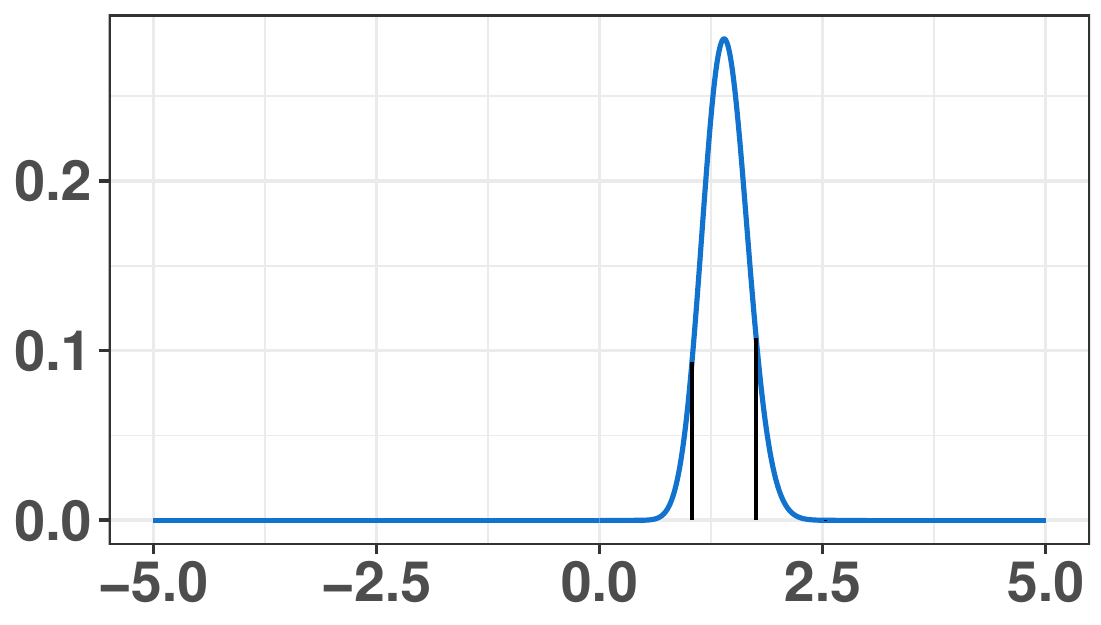}}}
	\subcaptionbox{\scriptsize{$\delta=0.2871$, $J=31$}}{{\includegraphics[scale=.39]{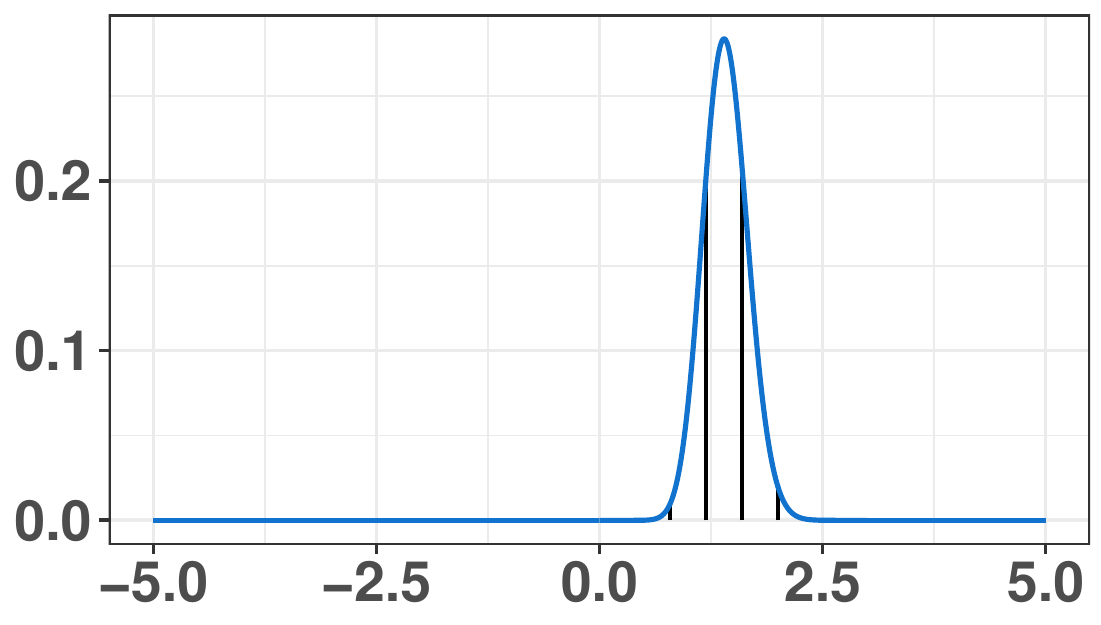}}}\\
	\subcaptionbox{\scriptsize{$\delta=0.9890$, $J=4$}}{{\includegraphics[scale=.39]{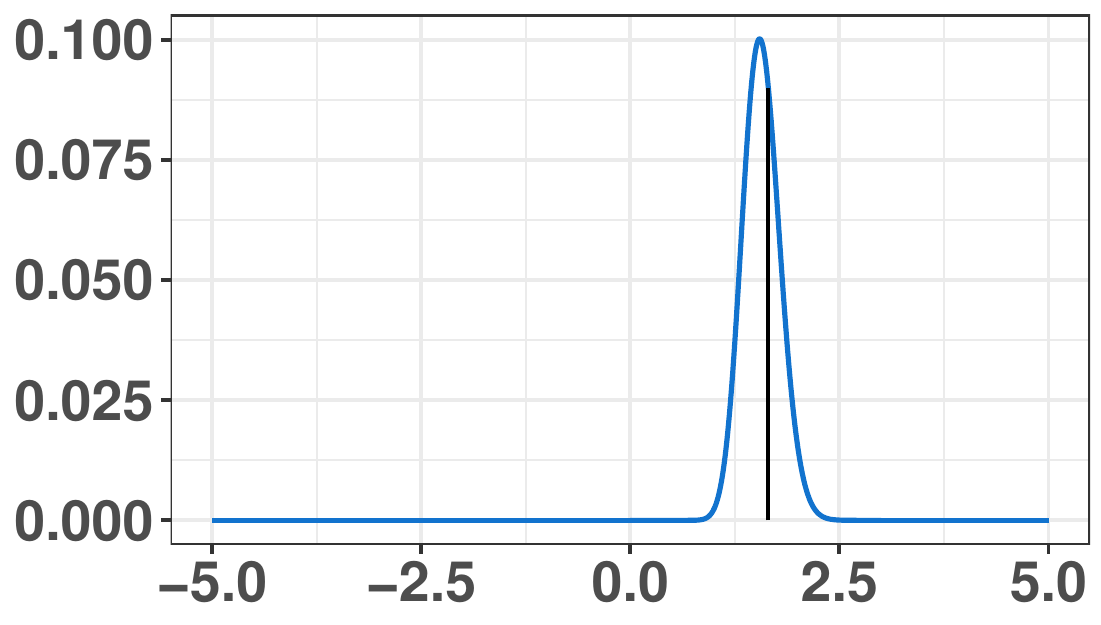}}}
	\subcaptionbox{\scriptsize{$\delta=0.9890$, $J=10$}}{{\includegraphics[scale=.39]{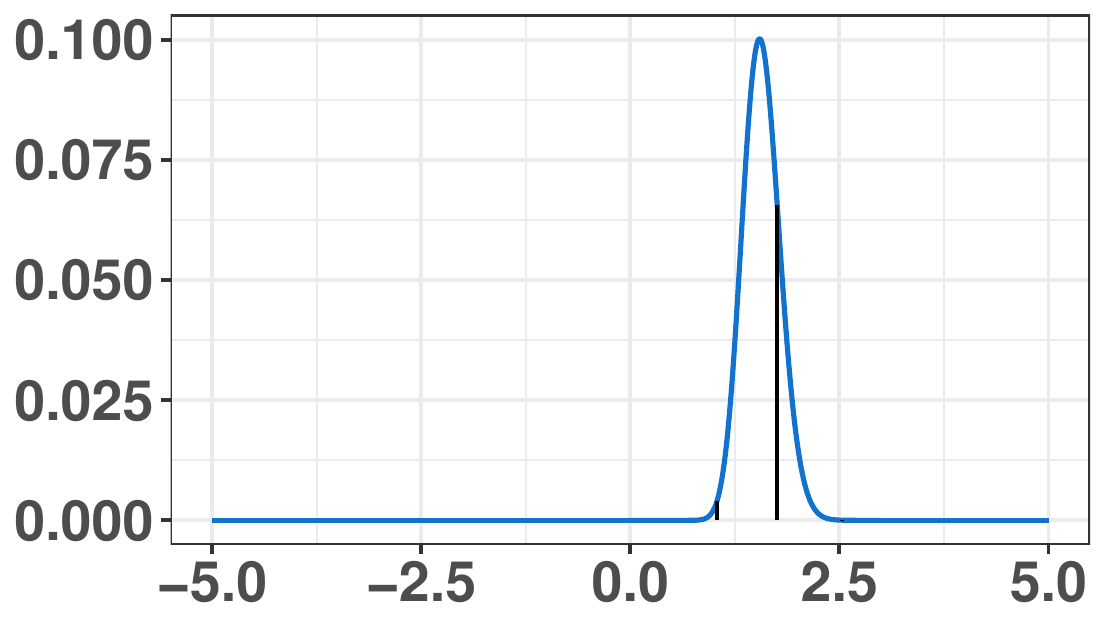}}}
	\subcaptionbox{\scriptsize{$\delta=0.9890$, $J=31$}}{{\includegraphics[scale=.39]{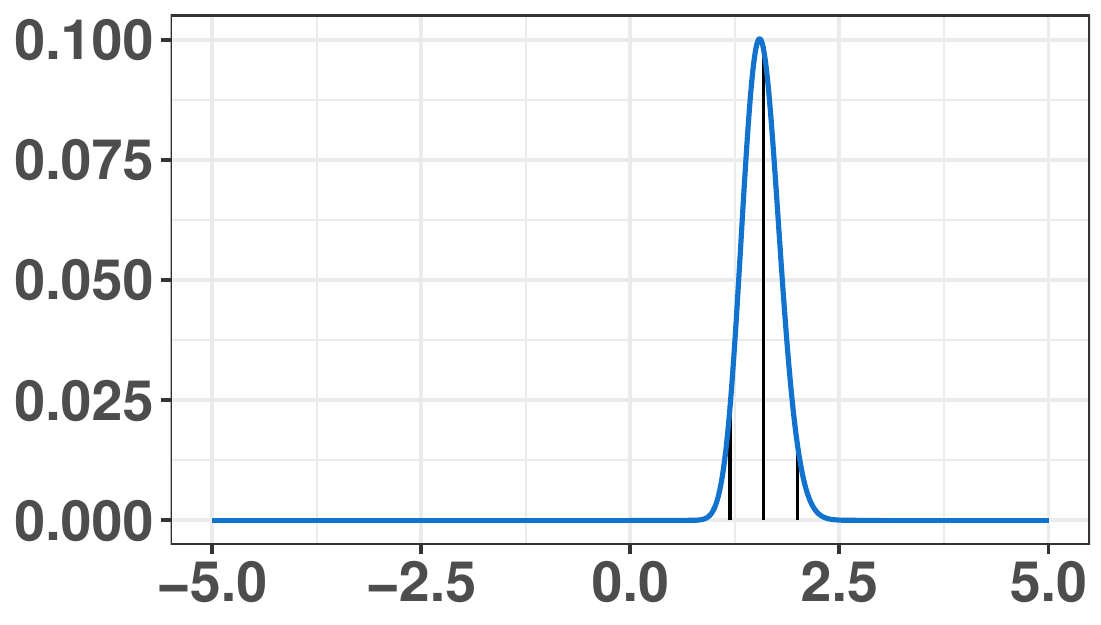}}}
	\caption[The effect of the number of integration points in Gauss-Hermite quadrature.]{The effect of the number of integration points in Gauss-Hermite quadrature. The blue line denotes %$ g(h) \exp(-h^2)$ 
	the integrand of $\bm{\beta}_n$ for the non-stationary SE kernel, with an SE kernel for $u(\cdot)$ with length-scale  $\lambda=0.1$. The vertical black lines depict the integrand evaluated at the node locations. }\label{fig:quad_effect}
\end{figure}

%Note that evaluation of the posterior of intereset requires the computation of intractable integrals; namely, the elements of $\bm{\beta}_n$ and $P_n$ for each data point $n=1,\ldots,N$. Specifically, we require $N(M+M(M+1)/2)$ {\textit{univariate}} intractable integrals, which will be approximated with Gauss-Hermite quadrature.
Clearly, the quadrature order $J$ can have a strong impact on the computational complexity; moreover, a large $J$ may be required for sufficient approximations. 
Here we analyse the integrand that corresponds to $\bm{\beta}_n$ 
%; however, similar behaviour will be encountered for $P_n$.
for the non-stationary squared exponential (SE) kernel. 
Figure~\ref{fig:quad_effect} illustrates how the shape of the integrand varies for different $\delta$ values, with $\delta$ representing the distance between two locations, and fixed hyperparameters. We can observe how the position of the nodes often misses the peak. Also, as we increase $\delta$, the function becomes sharper, and how fast this occurs is determined by the hyperparameters. Figure~\ref{fig:quad_effect}(f) illustrates clearly why the approximation can be poor, even with several nodes. We also highlight that in some cases, the node positions are located in regions where the function is flat, see for instance Figure~\ref{fig:quad_effect}(h), which should depict ten horizontal bars but only two can be observed.

\subsection{Pseudo-marginal approach}\label{SEC:SBP}
To address the shortcomings of the Gauss-Hermite approach, we employ the PM framework described in Section~\ref{section:PM}. In this case, we do not marginalize the inducing variables $\btz$, in order to maintain the factorized form of log-likelihood required. 
%Differently to the Gauss-Hermite approach, here we do not employ the marginal posterior in Eq.~\eqref{eq:Isopost} obtained by marginalising the function $\btz$ at the inducing locations. This is because we employ a variant of the PM scheme that employs a doubly stochastic estimator for the exponentiated expected log-likelihood, by also subsampling the data points. This provide computational gains, especially in larger settings; however, it requires us to write the log-likelihood as a sum over the observations, which is not possible for the marginal posterior $q(\btu,\btheta, \bvarphi)$.
In addition, to break the correlation,  whitening is applied for both $\btz$ and $\btu$ based on the transformations: $\btz=L(\btu)\btxi$ and $\btu=L(\bvarphi)\btzeta + \bm{\mu}_u,$ where $\btxi \sim \Norm(0,I_M)$, $\btzeta\sim \Norm(0,I_M)$, $L(\btu) L(\btu)^{T}=\CNS_{\bphi}$ and $L(\bvarphi) L(\bvarphi)^{T}=\CS_{\bvarphi}$. Thus, the whitened variational posterior has the form
\begin{equation}
\begin{split}
&q(\btxi,\btzeta,\btheta)   \propto \left[ \exp \left(  \sum_{n=1}^N \E_{u_n}[ l(y_n \mid u_n,\btz,\btheta)]\right) \right]_{_{\btz,\btu}} \Norm(\btxi\mid 0, I_M) \Norm(\btzeta\mid 0, I_M) \pi(\btheta) ,
%&\propto \prod_{n=1}^N \exp\left( -\frac{1}{2\sigma_\varepsilon^2}  \left[  \E_{\pi(u_n| \btzeta,\btheta, \bvarphi)} \left( (y_n- \E_{\pi(z_n|u_n, \btxi,\btzeta,\btheta, \bvarphi)}[z_n])^2 +  \V_{\pi(z_n| u_n,\btxi,\btzeta,\btheta, \bvarphi)}[z_n]\right) \right]\right) \\
%& \quad \quad  \quad  \quad \quad \quad  \quad  \quad  \quad  \quad \quad \quad  \quad  \quad  \quad  \quad  \quad \left(2\pi\sigma_\varepsilon^2\right)^{-\frac{N}{2}} \Norm(\btxi\mid 0, I_M)\pi(\btheta) \Norm(\btzeta\mid 0, I_M)\pi(\bvarphi),
\end{split} \label{eq:whitenedmarginal2level}
\end{equation} 
where  $\btz$ and $\btu$ are replaced by $L(\btu)\btxi$ and $L(\bvarphi)\btzeta + \bm{\mu}_u$, respectively, in the expression in brackets. 

To apply the PM scheme, we look for an unbiased estimator, $\widehat{E}$, of the first term in the right hand side of eq.~\eqref{eq:whitenedmarginal2level}. Such estimator can be found by employing the doubly stochastic block-Poisson estimator in Definition~\ref{def:BP}. In this case, the control variates, $\bar{\nu}_n(u_n)$, is defined through a first-order Taylor-expansion around the mean of $\pi(u_n \mid \btheta)$; such that,
\begin{equation*}%
\nu_n(u_n)=   l (y_n \mid \E[u_n], \btz,  \btheta )+(\ell_n- \exp(\E[u_n]))  l^{\prime}(y_n \mid \E[u_n],  \btz,  \btheta).
\label{eq:vn2level}\end{equation*}%.
Note that the computational cost of the control variates is $\mathcal{O}(NM^2)$ and consequently, evaluation of the doubly stochastic block estimator is of order $\mathcal{O}(NM^2 + \kappa B M^2 + M^3)$. We emphasize that one can define alternative control variates of reduced computational complexity (see Appendix ~\ref{app:details_estimator}).
Details of the sampling algorithm for this approach are provided in Appendix~\ref{app:PM_2level}.

\section{Simulation study}
In the following synthetic dataset we compare the performance of the Gauss-Hermite approach from Section~\ref{sec:GH} and the proposed pseudo-marginal scheme described in Section~\ref{SEC:SBP}. We aim to evaluate if the {S-BP-PM} algorithm performs better than the quadrature approach in (i) recovering the true parameters, (ii) predictive performance, and (iii) computational time.
In addition, we provide a comparison with the full (non-sparse) model, to highlight the computational benefits of the sparse scheme. 
%to understand better how sensitive the model is to these choices.

% On the other hand, treating the inducing points as variational parameters requires derivative calculations, which for our model are expensive and not straightforward to compute, and adds $M \times D$ parameters to the model. Consequently, we decide to use a strategy that is informed, yet simple to implement.
Following \citet{titsias2009variational}, we select the inducing points by maximising a lower bound to the sparse variational marginal likelihood of a stationary Gaussian process (which is available in closed form). More precisely, we employ a stationary Gaussian process with M\'atern $3/2$ covariance function and optimise the inducing points using the GPstuff toolbox implemented in MATLAB \citep{gpstuff}. The optimised locations are then employed in the proposed MCMC schemes.
\label{section:Simstudy}

\subsection{1-dimensional dataset}
    We simulate $N=1,000$ observations from the two-level non-stationary GP model with domain $[ 0,1 ]$, noise variance $\sigma_\varepsilon^{2}=0.02$, and stationary length-scale hyperparameter $\lambda=0.1$. 
    The results of the Gauss-Hermite scheme employing different numbers of inducing points ($M=30,45,60$) and different quadrature orders ($J=4,5,8,10,15$) are contrasted with those obtained with our proposed pseudo-marginal scheme.

\subsubsection{Posterior inference}\label{Sec:InferenceVMCMC}
 For the Gauss-Hermite approach, we employ the sampling scheme described in Appendix~\ref{app:approximateGH} %(Algorithm~\ref{algo:Isosampler}) 
 for $T=50,000$ iterations, employing the same initialisations across the varying number of inducing points $M=30, 45, 60$ and quadrature orders $J=4,5,8,10,15$. The first 30,000 iterations are discarded as burnin. 
 
 For the S-BP-PM scheme, we first determine the optimal tuning parameters needed for the algorithm; namely, $ a, \kappa,$ and $B$. To do so, we investigate the normality assumption of the estimator $\widehat{d}_{B=30}$ needed to employ Algorithm~\ref{algo:tuningpar_v2}. (Figure~\ref{fig:normality} in the Appendix shows histograms of $500$ estimators for $M=30,45,$ and $60$ confirming the assumption). The resulting optimal tuning parameters are also shown in Appendix~\ref{app:optimal_par_1d}. We run the sampling scheme discussed in Appendix~\ref{app:PM_2level} for $T=50,000$ with burnin of $30,000,$ for $M=30,60$, and $40,000$ for $M=45$.
 %The normality assumption is confirmed for $M=60$; however, both $M=30$ and $M=45$ appear to violate such assumption, as they depict heavy-tailed distributions. This suggests that we require a larger number of subsamples to ensure $\widehat{d}_{B}$ to be normally distributed.  For both $M=30$ and $M=45$, we found that we necessitate at least $B=300$ subsamples (see Figure~\ref{fig:normality_m30} and Figure~\ref{fig:normality_m45} in the Appendix) for normality to be a reasonable assumption.
% derive the employ Algorithm~\ref{algo:SBPM}. 
%We run the MCMC scheme described 

\begin{figure}[!htbp]	\subcaptionbox{\scriptsize{M=30}}{{\includegraphics[scale=.35]{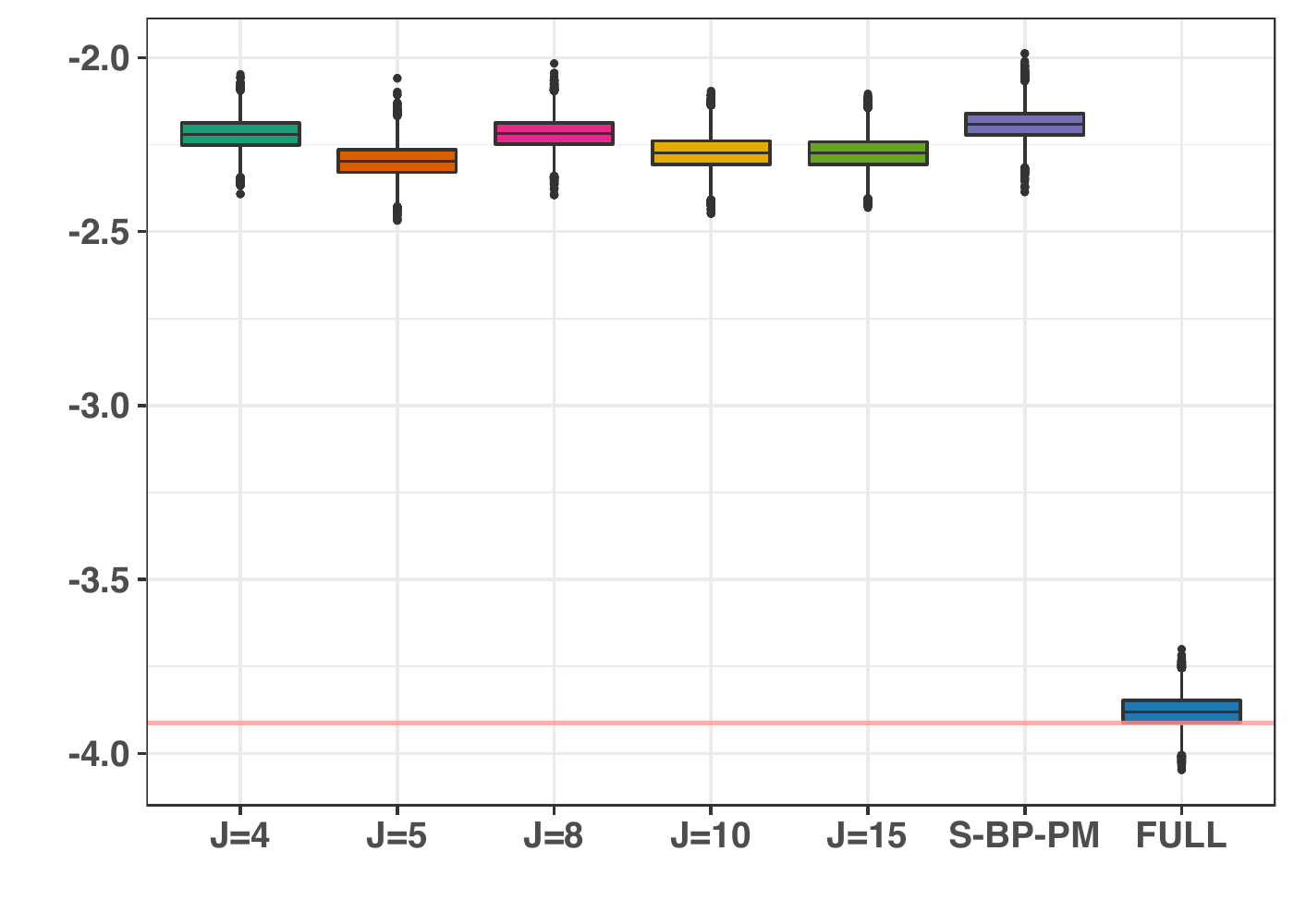}}}
	\subcaptionbox{\scriptsize{M=45}}{{\includegraphics[scale=.35]{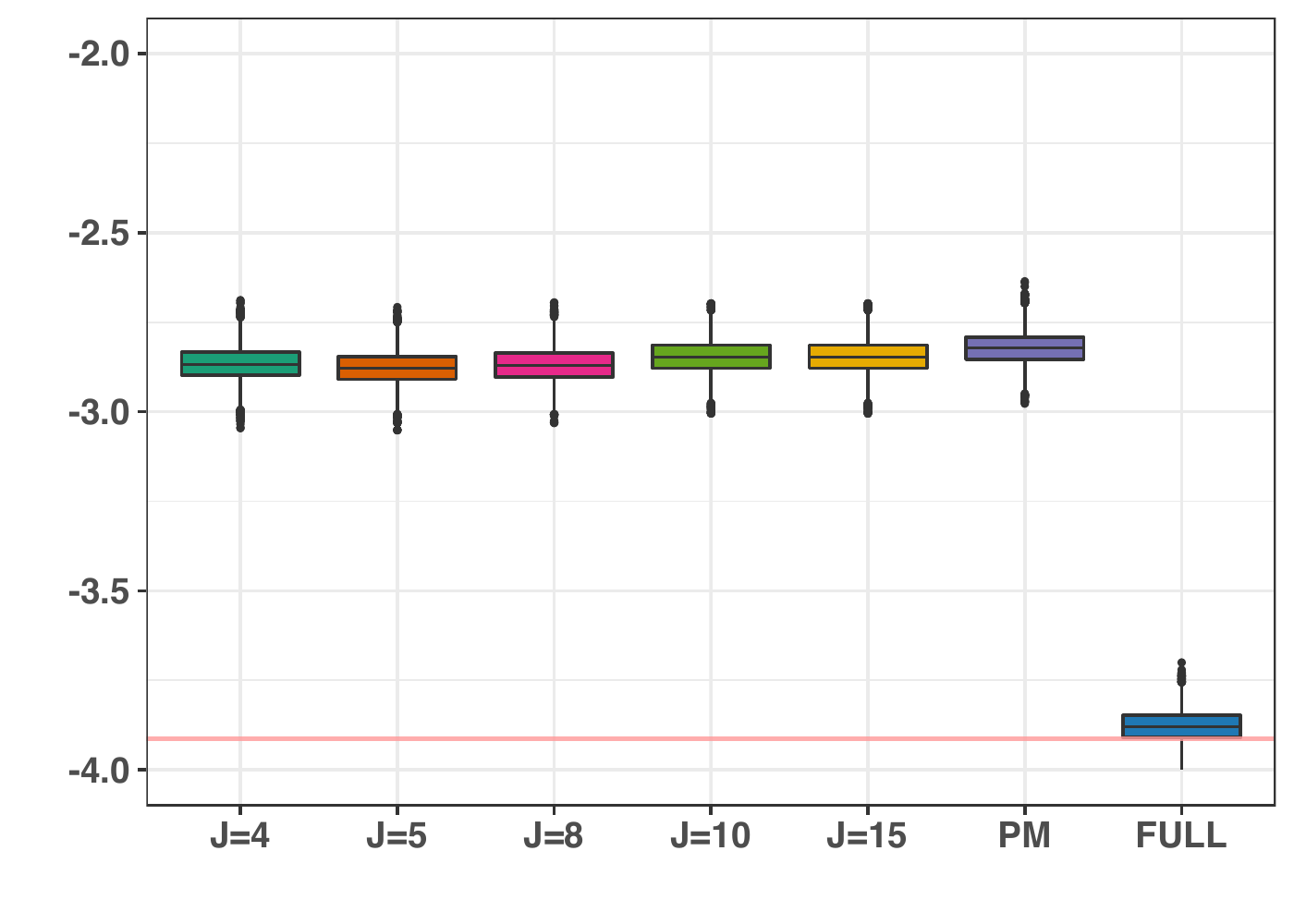}}}
	\subcaptionbox{\scriptsize{M=60}}{{\includegraphics[scale=.35]{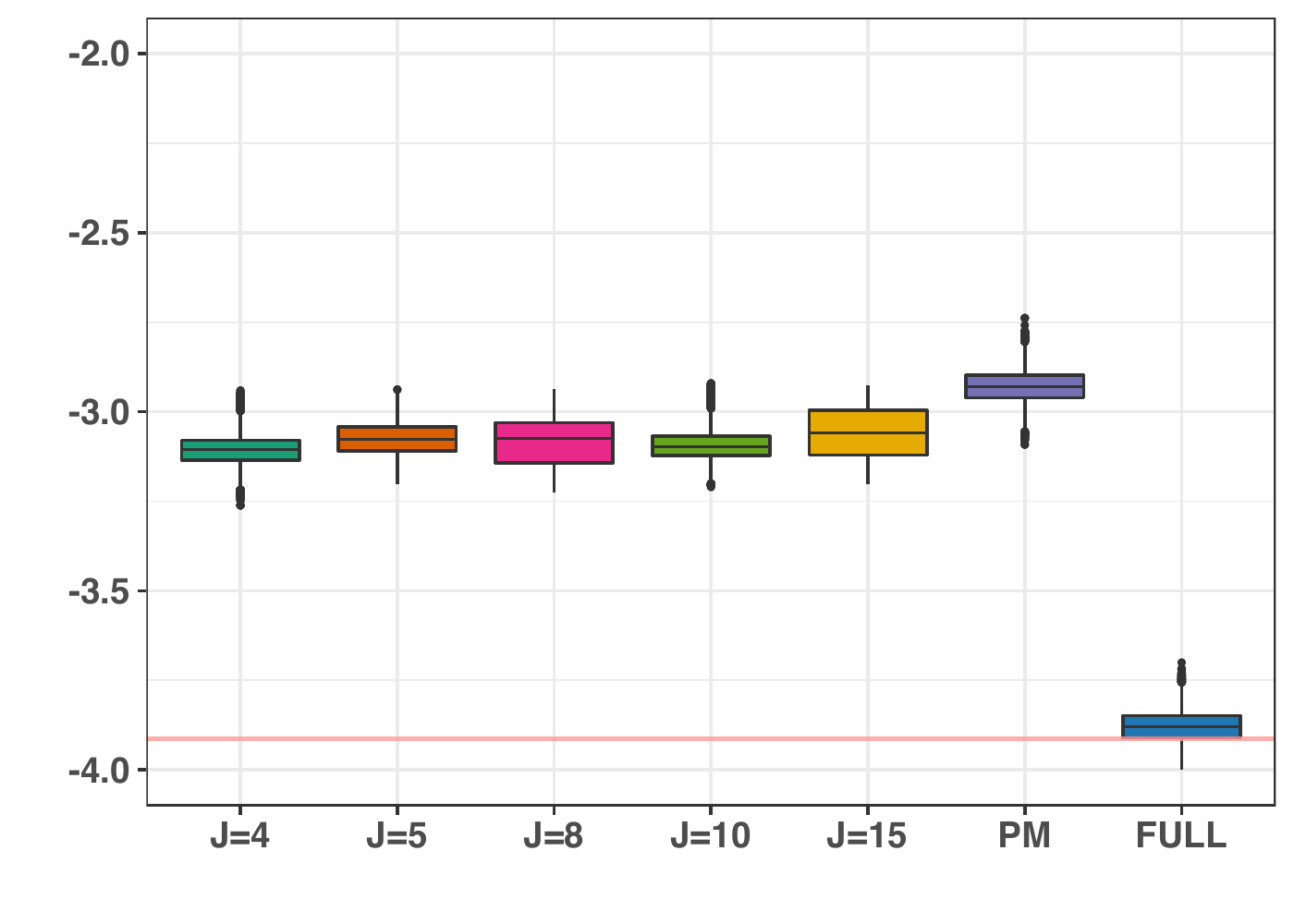}}}
	\caption{Boxplots of the MCMC samples for the logarithm of the noise variance parameter for different numbers of inducing points ($M=30,45,60$). Each plot shows results with different quadrature orders ($J=4,5,8,10,15$), the S-BP-PM scheme in purple and the full MCMC procedure in blue. The true parameter value is depicted in red.}% 
	\label{fig:bpnoise}
\end{figure}

\begin{figure}[!htbp]
\centering
	\subcaptionbox{\scriptsize{M=30}}{{\includegraphics[scale=.43]{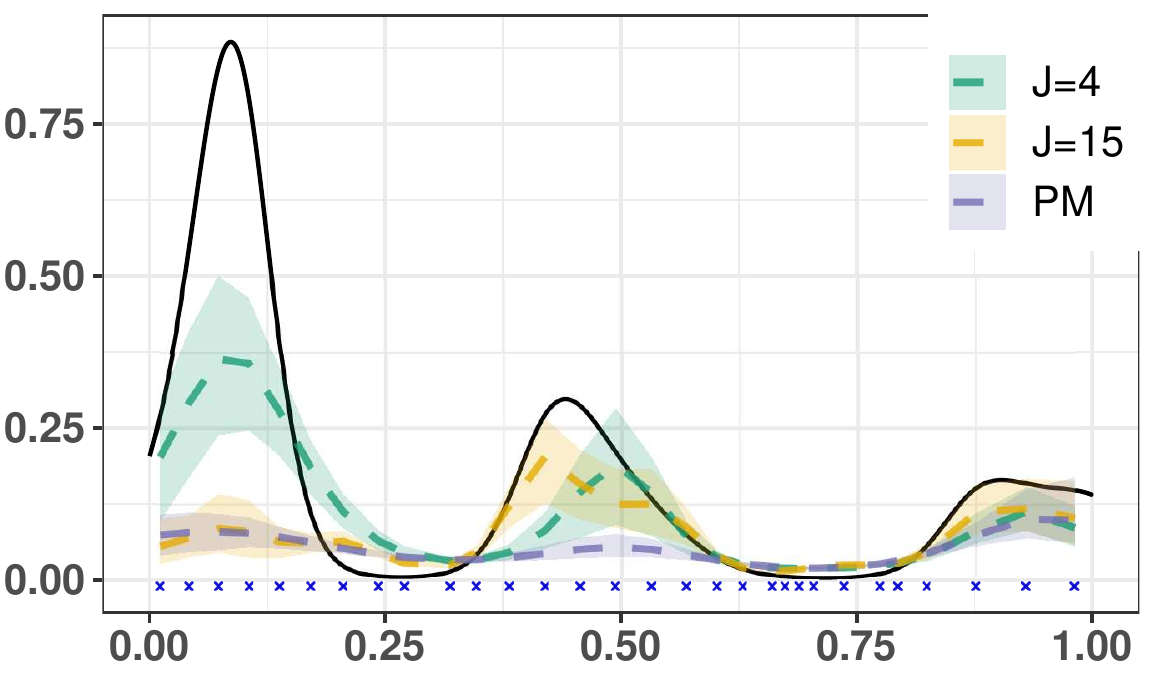}}}
	\subcaptionbox{\scriptsize{M=45}}{{\includegraphics[scale=.43]{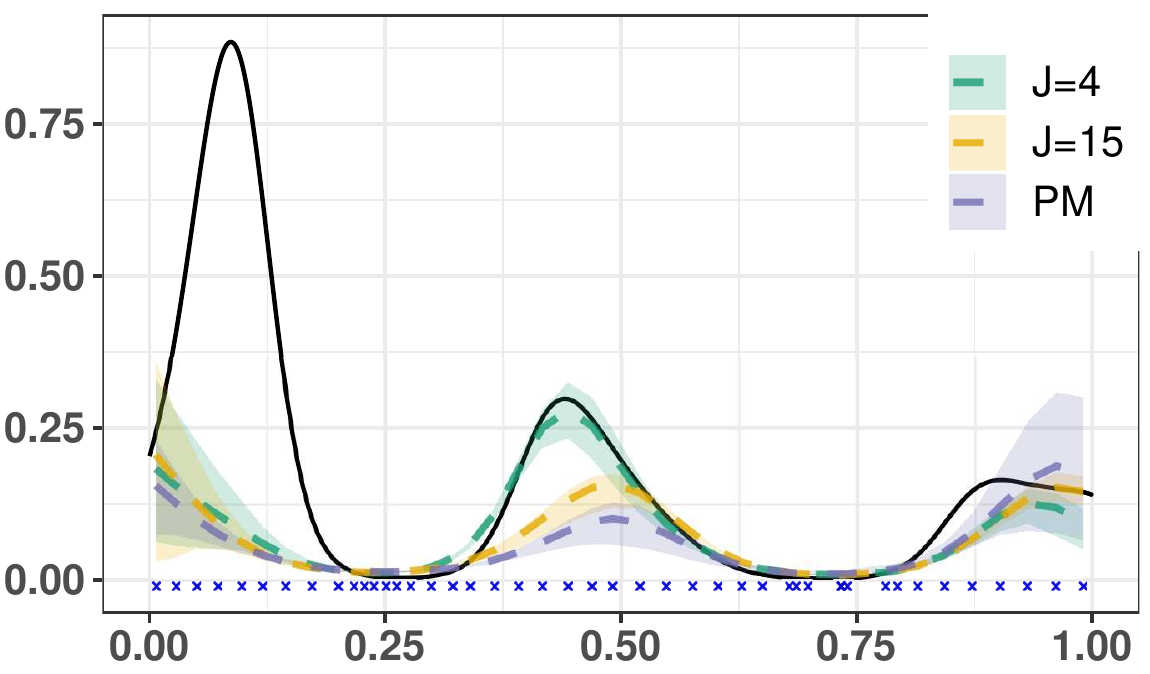}}}
	\subcaptionbox{\scriptsize{M=60}}{{\includegraphics[scale=.43]{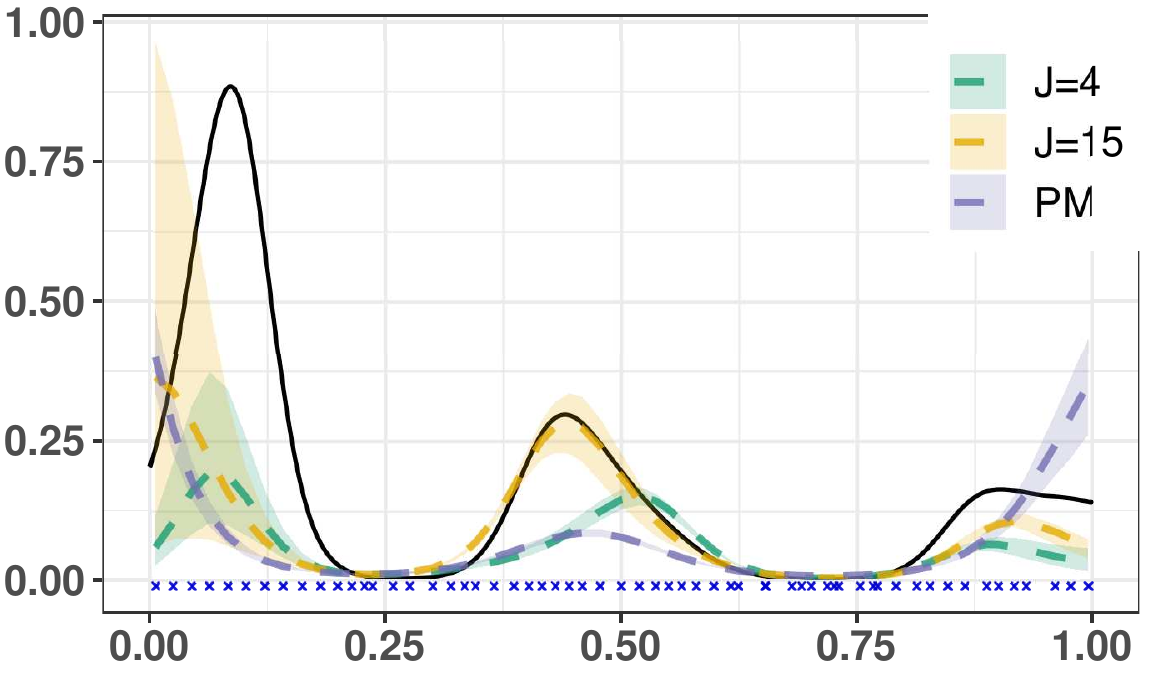}}}
	\caption{Posterior estimates for spatially varying parameter $\ell(\cdot)$. The black line denotes the true process. The colour-coded dashed lines show posterior estimates with 95\% HPD credible intervals for $J=4, 15$, and the S-BP-PM.}% 
	\label{fig:ell_posterior_sum}
\end{figure}

First, the posterior of the log noise variance indicate that the parameter can be greatly over-estimated, especially when using few inducing points (see Figure~\ref{fig:bpnoise}). This can be the result of underestimation of the posterior variance of the latent GP (and consequently, an overestimation of the noise variance to compensate \citep{gadd2018pseudo}). 
Second, posterior summaries of the length-scale process illustrated in Figure~\ref{fig:ell_posterior_sum} highlight that increasing the number of inducing points and the number of nodes in the quadrature approximation does not necessarily result in more accurate posterior estimates (see Appendix~\ref{app:plots_1d}, Figure~\ref{fig:ell_posterior} for more details).

Figure~\ref{fig:MAE_MSE_pm}(d) reports the computational time needed for inference with the S-BP-PM algorithm versus a full MCMC procedure and the Gauss-Hermite quadrature approach with the different number of nodes considered. %(see Table~\ref{tab:COMPARISON_time} for more details). 
Firstly, the figure highlights the clear computational benefit of employing the sparse variational posterior approximation. Secondly, 
we observe that our proposed S-BP-PM procedure with $M=60$ is cheaper in comparison to the Gauss-Hermite approximation utilising $M=30$ and only $J=4$. This is relevant because this computational advantage will permit us to increase the number of inducing points, which is crucial to efficiently recover non-stationarities.

\subsubsection{Predictions}\label{Sec:PredictionsVMCMC}
\begin{figure}[!htbp]
\centering
	\subcaptionbox{\scriptsize{M=30}}{{\includegraphics[scale=.44]{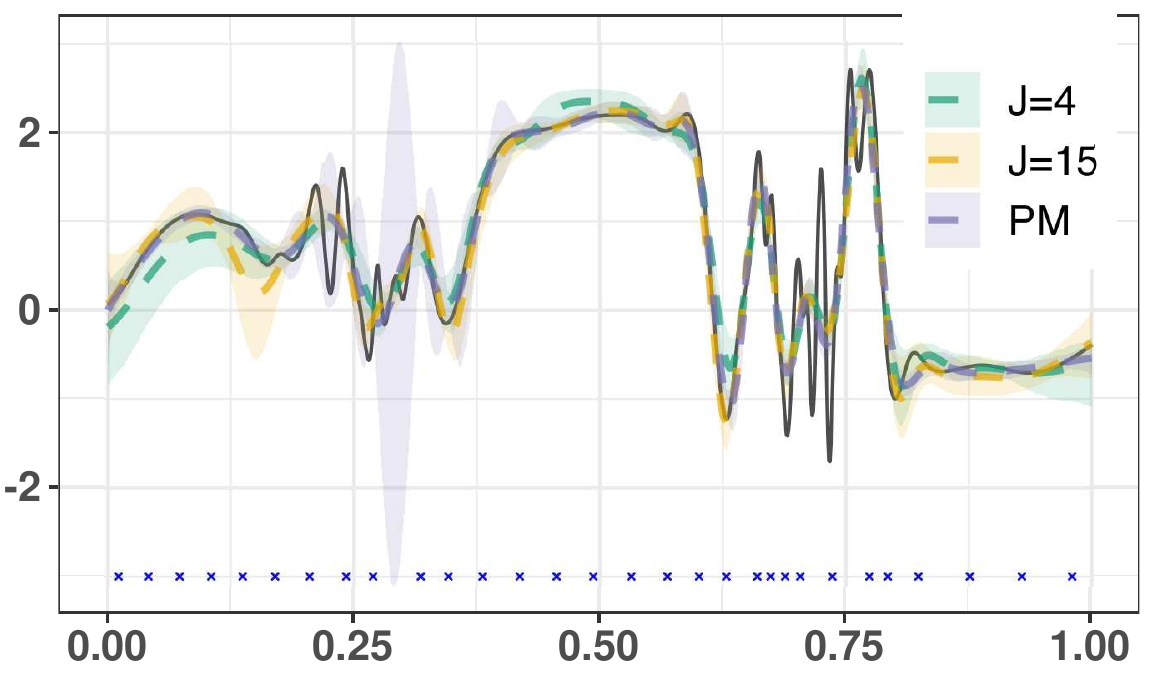}}}
	\subcaptionbox{ \scriptsize{M=45}}{{\includegraphics[scale=.44]{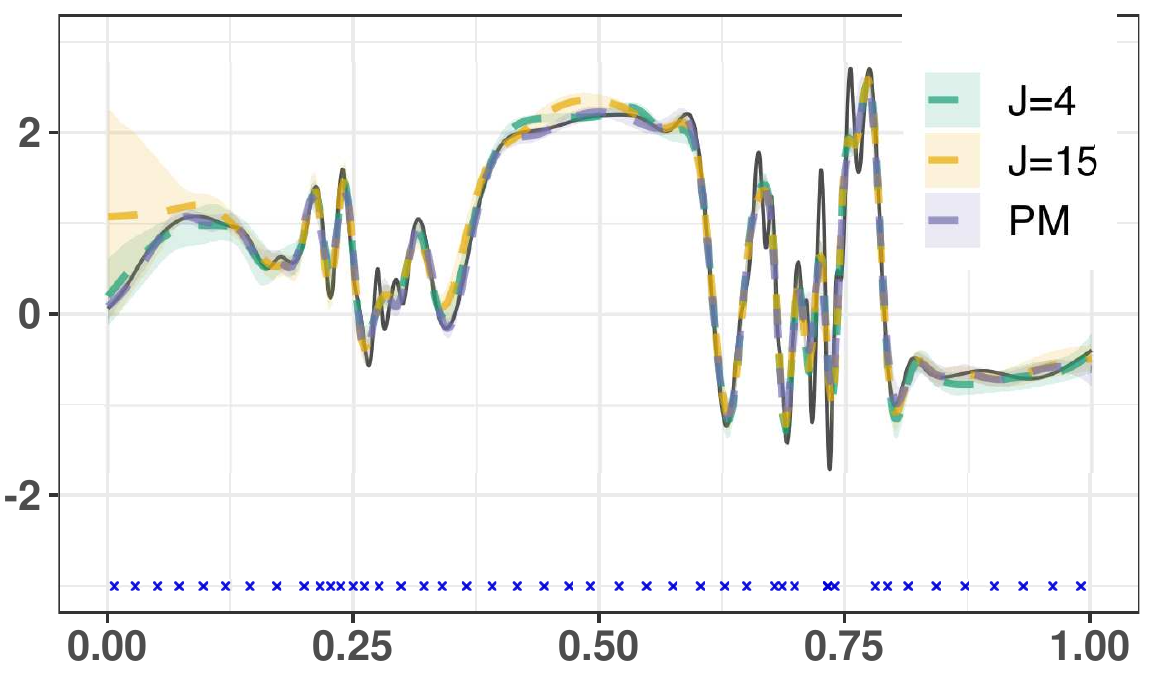}}}
	\subcaptionbox{\scriptsize{M=60}}{{\includegraphics[scale=.44]{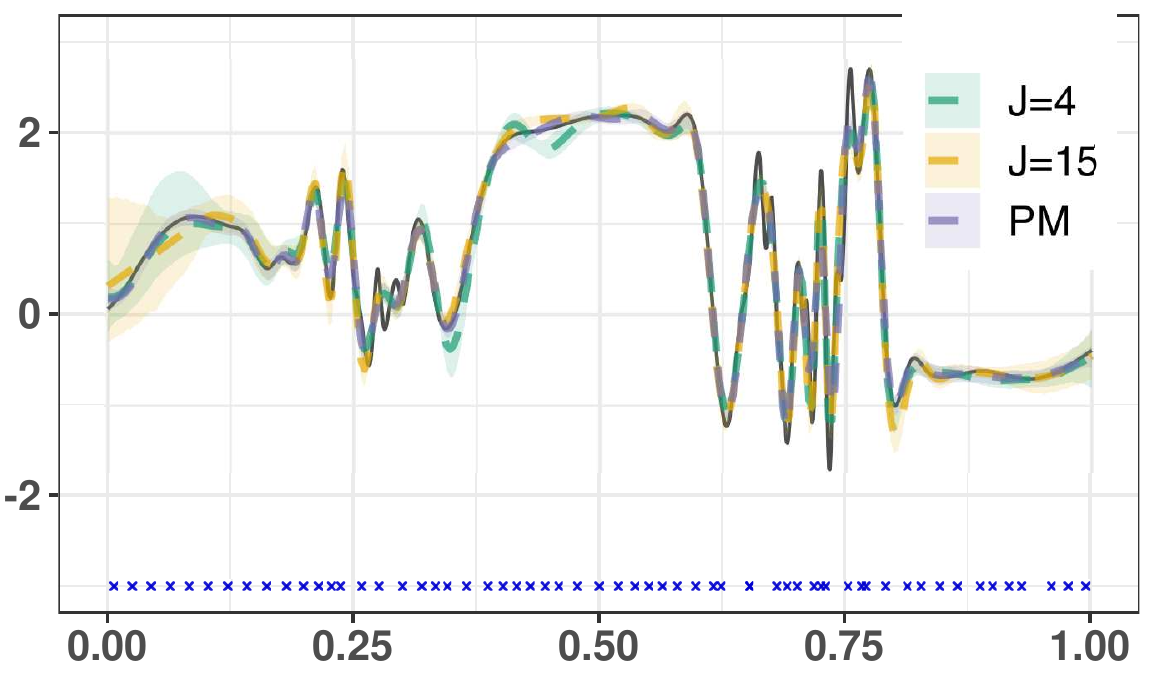}}}
	\caption{{Out-of-sample predictions with different number of inducing points. The color-coded dashed lines denote the predictive mean with shaded areas depicting 95\% HPD point-wise credible intervals. Solid black line denotes the true process.}}% 
	\label{fig:z_pred_pm}
\end{figure}

\begin{figure}[!htbp]
	\subcaptionbox{\scriptsize{MSE}}{{\includegraphics[width=7.3cm, height=3.7cm]{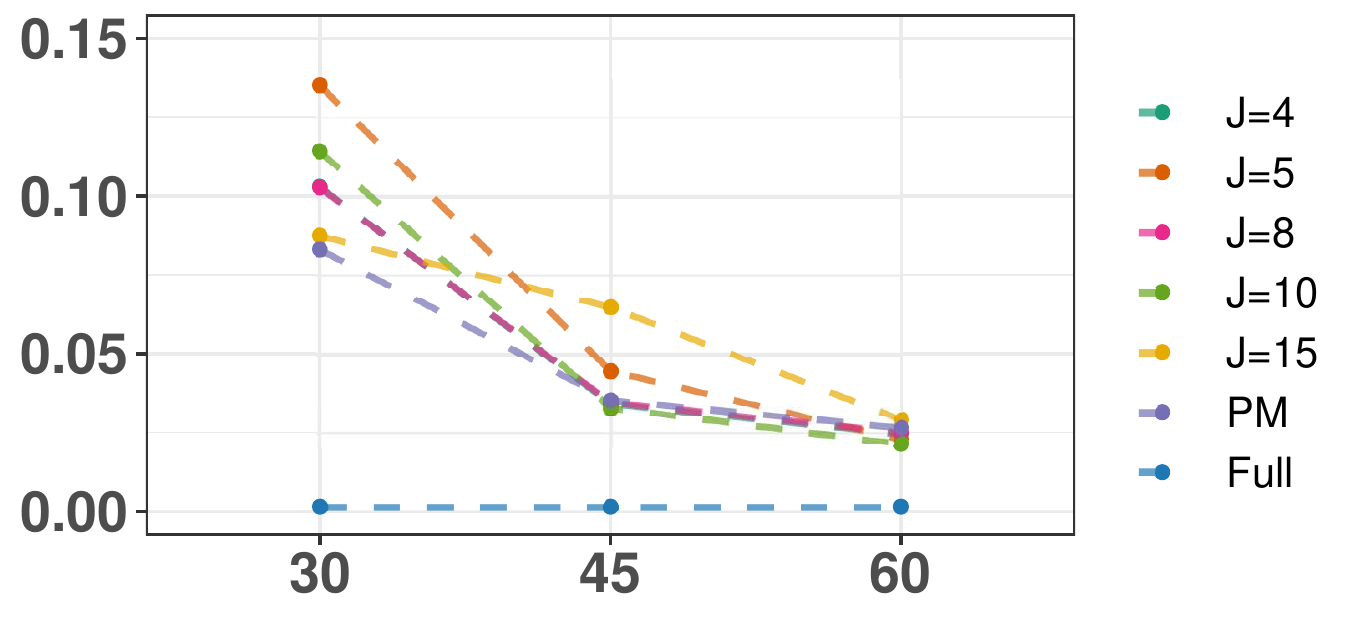}}}  \hspace{5mm}
	\subcaptionbox{ \scriptsize{MAE}}{{\includegraphics[width=7.3cm, height=3.7cm]{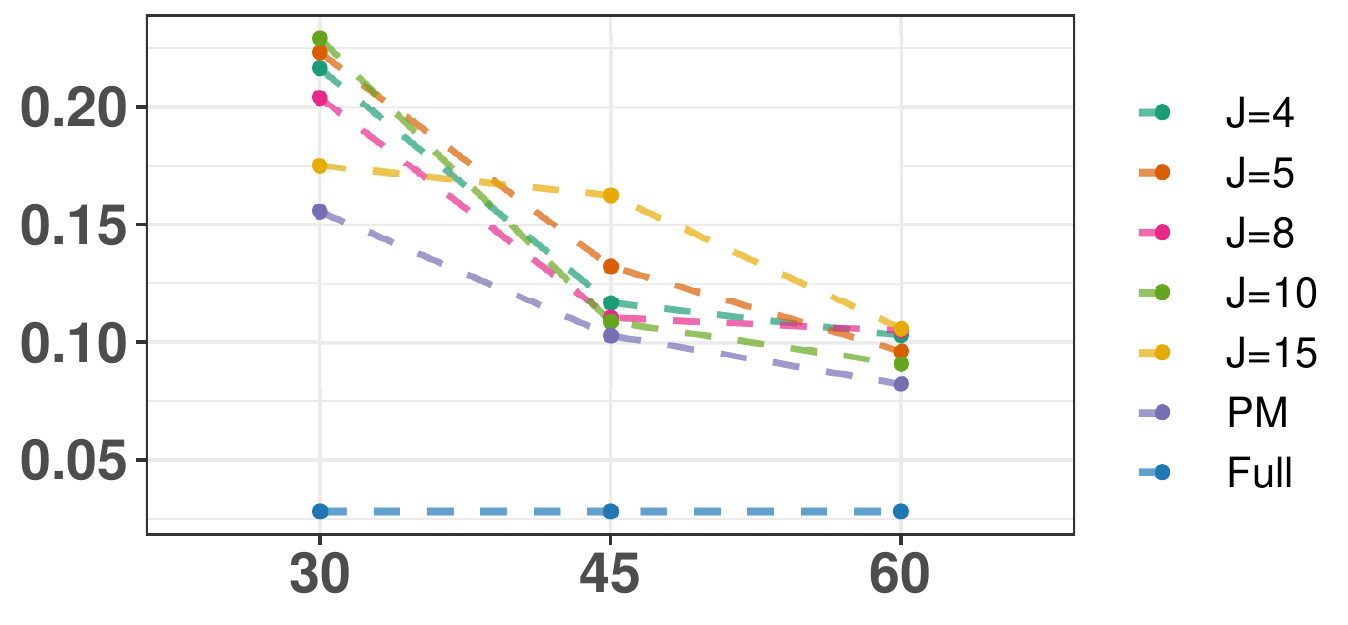}}}\\
	\subcaptionbox{\scriptsize{EC}}{{\includegraphics[width=7.3cm, height=3.7cm]{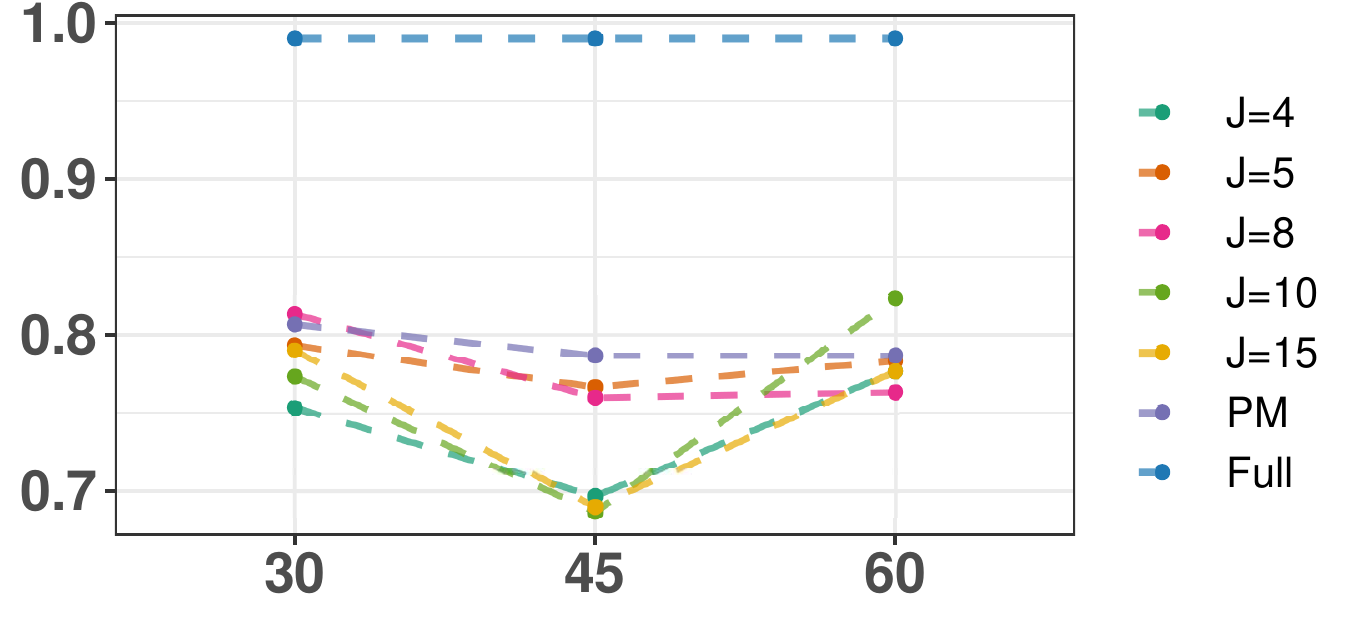}}} \hspace{5mm}
	\subcaptionbox{\scriptsize{Avg. time (min)}}{{\includegraphics[width=7.3cm, height=3.7cm]{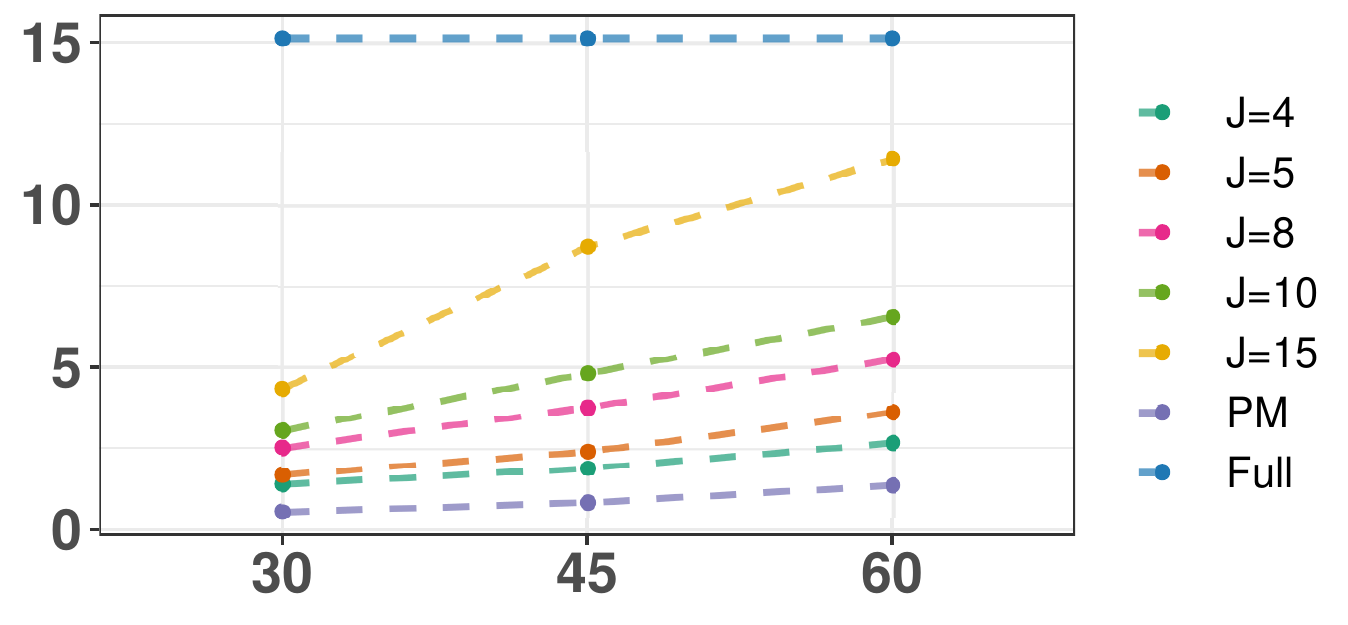}}}\\ 
	\caption[ Comparison of the predictive performance and computational time] { Comparison of the predictive performance and computational time. The results for S-BP-PM are shown in purple and compared with the Gauss-Hermite approximation approach with different number of nodes $(J=4,5,8,10,15)$ and the full (non-sparse) model. (a)-(c): Predictive performance comparison in terms of MSE, MAE and EC of out-of-sample predictions for different numbers of inducing points ($M=30,45,60$). (d): Average computational time (in minutes) required for $100$ MCMC iterations.}% 
	\label{fig:MAE_MSE_pm}
\end{figure}

To evaluate the predictive performance of the method, we make out-of-sample predictions at $300$ locations.
Figure~\ref{fig:z_pred_pm} shows predictive estimates of the non-stationary latent function. It is clear that when there are not enough inducing points or when they are not well located, important features of the function can be missed, see for instance Figures~\ref{fig:z_pred_pm}(a) in the range $[0.1,0.25]$ and compare with Figures~\ref{fig:z_pred_pm}(b)-(c). Moreover, for a fixed number and location of the inducing points, the order of the quadrature approximation also affects the results (see Appendix~\ref{app:plots_1d}, Figure~\ref{fig:z_pred} for more details). This is further emphasised in Figure~\ref{fig:MAE_MSE_pm}, where, in some cases, the predictive error increases as we increase the number of nodes employed in the Gauss-Hermite approximation (see e.g. $J=15$ for $M=45$ in Figure~\ref{fig:MAE_MSE_pm}(a)-(c)). Note also that with enough inducing points, e.g. $M=60$ in this example, the quadrature order has less of an effect on the prediction errors. Importantly, when comparing the attained predictive performance of the S-BP-PM with the GH quadrature approach, we observe a slight reduction in point-wise errors, with more significant differences for the MAE (see Figure~\ref{fig:MAE_MSE_pm}(a)-(b)). 
Finally, while we expect an underestimation in the posterior variance inherited from the variational distribution, we note that there is a small drop-off in EC when comparing the results of the GH quadrature (Figure~\ref{fig:MAE_MSE_pm}(c)).

\section{Discussion} \label{section:Conslusions}
We studied the MCMC for variationally sparse GPs approach introduced by \citet{hensman2015mcmc}, which due to its generality and computational benefits is an attractive framework to speed up GP inference. However, the necessity of approximating intractable expectations with Gauss-Hermite quadrature during likelihood evaluation can undermine the computational gains as some models might require high-order or multivariate GH quadrature approximations to explore the posterior efficiently. Instead, to avoid numerical integration, we propose a pseudo-marginal scheme, based on a doubly stochastic block-Poisson estimator, which permits asympotically exact inference on complex models and large datasets, while also reducing the computational cost. 
In this paper, we demonstrate the advantages of our method on a 2-level GP regression model.  However, we highlight that our approach applies to any GP based model, especially when the expected log-likelihood is not available in closed-form. Additionally, the proposed scheme can make use of parallel computations to further speed the inference procedure.
%The simulation study here presented highlight 
%An important advantage of the 2-level GP model here studied is the interpetability of the nested structure, where the second level GP dictates the range of dependence of the process of interest. We note that an heuristic choice of the inducing points based on length scale process can help to better learn non-stationarities.
Finally, we note that the computational benefits of the proposed pseudo-marginal scheme will be more evident in bigger datasets, higher dimensions, and/or deeper architectures.

\bibliographystyle{plainnat}
\bibliography{biblio.bib}

\newpage
\appendix
\appendixpage
\section{Derivations} \label{AppA}
\subsection{Optimal variational posterior}

We seek the variational posterior distribution which minimizes the Kullback-Leibler (KL) divergence between the approximate and true posterior,
\begin{align*}
&\KL(q(\bz,\btz, \bu,\btu,\btheta) {\lVert} \pi(\bz,\btz, \bu,\btu,\btheta \mid \by,X, \tilde{X}))\\
&= - \E_q \left[  \log \left(\frac{\Norm(\by | \bz,\sigma_\varepsilon^{2} I) \pi(\bz| \btz, \bu,\btu, \btheta) \pi(\btz | \btu, \btheta)  \pi(\bu | \btu, \btheta)  \pi(\btu |\btheta) \pi(\btheta)}{p(\by | X) \pi(\bz| \btz, \bu,\btu, \btheta) \pi(\bu | \btu, \btheta) q(\btz, \btu, \btheta)} \right)\right]\\
&=- \E_q \left[ \log \left( \frac{\Norm(\by|\bz,\sigma_\varepsilon^{2} I)  \pi(\btz | \btu, \btheta) \pi(\btu |\btheta)  \pi(\btheta)}{ q(\btz, \btu, \btheta)} \right)\right]+\log (p(\by|X) )\\
\begin{split}
&= - \E_{q(\btz, \btu, \btheta)} \left[  \log \left( \frac{\exp\left( \E_{\pi(\bz, \bu| \btz,\btu,\btheta)}[ \log(\Norm(\by|\bz,\sigma_\varepsilon^2 I)) ] \right) \pi(\btz| \btu, \btheta) \pi(\btu |\btheta)  \pi(\btheta)}{ q(\btz, \btu, \btheta)} \right)\right] \\
  &\quad  \quad + \log (p(\by|X) ).
\end{split}
\end{align*}
Thus, minimisation reveals that the optimal variational posterior is
\begin{equation}
\begin{split}
q(\btz, \btu,\btheta) \propto \exp\left( \sum_{n=1}^N \E_{(z_n,u_n)}[ \log(\Norm(y_n \mid z_n,\sigma^2_\varepsilon)) ] \right) \pi(\btz \mid \btu, \btheta)  \pi(\btu \mid \btheta)  \pi(\btheta). \label{eq:optimalvar}
\end{split}
\end{equation}

%	\begin{align}
%	\widehat{E}  = \exp\left( \sum_{n=1}^N  l (y_n \mid \E[z_n], \brho) \right ) \prod_{k=1}^\kappa \exp \left(  \frac{a+\kappa}{\kappa}\right) \prod_{h=1}^{\mathcal{H}_k} \left(  \frac{ \widehat{d}_B^\newexp -a }{\kappa}\right). 
%	\end{align}
%	Note that $a = \bar{d} - \kappa$, so that:
%		\begin{align}
%	\widehat{E}  = \exp\left( \sum_{n=1}^N  l (y_n \mid \E[z_n], \brho) \right )  \exp \left(  \bar{d}\right) \prod_{k=1}^\kappa \prod_{h=1}^{\mathcal{H}_k} \left(  \frac{ \widehat{d}_B^\newexp -\bar{d} + \kappa }{\kappa}\right), 
%	\end{align}
%	Since only the last term can be negative:
%		\begin{align}
%	\log(|\widehat{E}|)  =  \sum_{n=1}^N  l (y_n \mid \E[z_n], \brho)  +   \bar{d} + \sum_{k=1}^\kappa \sum_{h=1}^{\mathcal{H}_k} \log\left( \abs*{ \left(  \frac{ \widehat{d}_B^\newexp -\bar{d} + \kappa }{\kappa}\right)}\right). \label{eq:logE}
%	\end{align}

\subsection{Marginal variational posterior}

Consider the optimal variational posterior in eq.~\eqref{eq:optimalvar}. Our goal is to marginalise $\btz$ to obtain $q(\btu,  \btheta)$ and consequently derive the conditional variational posterior $q(\btz \mid \btu,\btheta)$. 
First, let us rewrite eq.~\eqref{eq:optimalvar} as:
\begin{equation*}
\begin{split}
q(\btz, \btu,\btheta) \propto \exp\left( \sum_{n=1}^N \E_{u_n}[ l(y_n \mid u_n, \btz,\btheta ] \right) \pi(\btz \mid \btu, \btheta)  \pi(\btu \mid \btheta)  \pi(\btheta),
\end{split}
\end{equation*}
where 
\begin{align*}
&\l(y_n|u_n,\btz,\btheta) = \E_{z_n}\left[ \log(\Norm(y_n \mid z_n,\sigma^2_\varepsilon)) \mid u_n\right] \\
&= -\frac{1}{2} \log(2\pi \sigma_\varepsilon^{2}) - \frac{1}{2\sigma^2_\varepsilon} \left[(y_n-\E_{z_n}[z_n|u_n])^2 + V_{z_n}(z_n|u_n)\right]\\
&= -\frac{1}{2} \log(2\pi \sigma_\varepsilon^{2}) - \frac{1}{2\sigma^2_\varepsilon} \left[ (y_n-\bCNS_{z_n, \btz} (\bCNS_{\btz,\btz})^{-1} 	\btz)^2 + \tau^2_z- \bCNS_{z_n, \btz} (\bCNS_{\btz, \btz})^{-1} \bCNS_{ \btz,z_n} \right].
\end{align*} Then, by expanding the terms inside the exponent and defining
\begin{equation*}%
\begin{split}%
\boldsymbol{\beta}_n =  \E_{u_{n} } \left[ \bCNS_{z_n, \btz}\right], \quad 
\bm{P}_n = \E_{u_{n} } \left[ \bCNS_{ \btz,z_n} \bCNS_{z_n, \btz} \right], \quad \quad \text{and} \quad  \alpha_n =  \E_{u_n}[\CNS_{z_n, \btz} (\CNS_{\btz, \btz})^{-1} \CNS_{ \btz,z_n}]
\end{split} 
\end{equation*}
we obtain:
\begin{align}
\begin{split}
&q(\btz, \btu,\btheta) \propto \left(\frac{\sigma^{-2}_\varepsilon}{2\pi} \right)^{\frac{N}{2}}   \pi(\btz \mid \btu, \btheta)  \pi(\btu \mid \btheta)  \pi(\btheta) \times \\
	& \exp \left( -\frac{1}{2\sigma_\varepsilon^2} \sum_{n=1}^N y_n^2 -2 y_n\boldsymbol{\beta}_n(\CNS_{\btz,\btz})^{-1} \btz +  \cancel{\boldsymbol{\beta}_n(\CNS_{\btz,\btz})^{-1} \btz \left( \boldsymbol{\beta}_n(\CNS_{\btz,\btz})^{-1} \btz\right)^T} +\tau^2_z - \alpha_n  + \right. \\
 	&   \quad  \quad \quad \quad \quad \quad \quad \quad \quad \quad \quad  \left. \vphantom{ -\frac{1}{2\sigma_\varepsilon^2} \sum_{n=1}^N y_n^2} \btz^T (\CNS_{\btz, \btz})^{-1}P_n (\CNS_{\btz, \btz})^{-1}\btz  - \cancel{\btz^T (\CNS_{\btz, \btz})^{-1}(\boldsymbol{\beta}_n^T\boldsymbol{\beta}_n) (\CNS_{\btz, \btz})^{-1}\btz} \right),
\end{split} \label{eq:q_expand}
\end{align}
Now, let us collect the terms that do not depend on $\btz$ in 
$$\Xi := \left( \frac{\sigma^{-2}_\varepsilon}{2\pi} \right)^{\frac{N}{2}}   \pi(\btu \mid \btheta)  \pi(\btheta) \exp \left( -\frac{1}{2\sigma_\varepsilon^2} \sum_{n=1}^N y_n^2 + \tau_z^2 - \alpha_n \right),$$
and re-write Eq.~\eqref{eq:q_expand} as
\begin{align*}
\begin{split}
&q(\btz,\btu, \btheta)\\
&  \propto \Xi   \exp \left( -\frac{1}{2\sigma_\varepsilon^2} \sum_{n=1}^N -2y_n\boldsymbol{\beta}_n(\CNS_{\btz,\btz})^{-1} \btz+\btz^T (\CNS_{\btz, \btz})^{-1}\bm{P}_n (\CNS_{\btz, \btz})^{-1}\btz   \right) \pi(\btz|\btu, \btheta)\\
&  \propto \Xi   \exp \left( -\frac{1}{2\sigma_\varepsilon^2} \sum_{n=1}^N -2y_n\boldsymbol{\beta}_n(\CNS_{\btz,\btz})^{-1} \btz+\btz^T (\CNS_{\btz, \btz})^{-1}\bm{P}_n (\CNS_{\btz, \btz})^{-1}\btz   \right) \Norm(\btz \mid 0, \CNS_{\btz, \btz})\\
&  \propto \Xi   \exp \left( {\sigma_\varepsilon^{-2}}  \by^T
\bm{B}(\CNS_{\btz,\btz})^{-1} \btz- \frac{1}{2}\btz^T (\CNS_{\btz, \btz})^{-1} \sigma_\varepsilon^{-2}  \bm{P}  (\CNS_{\btz, \btz})^{-1}\btz   \right) \Norm(\btz \mid  0, \CNS_{\btz, \btz}),
\end{split}
\end{align*} where $\bm{P}=  \sum_{n=1}^N \bm{P}_n$ and $\bm{B}$ an $(N \times M)$ matrix with rows $\bm{\beta}_n$.

The marginal variational posterior $q(\btu,  \btheta) = \int q(\btz,\btu, \btheta)  \,d\btz$ is therefore:
\begin{align}
\begin{split}
&q(\btu,  \btheta )\\
&  \propto \int   \Xi   \exp \left( {\sigma_\varepsilon^{-2}} \by^T \bm{B}(\CNS_{\btz,\btz})^{-1} \btz- \frac{1}{2}\btz^T (\CNS_{\btz, \btz})^{-1}  \sigma_\varepsilon^{-2}  \bm{P}  (\CNS_{\btz, \btz})^{-1}\btz   \right) \Norm(\btz \mid  0, \CNS_{\btz, \btz}) \,d\btz\, \\
& \propto  \Xi  |\CNS_{\btz, \btz} |^{-\frac{1}{2}} \int   \exp \left( {\sigma_\varepsilon^{-2}}  \by^T \bm{B}(\CNS_{\btz,\btz})^{-1} \btz- \frac{1}{2}\btz^T (\CNS_{\btz, \btz})^{-1}  \sigma_\varepsilon^{-2}  \bm{P}  (\CNS_{\btz, \btz})^{-1}\btz  \right. \\
&  \quad \quad \quad \quad \quad \quad  \quad \quad \quad \quad \quad \quad \quad \quad \quad  \quad  \quad \quad \quad \quad \quad \quad \quad \quad \quad  \quad \quad      \left.   -\frac{1}{2}\btz^T (\CNS_{\btz, \btz})^{-1} \btz \right)\,d\btz\\
& \propto  \Xi  |\CNS_{\btz, \btz} |^{-\frac{1}{2}} \int   \exp \left( {\sigma_\varepsilon^{-2}}\by^T \bm{B}(\CNS_{\btz,\btz})^{-1} \btz- \frac{1}{2}\btz^T  \left[ (\CNS_{\btz, \btz})^{-1} \sigma_\varepsilon^{-2}  \bm{P}  (\CNS_{\btz, \btz})^{-1}  + \CNS_{\btz, \btz} \right] \btz  \right) \,d\btz.
%& \propto  \Xi  |\CNS_{\btz, \btz} |^{-\frac{1}{2}}  \int \exp \left( -\frac{1}{2}  \left( -2\sigma_\varepsilon^{-2}\by^T B(\CNS_{\btz,\btz})^{-1} \btz+ \btz^T  \left[ (\CNS_{\btz, \btz})^{-1} \sigma_\varepsilon^{-2}  \bm{P}  (\CNS_{\btz, \btz})^{-1}  + \CNS_{\btz, \btz} \right] \btz  \right) \right)  \,d\btz.
\end{split} \label{eq:q}
\end{align}
We now notice that the terms inside the integral resemble the kernel of Gaussian with mean and variance
 \begin{align*}
\begin{split}
\hat{\bmu}_{\btz} &= \sigma_\varepsilon^{-2} \CNS_{\btz, \btz}  \left( \sigma_\varepsilon^{-2}  \bm{P} +  \CNS_{\btz, \btz} \right)^{-1}  \bm{B}^T \by,\\
\hat{\Sigma}_{\btz} &= \left( (\CNS_{\btz, \btz})^{-1} \sigma_\varepsilon^{-2}  \bm{P}  (\CNS_{\btz, \btz})^{-1}  + \CNS_{\btz, \btz} \right)^{-1}\\
&= \CNS_{\btz, \btz} \left( \sigma_\varepsilon^{-2}  \bm{P} +  \CNS_{\btz, \btz} \right)^{-1} \CNS_{\btz, \btz}.
\end{split} 
\end{align*}
Thus, by completing the square, we re-write Eq.~\eqref{eq:q} as
 \begin{align*}
\begin{split}
%&q(\tilde{U},  \btheta, \bvarphi ) \\
%\propto & \hspace{.2mm} \Xi  |\CNS_{\btz, \btz} |^{-\frac{1}{2}}  \cancelto{1}{ \int \Norm(\btz|\hat{\bmu}_{\btz}, \hat{\Sigma}_{\btz} ) \,d\btz} \exp \left( \frac{\sigma_\varepsilon^{-4}}{2}  \left(  \by^TB ( \sigma_\varepsilon^{-2}  \bm{P} +  \CNS_{\btz, \btz})^{-1} B^T \by   \right) \right) | \CNS_{\btz, \btz} ( \sigma_\varepsilon^{-2}  \bm{P} +  \CNS_{\btz, \btz})^{-1} \CNS_{\btz, \btz} |^{\tfrac{1}{2}}  \\ 
%& \quad \quad  \quad \quad  \quad \quad   \quad \quad  \quad \quad  \quad \quad   \quad \quad  \quad \quad  \quad \quad  \quad \quad  | \CNS_{\btz, \btz} ( \sigma_\varepsilon^{-2}  \bm{P} +  \CNS_{\btz, \btz})^{-1} \CNS_{\btz, \btz} |^{\tfrac{1}{2}} \\ 
q(\btu,  \btheta ) \propto \hspace{.2mm}   \Xi   \hspace{.1mm} |\CNS_{\btz, \btz} |^{\frac{1}{2}}  | \sigma_\varepsilon^{-2}  \bm{P} +  \CNS_{\btz, \btz}|^{\tfrac{1}{2}}  \exp \left( \frac{\sigma_\varepsilon^{-4}}{2}  \left(  \by^T \bm{B} ( \sigma_\varepsilon^{-2}  \bm{P} +  \CNS_{\btz, \btz})^{-1} \bm{B}^T \by   \right) \right) \\ \times \cancelto{1}{ \int \Norm(\btz|\hat{\bmu}_{\btz}, \hat{\Sigma}_{\btz} ) \,d\btz,} 
\end{split} 
\end{align*}
where by plugging the values of $\Xi $ and re-arranging terms, we obtain 
\begin{align*}
\begin{split}
&q(\btu,  \btheta)\propto   \sigma_\varepsilon^{-N} \pi(\btu \mid \btheta)  \pi(\btheta)   \exp \left(  \frac{1}{2\sigma_\varepsilon^4} \by^T \bm{B}\left(  \CNS_{\btz,\btz}+\sigma^{-2}_\varepsilon\bm{P} \right)^{-1} \bm{B}^T \by \right) \\
&\quad \quad  \quad  \quad  | \CNS_{\btz, \btz}+\sigma^{-2}_\varepsilon\bm{P}|^{-\frac{1}{2}}  |\CNS_{\btz, \btz} |^{\frac{1}{2}}\exp \left( -\frac{1}{2\sigma^2_\varepsilon} \sum_{n=1}^N (y_n^2+\tau^2_z) +\frac{1}{2\sigma_\varepsilon^2} \sum_{n=1}^N \alpha_n \right),
\end{split}
\end{align*}
with 
\begin{align*}
\begin{split}
\sum_{n=1}^N \alpha_n &=\sum_{n=1}^N   \E_{u_n}[\CNS_{z_n, \btz} (\CNS_{\btz, \btz})^{-1} \CNS_{ \btz,z_n}]\\
 &= \sum_{i,j=1}^{M}  (\CNS_{\btz, \btz})^{-1}_{ij} \sum_{n=1}^N \E_{u_n} \left[\CNS_{\tilde{z}_i, z_n} \CNS_{z_n, \tilde{z}_j} \right] \\
    &= \sum_{i,j=1}^{M}  (\CNS_{\btz, \btz})^{-1}_{ij} \bm{P}_{ij}\\
     &= \  \sum_{i,j=1}^{M} \left( \CNS_{\btz, \btz})^{-1}  \odot \bm{P} \right)_{ij}.
\end{split}
\end{align*}

The conditional variational posterior is Gaussian with mean $\hat{\bmu}_{\btz}$ and variance $\hat{\Sigma}_{\btz}$, such that
\begin{align*}
\begin{split}
q(\btz \mid \btu, \btheta)= \Norm\left(     \sigma_\varepsilon^{-2} \CNS_{\btz, \btz}  \left( \sigma_\varepsilon^{-2}  \bm{P} +  \CNS_{\btz, \btz} \right)^{-1}  \bm{B}^T \by    ,\sigma_\varepsilon^{-2} \CNS_{\btz, \btz}  \left( \sigma_\varepsilon^{-2}  \bm{P} +  \CNS_{\btz, \btz} \right)^{-1}  \bm{B}^T \by  \right).
\end{split}
\end{align*}

\subsection{Derivative} \label{app:details_estimator}
To compute the difference estimator for the two-level GP model, we need to calculate the first-order Taylor expansion $\nu_n(u_n)$, which requires the derivative 
\begin{equation*}
\begin{split}
\frac{d}{d {{u_n}}}l(y_n \mid u_n, \btz, \btheta) =\frac{1}{ \sigma^2_\varepsilon} \left[  \left(y_n- \CNS_{z_n, \btz}(\CNS_{\btz,\btz})^{-1} \btz \right)  {\frac{\partial \CNS_{z_n, \btz} }{\partial {u_n}}} (\CNS_{\btz,\btz})^{-1}\btz  \right. +\\
\left. {\frac{\partial \CNS_{z_n, \btz} }{\partial {u_n}}}(\CNS_{\btz,\btz})^{-1} \CNS_{\btz,z_n}\right] 
\end{split}
\end{equation*}
where the $m$th entry of \( \tfrac{\partial \CNS_{z_n, \btz} }{\partial {u_n}} \) is given by $\tfrac{\partial \CNS_{z_n, z_m}}{\partial {u_n}}$ with $\CNS_{z_n, \tilde{z}_m}$ obtained through Eq.~\eqref{eq:non_stat_iso}. In this case, the derivative for the non-stationary isotropic SE kernel is:
%\SW{Karla: check this derivative, I got a different result (note first $\ell_n = d/d u_n \exp(u_n)$ is simply the additional term due to taking the derivative wrt $u_n$ .}
\begin{equation*}
\begin{split}
%\frac{\partial \CNS_{z_n, \tilde{z}_m} }{\partial \ell_n}= \left( \frac{\ell(\bx_{n}) }{2 \ell (\btx_m)} + \frac{\ell(\btx_{m})}{ 2\ell(\bx_{n})}\right)^{-\tfrac{D}{2}}  \exp \left( \frac{ -\sum_{d=1}^D  (x_{nd}-\tilde{x}_{md})^2}{ \ell^2(\bx_{n}) +\ell^2(\btx_{m})} \right) \left[  \left(  \vphantom{\frac{\ell(\btx_{m})}{\ell^2(\bx_{n})}}  \frac{-D}{2 \ell (\btx_m)} +\right. \right. \\
%\left. \left. \frac{D\ell(\btx_{m})}{2\ell^2(\bx_{n})}\right) +  \frac {2\ell(\bx_{n})}{ ( \ell^2(\bx_{n}) +\ell^2(\btx_{m}) )^2} {\sum_{d=1}^D (x_{nd}-\tilde{x}_{md})^2 } \right].
\frac{\partial \CNS_{z_n, \tilde{z}_m} }{\partial {u_n}}&= {\ell_n} \left( \frac{\ell_n }{2 \tell_m} + \frac{\tell_m}{ 2\ell_n}\right)^{-\tfrac{D}{2}}  \exp \left(- \frac{ ||\bx_n-\btx_m||^2}{ \ell^2_n +\tell^2_m} \right)  \times \\ 
& \left[ {\left(\frac{\ell_n}{\tell_m} +\frac{\tell_m}{\ell_n}\right)^{-1}} \left(    \frac{-D}{2 \tell_m}+  \frac{D\tell_{m}}{2\ell^2_{n}}\right) +  \frac {2\ell_n}{ ( \ell^2_n  +\tell^2_m )^2} ||\bx_n-\btx_m||^2  \right].
\end{split}
\end{equation*}

\subsection{Control variates} \label{app:control variates}
Note that computing $l(y_n|u_n,\btz,\btheta)$ is $\mathcal{O}(NM^2)$. Thus, computing the control variates is $\mathcal{O}(NM^2)$ and the cost of block-Poisson estimator is $\mathcal{O}(NM^2 +\kappa B M^2)$. We can reduce this to $\mathcal{O}(NM +\kappa B M^2)$ by defining:
\begin{align*}
l(y_n|u_n,\btz,\btheta)  &= l_1(y_n|u_n,\btz,\btheta) + l_2(y_n|u_n,\btz,\btheta),
\end{align*}
where
\begin{align*}
\l_1(y_n|u_n,\btz,\btheta) &=  -\frac{1}{2} \log(2\pi \sigma_\varepsilon^{2}) - \frac{1}{2\sigma^2_\varepsilon} \left[ (y_n-\bCNS_{z_n, \btz} (\bCNS_{\btz,\btz})^{-1} 	\btz)^2  \right],\\
\l_2(y_n|u_n,\btheta) &=  - \frac{1}{2\sigma^2_\varepsilon} \left[ \tau^2_z- \bCNS_{z_n, \btz} (\bCNS_{\btz, \btz})^{-1} \bCNS_{ \btz,z_n} \right].
\end{align*}
and defining the control variates as the expectation of $\nu_n(u_n)$, which is a first order Talyor expansion of $l_1(y_n|u_n,\btz,\btheta)$ around $\E[u_n]$. Thus, the control variates are:
$$ \bar{\nu}_n = l_1(y_n|\E[u_n],\btz,\btheta),  $$
which is $\mathcal{O}(NM)$. This would only be beneficial if $\kappa B < N$. 
%(which is not the case in our simulation 1).

%\section*{Appendix A. Derivations}

%	\begin{align}
%	\widehat{E}  = \exp\left( \sum_{n=1}^N  l (y_n \mid \E[z_n], \brho) \right ) \prod_{k=1}^\kappa \exp \left(  \frac{a+\kappa}{\kappa}\right) \prod_{h=1}^{\mathcal{H}_k} \left(  \frac{ \widehat{d}_B^\newexp -a }{\kappa}\right). 
%	\end{align}
%	Note that $a = \bar{d} - \kappa$, so that:
%		\begin{align}
%	\widehat{E}  = \exp\left( \sum_{n=1}^N  l (y_n \mid \E[z_n], \brho) \right )  \exp \left(  \bar{d}\right) \prod_{k=1}^\kappa \prod_{h=1}^{\mathcal{H}_k} \left(  \frac{ \widehat{d}_B^\newexp -\bar{d} + \kappa }{\kappa}\right), 
%	\end{align}
%	Since only the last term can be negative:
%		\begin{align}
%	\log(|\widehat{E}|)  =  \sum_{n=1}^N  l (y_n \mid \E[z_n], \brho)  +   \bar{d} + \sum_{k=1}^\kappa \sum_{h=1}^{\mathcal{H}_k} \log\left( \abs*{ \left(  \frac{ \widehat{d}_B^\newexp -\bar{d} + \kappa }{\kappa}\right)}\right). \label{eq:logE}
%	\end{align}

\subsection{Inducing points} \label{app:inducing}
As alternative, we propose to use the pilot MCMC to select the number of inducing points by optimizing a measure that combines both accuracy and computational cost. To measure accuracy, we consider the log-pseudo marginal likelihood \citep[LPML,][]{geisser1979predictive}:
\begin{align*}
    \text{LPML} &= \sum_{n=1}^N \log(\text{CPO}_n),\\
    \text{CPO}_n &\approx \left( \frac{1}{S} \sum_{s=1}^S \left[ p(y_n| \mathbb{E}[z_n|\btxi^s, \bphi^s], \brho^s) \right]^{-1}  \right)^{-1},
\end{align*}
where the conditional predictive ordinate (CPO) is approximated by the harmonic mean \citep{gelfand1994bayesian} and to reduce computational cost, we plug-in the expectation of $z_n$. The number of inducing points can then by selected to maximize $\text{LMPL}/\text{CT}$, to explicitly account for the desire to maximize the marginal likelihood and also minimize computational cost.

\subsection{Approximate MCMC for two-level GP regression} \label{app:approximateGH}
We discuss the algorithm employ to sample from the whitened marginal approximated posterior. The sampler is detailed for the case when a squared exponential covariance function is employed for both the stationary and non-stationary processes. In addition, to improve parameter identifiability, we make use of the empirical prior approach discussed in \cite{monterrubio2020} to fix the magnitude and mean of the length-scale processes. Furthermore, we standarize the observations, $\bm{y}$, to have zero mean and unit variance, such that fixing $\tau^2_z=1$ is an appropriate assumption. %In this case, the target distribution in Eq.~\eqref{eq:Isopost} reduces to
%\begin{small}
%	\begin{align*}
%	\begin{split}
%	q(\btzeta,\sigma^2_\varepsilon, \bm{\lambda}) \propto &  \Norm(\btzeta \mid 0,I_M) \prod_{d=1}^{D} \pi(\lambda_d)\pi(\sigma_\varepsilon^2) \sigma_\varepsilon^{-N} \exp \left( -\frac{1}{2\sigma_\varepsilon^2} \sum_{n=1}^N  (y_n^2)   -\frac{N}{2 \sigma_\varepsilon^2} \right) \\
%	&\left[ \exp \left( \frac{1}{2\sigma_\varepsilon^2} \sum_{i,j=1}^{M} \left( (\CNS_{\btz, \btz})^{-1}  \odot \bm{P}   \right)_{ij}  + \frac{1}{2\sigma_\varepsilon^4} \by^TB\left( \CNS_{\btz,\btz} +\sigma^{-2}_\varepsilon\bm{P}\right)^{-1} B^T \by  \right) 	\right]_{_{\btu}} \\ & \left[ \frac{|\CNS_{\btz, \btz} |^{\frac{1}{2}}}{| \CNS_{\btz, \btz}+\sigma^{-2}_\varepsilon \bm{P}|^{\frac{1}{2}} }
%	\right]_{_{\btu}},
%	\end{split}
%	\end{align*}
%\end{small}
We let $\bm{\lambda}=(\lambda_1 \ldots \lambda_D)$ denote the length-scale parameters for the second level stationary GP prior.

The proposed algorithm uses a MwG scheme, where the whitened spatially varying parameters are sampled employing elliptical slice sampling \citep[ELLSS,][]{murray2010elliptical} and the remaining parameters are drawn with an adaptive random-walk MH procedure \citep[Section 3]{roberts2009examples}. 
%Algorithm~\ref{algo:Isosampler} presents the employed sampler. 
With exception of the noise variance, the sampling mechanism requires to approximate the relevant quantities in Eq.~\eqref{eq:expectations_iso} when iterating over $(\btzeta, \bm{\lambda})$.
%Note that Algorithm~\ref{algo:Isosampler} can also be easily adapted to the case of an isotropic GP prior on the log length-scale process.
For efficiency, the positions and associated weights required in the quadrature schemes can be precomputed, prior to running the MCMC, and passed to the samplers. The R package fastGHQuad \citep{fastgh} is employed to compute the weights and nodes.

\subsection{Exact MCMC for two-level GP regression} \label{app:PM_2level}
We present details on how to sample from the whitened approximate posterior distribution for a variationally sparse 2-level GP model with an isotropic assumption for the non-stationary kernel. Again, here, we fix some of the parameters employing empirical priors.
For the optimal tuning parameters we follow the approach described in Section~\ref{subsec:Alg_parameters}, resulting in a slight modification of Algorithm~\ref{algo:tuningpar_v2}, where we add an extra step after generating the pilot MCMC samples. This extra step aims to draw $S$ samples from $\btz \mid  \btu,\btheta$.
For the pilot MCMC we utilise the already implemented algorithm based on Gauss-Hermite quadrature with $J=10$.

In a similar fashion to Algorithm~\ref{algo:SBPM_v2}, the proposed scheme uses a MwG sampler to iterate over $(\btxi,\btzeta,\sigma^2_\varepsilon, \bm{\lambda})$. The noise variance $\sigma_\varepsilon^2$ and second level length-scale $\bm{\lambda}$, are sampled with adaptive random-walk MH steps \citep[Section 3]{roberts2009examples}. For the whitened spatially varying length-scale, we employ ELLSS \citep[]{murray2010elliptical}, and for the non-stationary function $\btz$, we use an independent MH step with a Gaussian proposal that approximates the true variational conditional posterior; more precisely, the proposal is:
$$
\Norm \left(  \sigma^{-2}_\varepsilon\CNS_{\btz,\btz} \left(\CNS_{\btz,\btz}+\sigma^{-2}_\varepsilon \hat{\bm{P}} \right)^{-1} \hat{\bm{B}}^T\by,\CNS_{\btz, \btz}\left( \CNS_{\btz,\btz} + \sigma^{-2}_\varepsilon \hat{\bm{P}} \right)^{-1} \CNS_{\btz,\btz}\right), \label{eq:proposalz}
$$
with $\hat{\bm{B}}$ an $N \times M$ matrix with rows $\hat{\bm{\beta}}_n=[ \CNS_{z_n, \btz}]_{\mu_{\ell_n}}$ and  $\hat{\bm{P}}= \sum_{n=1}^N \hat{ \bm{P}}_n,$ with $\hat{ \bm{P}}_n=[ \CNS_{ \btz,z_n} \CNS_{z_n, \btz}]_{\mu_{\ell_n}}, $ where we use $\mu_{\ell_n}=\exp(c_n+w_n^2/2)$ to evaluate the expressions in square brackets with
\begin{equation*}%
\begin{split}%
c_n &= {\mu}_u+ \CS_{u_n, \btu}(\CS_{\btu,\btu})^{-1}(L(\bvarphi)\btzeta +{\bmu}_u ), \\
w^2_n &=    \tau^2_u -  \CS_{u_n, \btu}(\CS_{\btu,\btu})^{-1} \CS_{\btu,u_n}.
\end{split}
\end{equation*}

The computational complexity of the algorithm is $\mathcal{O}( (\sum_{k=1}^{\kappa} \mathcal{H}_k )B M^2+M^3+NM^2))$. We emphasise that for all the parameter updates, we can compute the difference estimator
$\widehat{d}_{\alpha_b^\newexp}$ for all $h$ and $k$, in parallel.
While our current implementation does not make use of parallel computing, this can be adapted to vectorise some of the operations; for instance, $c_n$, $w_n$, $ l (y_n \mid u_n, \sigma^2_\varepsilon, \bm{\lambda}, \btxi, \btzeta)$, and $ l^\prime (y_n \mid c_n, \sigma^2_\varepsilon, \bm{\lambda}, \btxi, \btzeta)$ can be easily vectorised.

\section{Simulation study}
\subsection{1-dimensional dataset}

\subsubsection{Optimal tuning parameters and $\text{CT}^*$} \label{app:optimal_par_1d}
To find the optimal number of Poisson estimators, $\kappa$, for a fixed value of $B=30$, we utilise a slight modification of Algorithm~\ref{algo:tuningpar_v2}, where we add an extra step after generating $S=1000$ MCMC samples from a pilot run which employs a GH quadrature approximation. This extra step aims to draw samples from $\btz \mid  \btu,\btheta$, which will be employed to compute the difference estimator.
%We run Algorithm~\ref{algo:tuningpar} for the three different number of inducing points previously employed, i.e. $M=30,45,$ and $60$.  
%Figure~\ref{fig:CT}, illustrates the results when $B=30$ for the different values of $M$ by depicting the logarithm of $\text{CT}^\star$ as a function of $\kappa$ for two different samples of the {mcmc} chains with five random subsamples each. 
As discussed in Section~\ref{subsec:Alg_parameters}, a conservative approach is to set the value of $\gamma$ at its maximum across different subsamples to do a grid search of $\kappa$ over $\{4,5,\ldots,100\}$. Once the optimal $\kappa$ value is computed, the lower bound is set to $a= \bar{d}-\kappa$.
The results of this approach are shown in Table~\ref{tab:optimal_tun1d} for the different number of inducing points studied. According to this, we employ $B=30$ subsamples and we set the optimal $\kappa =4$ for all $M=30,45,$ and $60$. In addition, this table also shows $\text{CT}^\star$, as well as our proposed measure to select the number of inducing points introduced in Section~\ref{app:inducing}. According to LPLM/$\text{CT}^*$, the maximum value is attained with $M=60$.
%an initial selection of the number of inducing points is $M=60$
% Thus, an initial selection of the number of inducing points based on $\text{CT}^\star$ is $M=60$.
\begin{table}[htbp]
  \centering
    \begin{tabular}{lcccccc}
    	\toprule
          & \multicolumn{1}{c}{$\gamma_{\text{max}}$} &  \multicolumn{1}{c}{$\kappa$} &
         \multicolumn{1}{c}{$\bar{d}$ ($\mathrm{e}{-05}$)} &
          \multicolumn{1}{c}{$\text{CT}^*$ ($\mathrm{e}{+06}$)} &
          \multicolumn{1}{c}{LPLM }& \multicolumn{1}{c}{LPLM/$\text{CT}^*$}  \\
        	\cmidrule(lr){2-7}   
    $M=30$  & $2.1755$ & $4$    & $-2.72 $ &    $1.12$  & $-7,088.8 $& $-0.00634$ \\
   $ M=45$  & $0.5352$ & $4$     & $6.13$ &
    $2.49$& $-9,145.5$ &$-0.00367$  \\
   $ M=60 $ & $1.8759$ &  $4$     & $172$ &  $4.61$ & $-16,874.7$ & $-0.00366$ \\
    \bottomrule \\
    \end{tabular}%
     \caption{Optimal tuning parameters values and $\text{CT}^*$ values for $M=30,45,$ and $60$.}
  \label{tab:optimal_tun1d}%
\end{table}%

%The results for the grid search with $B=300$ indicate that for $M=30$, $\kappa=5$ (see (Figure~\ref{fig:CT_B300}(a)~and~\ref{fig:CT_B300}(c)), which matches the obtained value when $B=30$ with a significant increase in the $\text{CT}^*$ measure. For $M=45$, the optimal $\kappa$ is found around $1800$, attaining the maximum $\text{CT}^*$ (Figure~\ref{fig:CT_B300}(b) and~\ref{fig:CT_B300}(d)).

%Because both $\text{CT}^*$ and the overall computational complexity of the algorithm depend on $B$ as well as $\kappa$, one must be careful with setting such parameters to very large values, as this can undermine the computational gains of the method. 
%In addition, \citet[Appendix S2]{quiroz2016block} point out that when $\widehat{d}_{B}$ follows a Student $t$ distribution, the optimal value of $\kappa$ obtained under a normality assumption and $B=30$ is comparable to that obtained when the estimator's distribution is approximated with a mixture of Gaussians (the guidelines under that assumption are not here investigated). 

\begin{figure}[!htbp]
	\centering
	\subcaptionbox{\scriptsize{M=30}}{{\includegraphics[scale=.33]{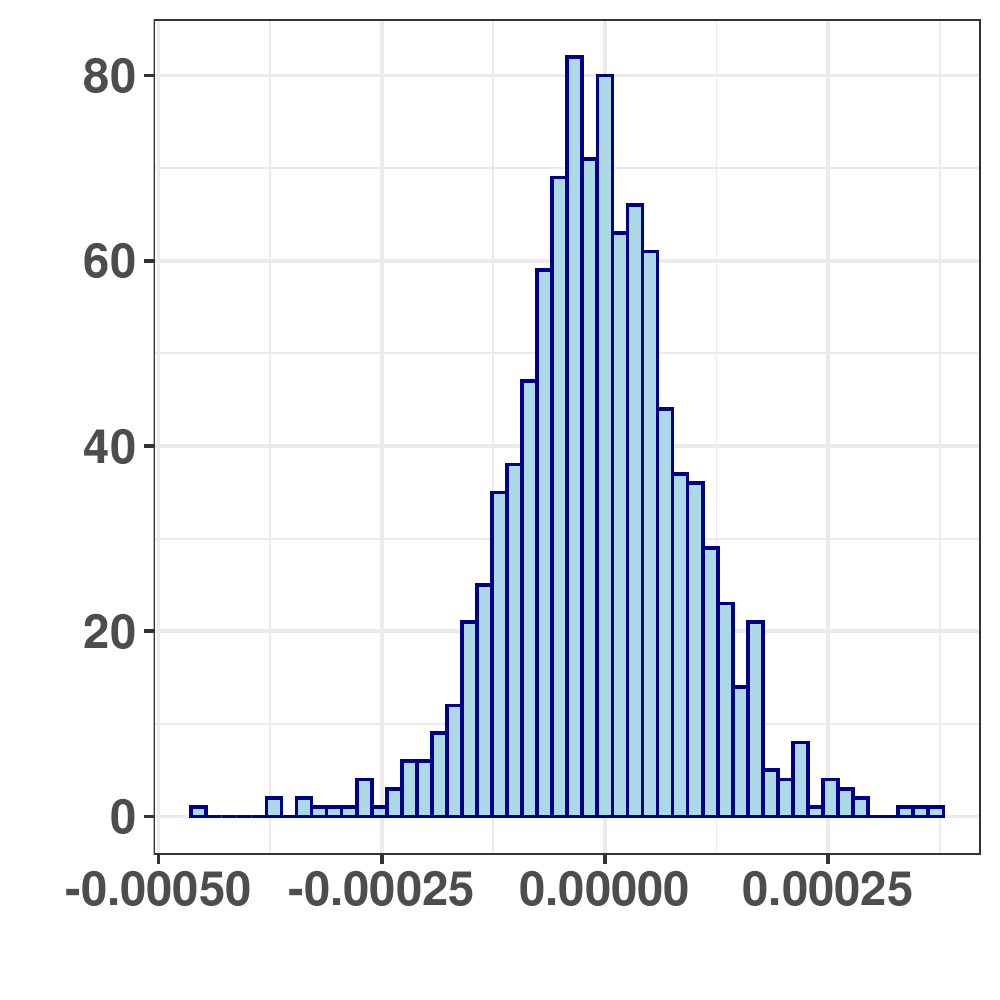}}}
	\subcaptionbox{ \scriptsize{M=45}}{{\includegraphics[scale=.33]{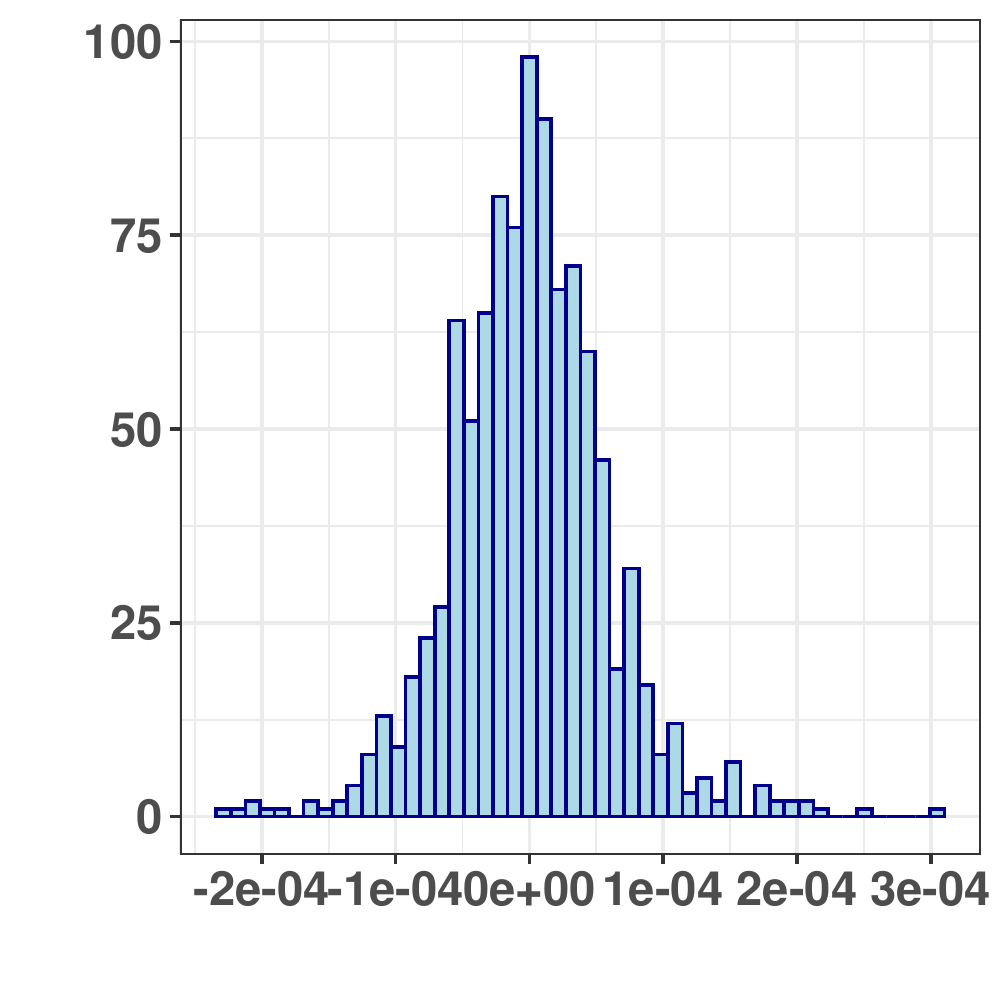}}}
	\subcaptionbox{\scriptsize{M=60}}{{\includegraphics[scale=.33]{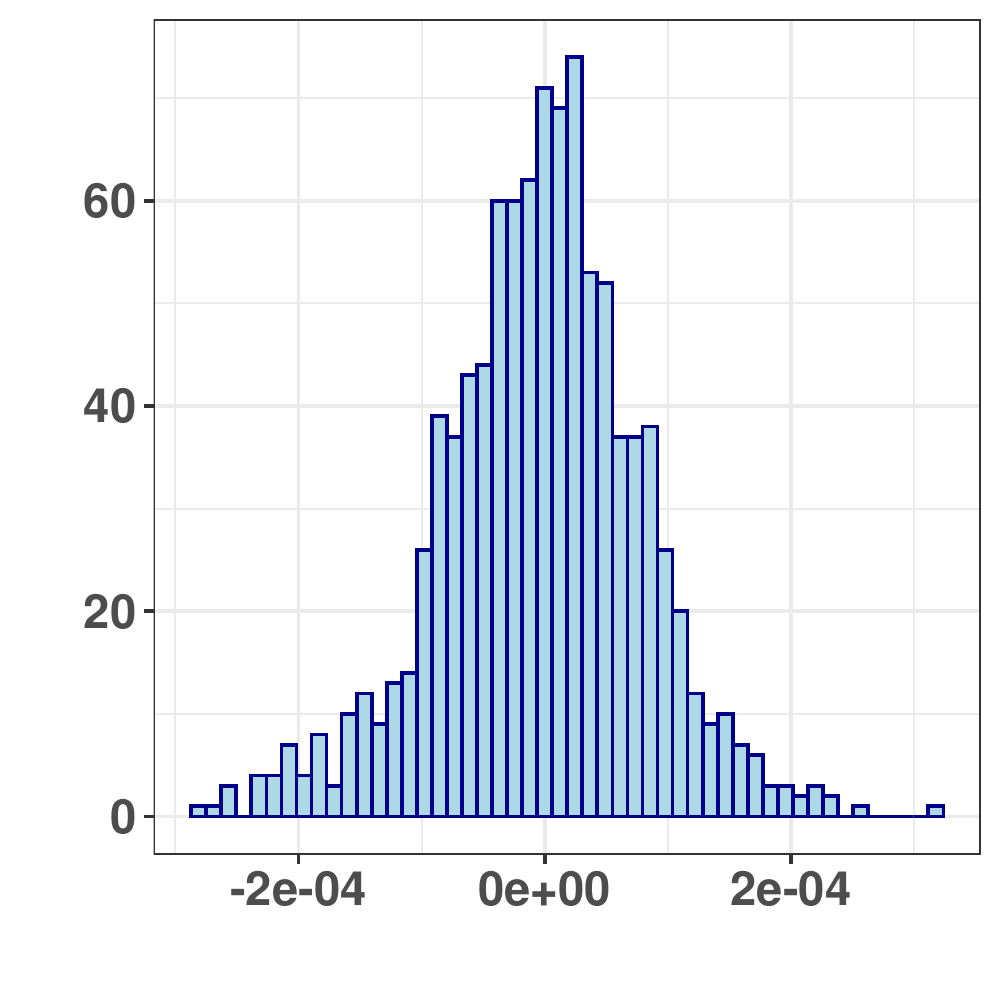}}}
	\caption[Histograms of $\widehat{d}_{B}$ with $B=30$ for different numbers of inducing points ($M=30,45,60$) ]{Histograms of $\widehat{d}_{B}$ with $B=30$ for different numbers of inducing points ($M=30,45,60$).}% 
	\label{fig:normality}
\end{figure}

Finally, Figure~\ref{fig:normality} confirms the normality assumption required by the difference estimate to find the optimal tuning parameters using Algorithm~\ref{algo:tuningpar_v2}.

\subsubsection{Posterior inference and predictions}\label{app:plots_1d}

\begin{figure}[!htbp]
	\subcaptionbox{\scriptsize{M=30, J=4}}{{\includegraphics[scale=.44]{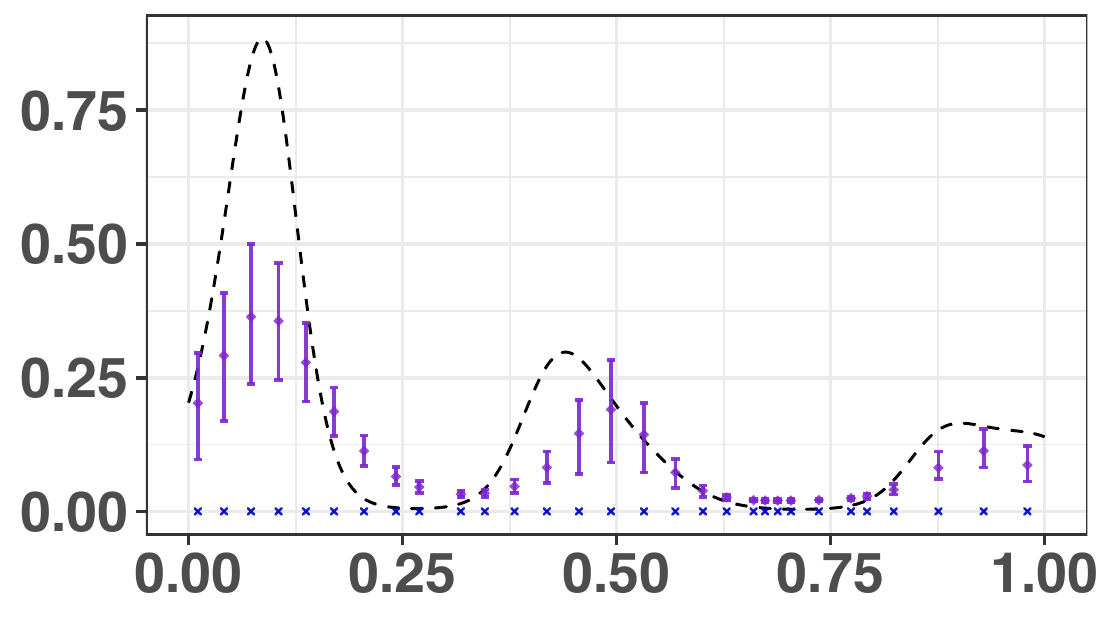}}}
	\subcaptionbox{\scriptsize{M=45, J=4}}{{\includegraphics[scale=.44]{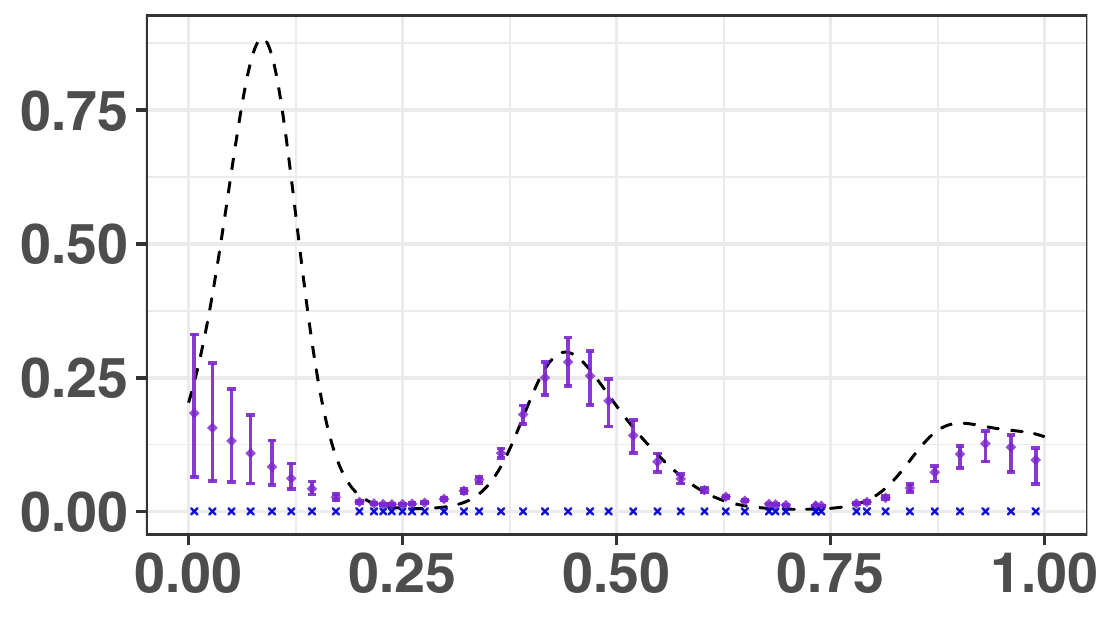}}}
	\subcaptionbox{ \scriptsize{M=60, J=4}}{{\includegraphics[scale=.44]{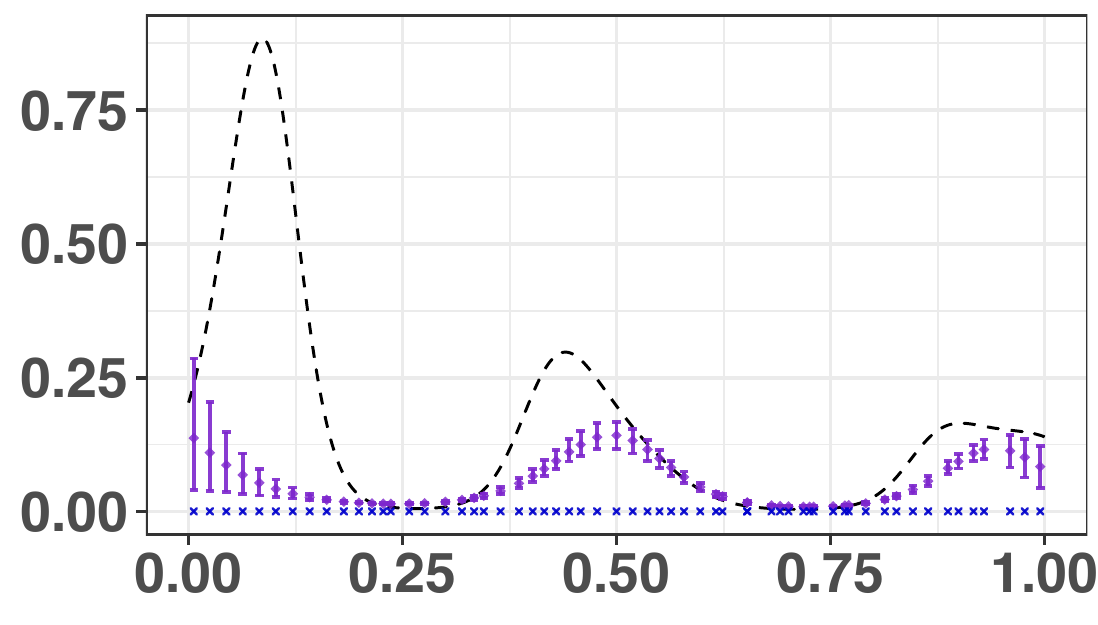}}}
	\\
	\subcaptionbox{\scriptsize{M=30, J=5}}{{\includegraphics[scale=.44]{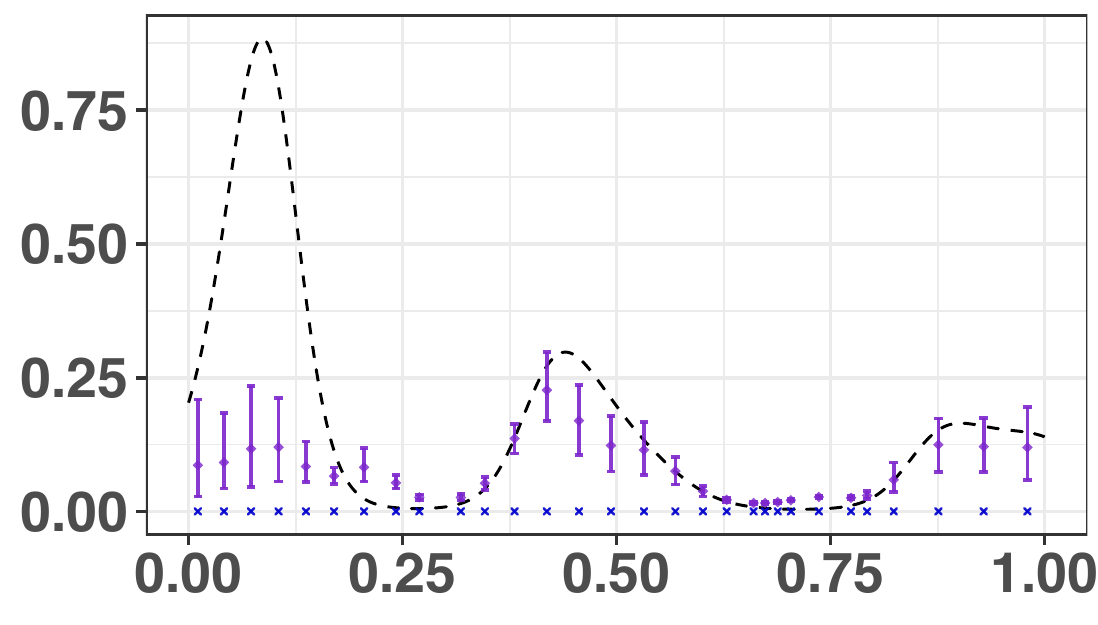}}}
	\subcaptionbox{\scriptsize{M=45, J=5}}{{\includegraphics[scale=.44]{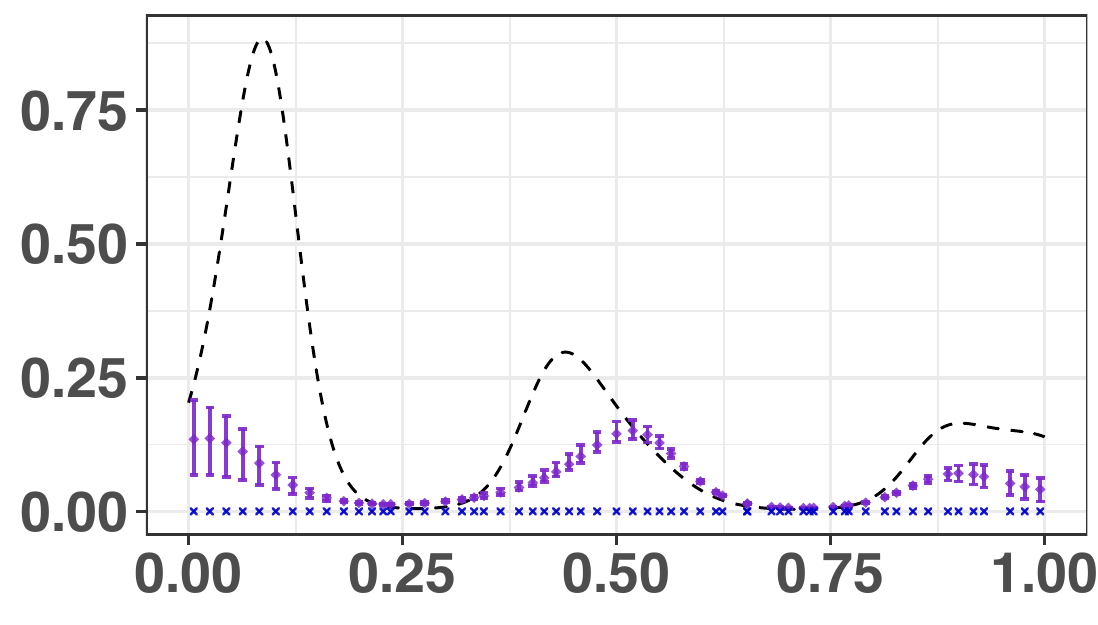}}}
	\subcaptionbox{\scriptsize{M=60, J=5}}{{\includegraphics[scale=.44]{SimstudyMatern/ell_m60_5_b30000.pdf}}}
	\\
	\subcaptionbox{\scriptsize{M=30, J=8}}{{\includegraphics[scale=.44]{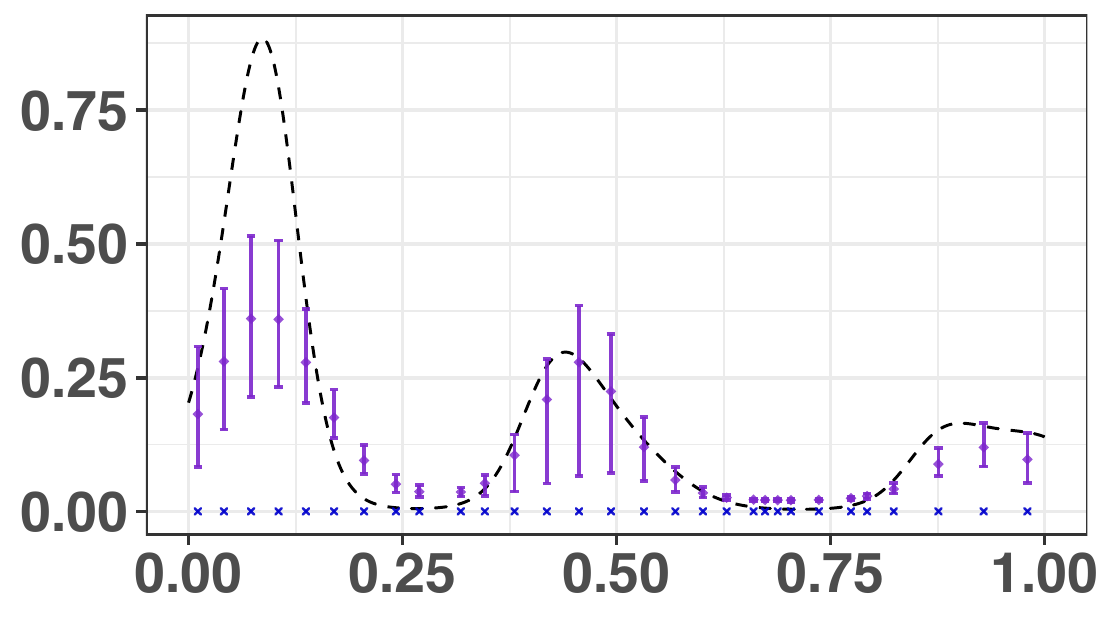}}}
	\subcaptionbox{ \scriptsize{M=45, J=8}}{{\includegraphics[scale=.44]{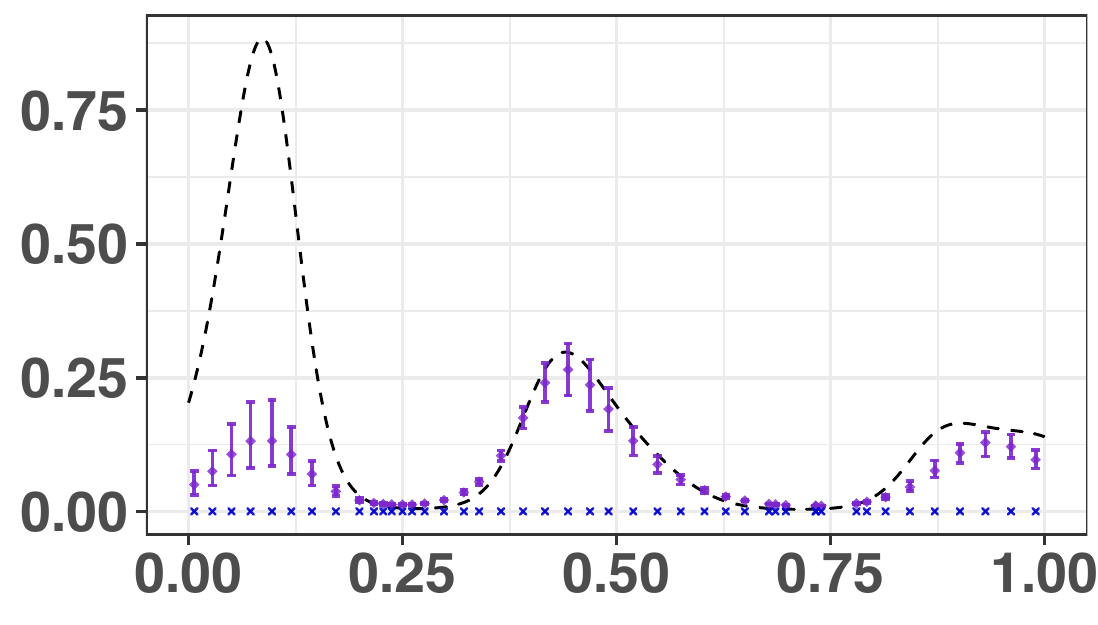}}}
	\subcaptionbox{ \scriptsize{M=60, J=8}}{{\includegraphics[scale=.44]{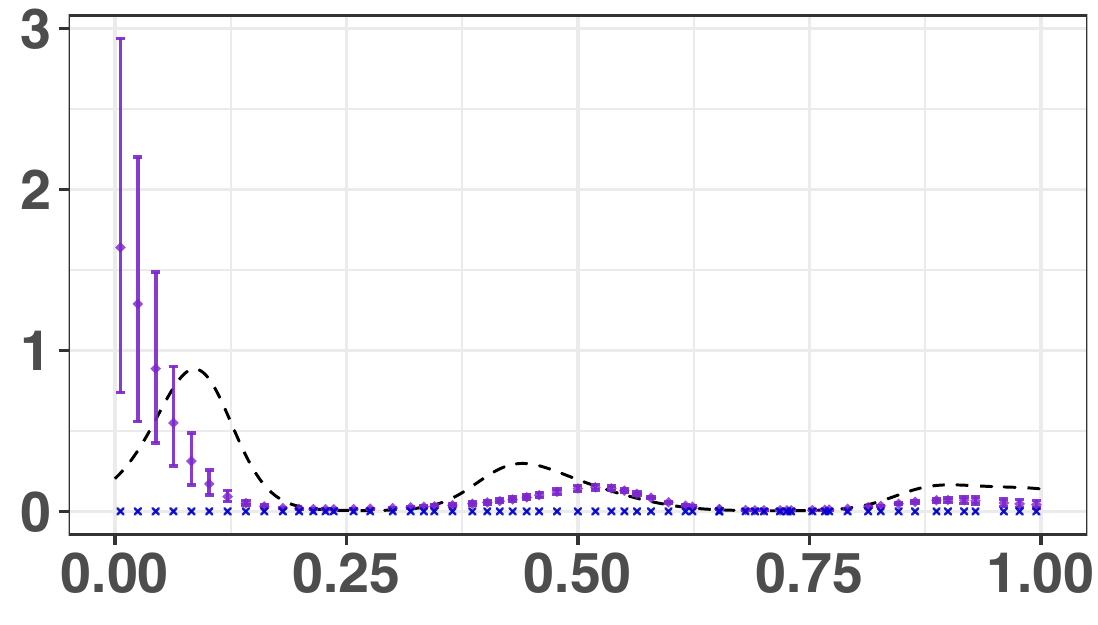}}}
	\\
	\subcaptionbox{\scriptsize{M=30, J=10}}{{\includegraphics[scale=.44]{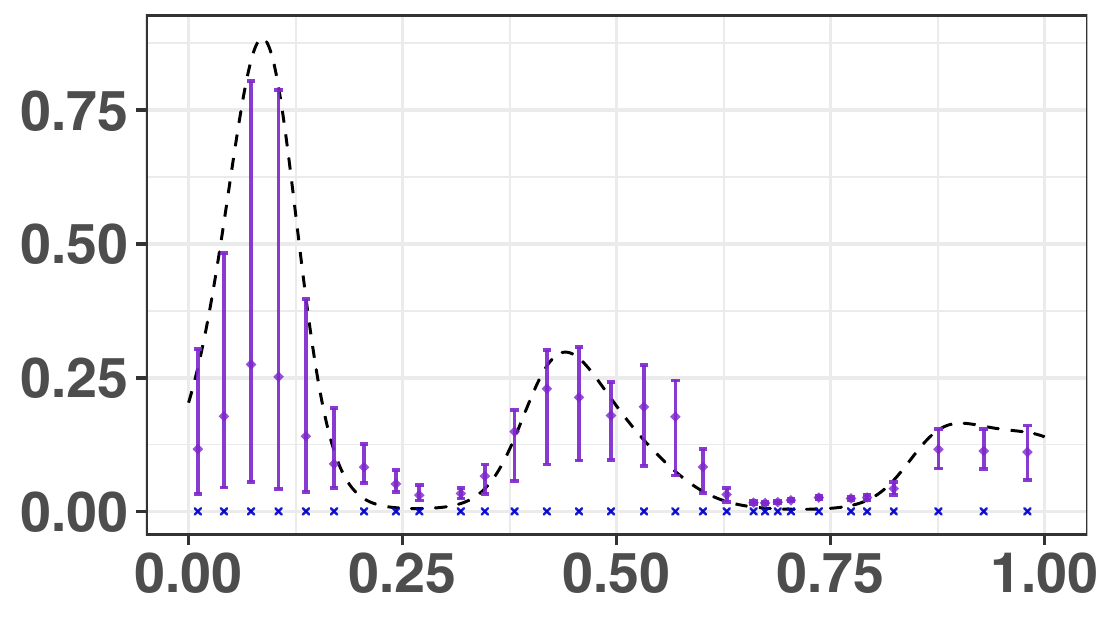}}}
	\subcaptionbox{ \scriptsize{M=45, J=10}}{{\includegraphics[scale=.44]{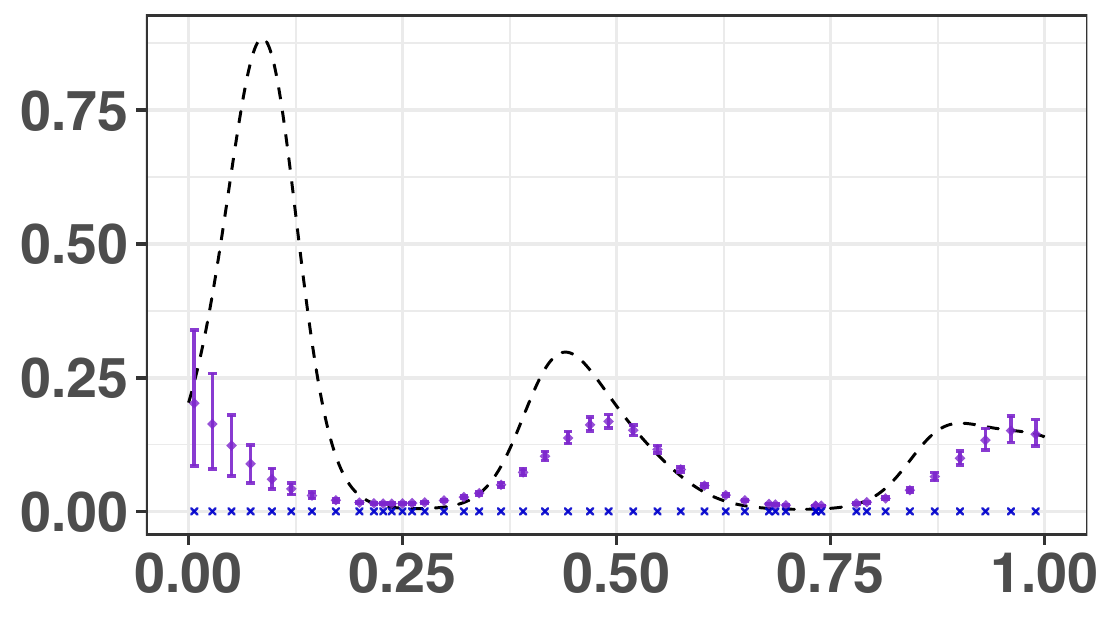}}}
	\subcaptionbox{ \scriptsize{M=60, J=10}}{{\includegraphics[scale=.44]{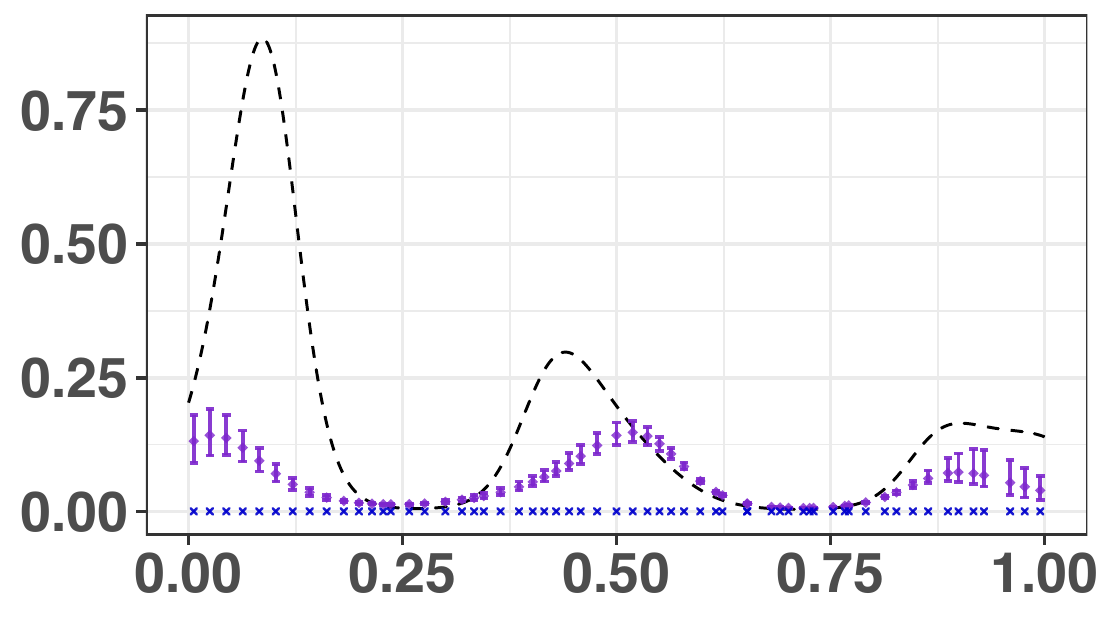}}}
\\
\subcaptionbox{\scriptsize{M=30, J=15}}{{\includegraphics[scale=.44]{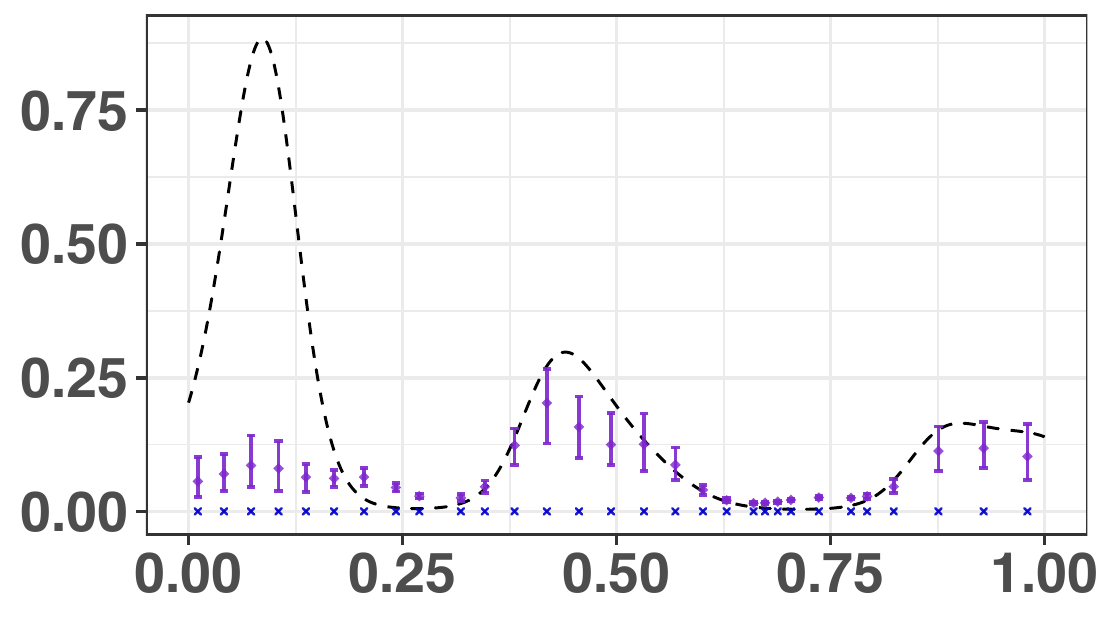}}}
\subcaptionbox{ \scriptsize{M=45, J=15}}{{\includegraphics[scale=.44]{SimstudyMatern/ell_m45_10_b30000.pdf}}}
	\subcaptionbox{ \scriptsize{M=60, J=15}}{{\includegraphics[scale=.44]{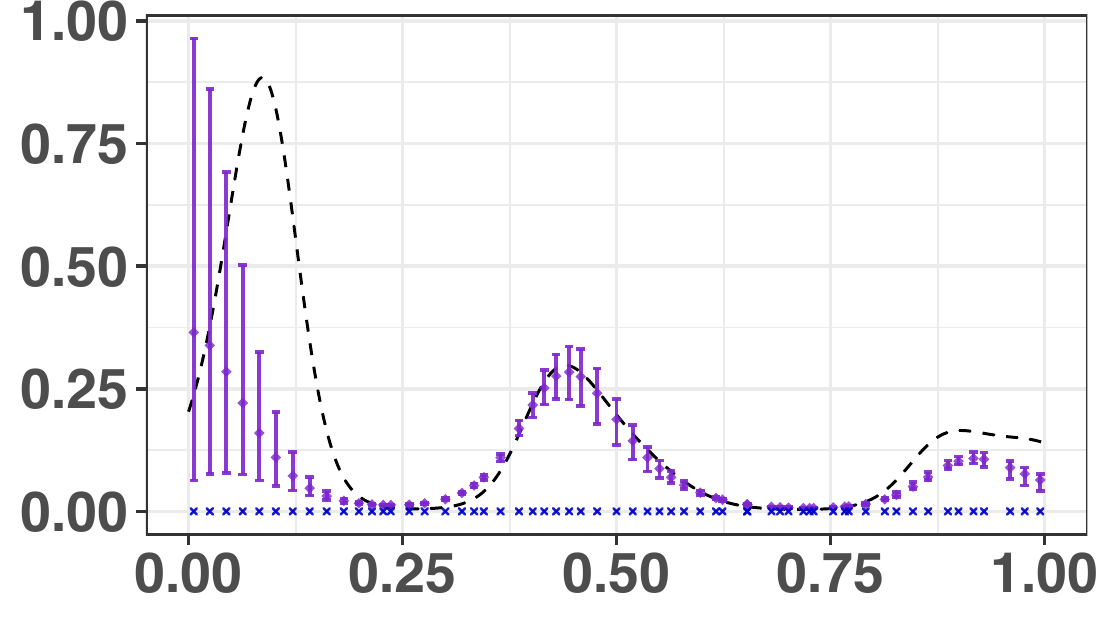}}}
\\
\subcaptionbox{\scriptsize{M=30, S-BP-PM}}{{\includegraphics[scale=.44]{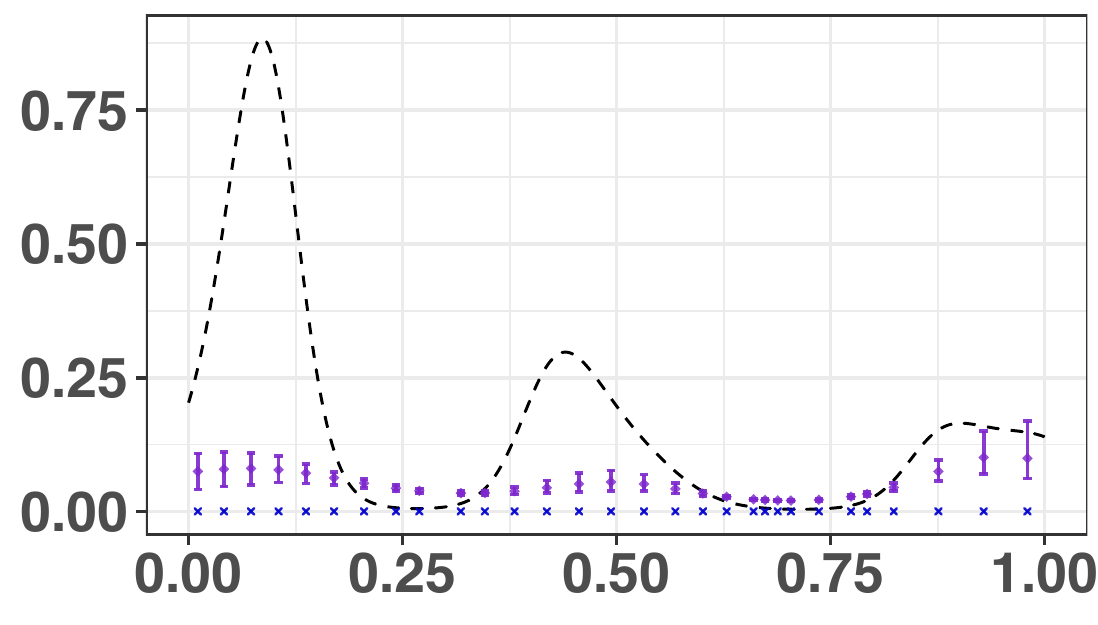}}}
\subcaptionbox{ \scriptsize{M=45, S-BP-PM}}{{\includegraphics[scale=.44]{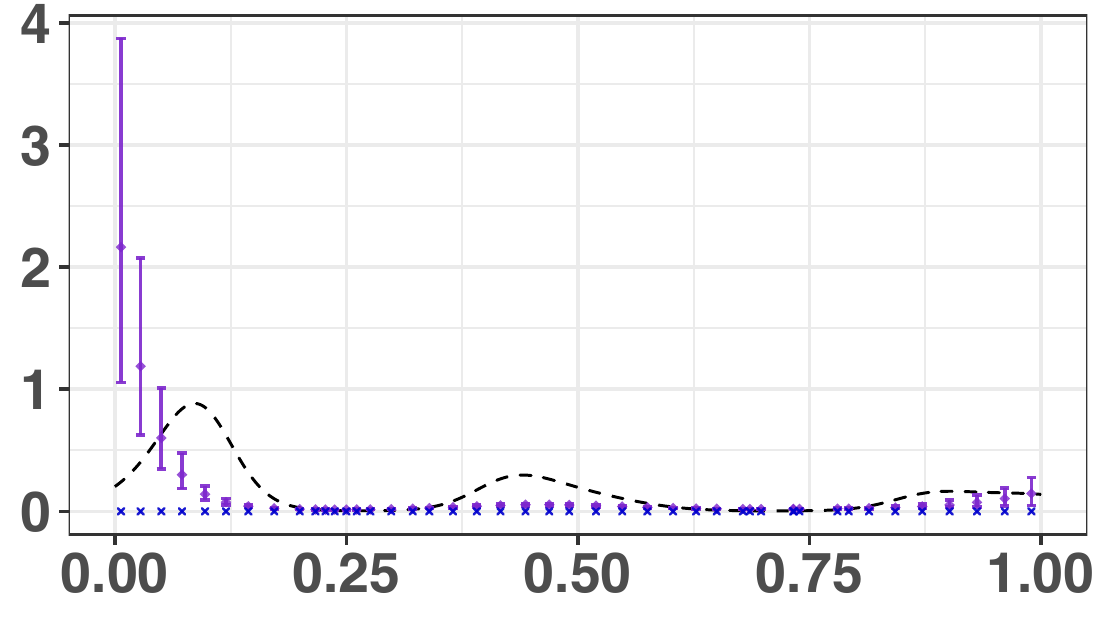}}}
\subcaptionbox{ \scriptsize{M=60, S-BP-PM}}{{\includegraphics[scale=.44]{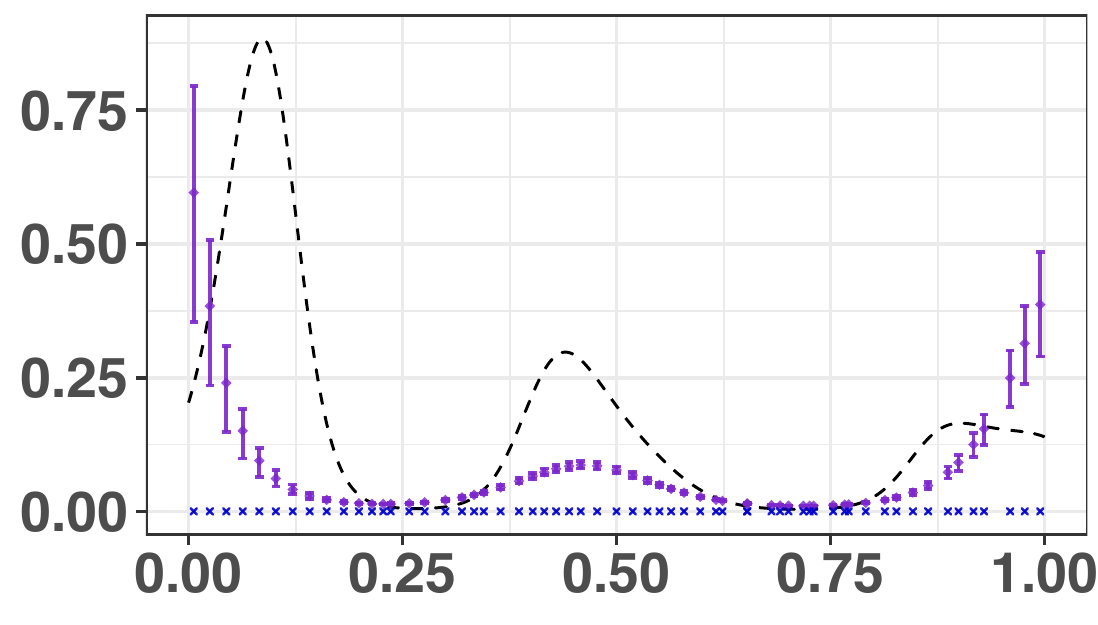}}}
	\caption[Posterior estimates for spatially varying parameter $\ell(\cdot)$]{ Posterior estimates for spatially varying parameter $\ell(\cdot)$. The dashed line denotes the true process. Purple dots and bars show posterior estimates at the inducing locations with 95\% HPD credible intervals.}% 
	\label{fig:ell_posterior}
\end{figure}

\begin{figure}[!htbp]
	\subcaptionbox{\scriptsize{M=30, J=4}}{{\includegraphics[scale=.444]{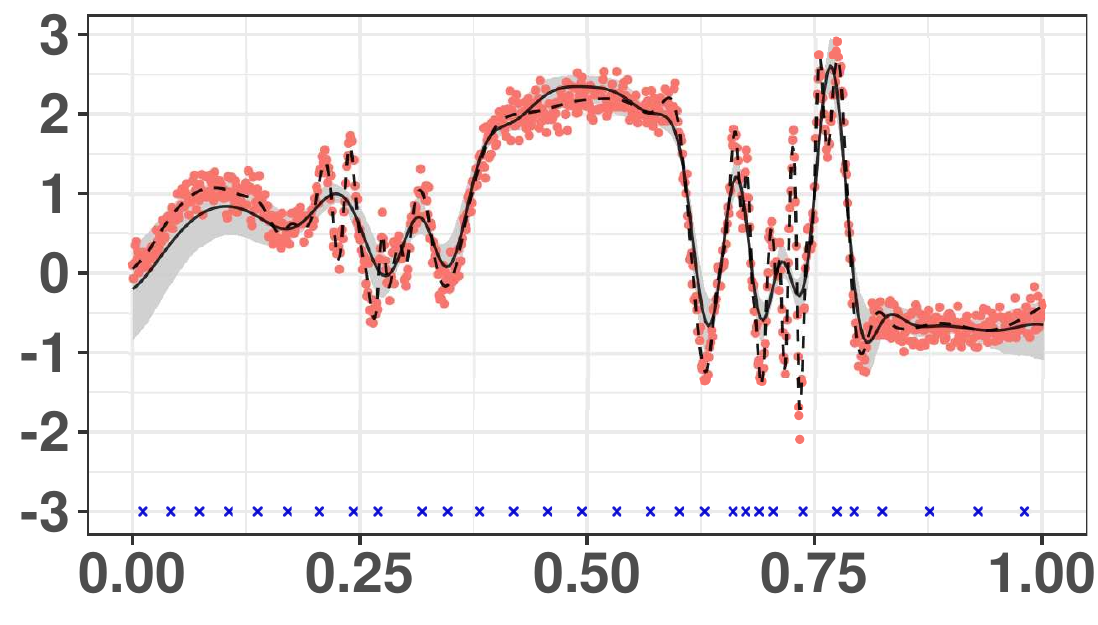}}}
	\subcaptionbox{ \scriptsize{M=45, J=4}}{{\includegraphics[scale=.444]{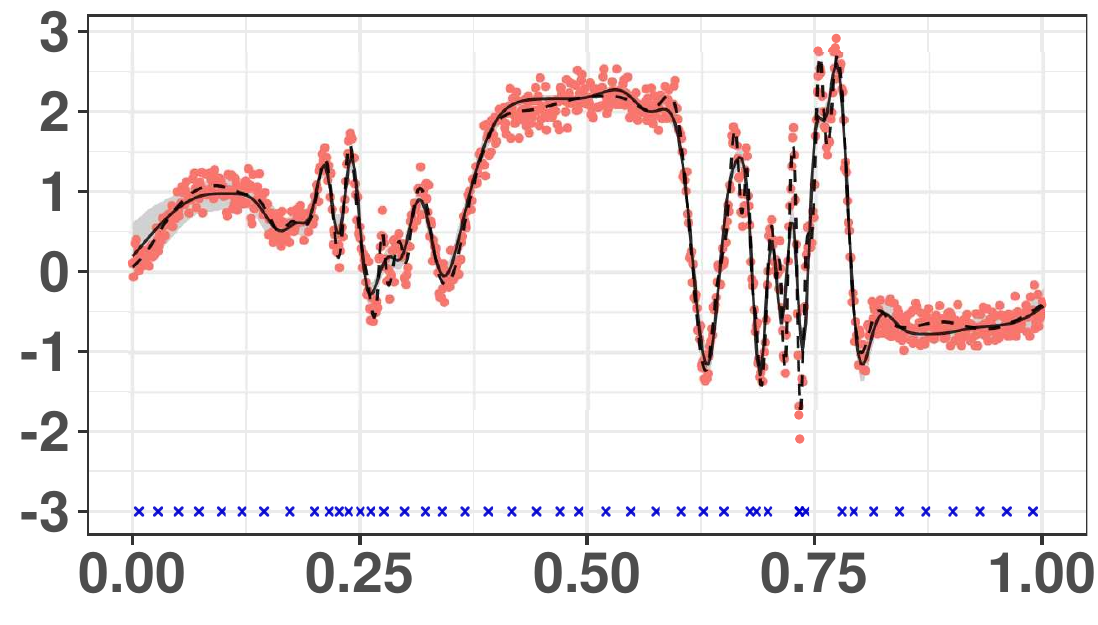}}}
    \subcaptionbox{ \scriptsize{M=60, J=4}}{{\includegraphics[scale=.444]{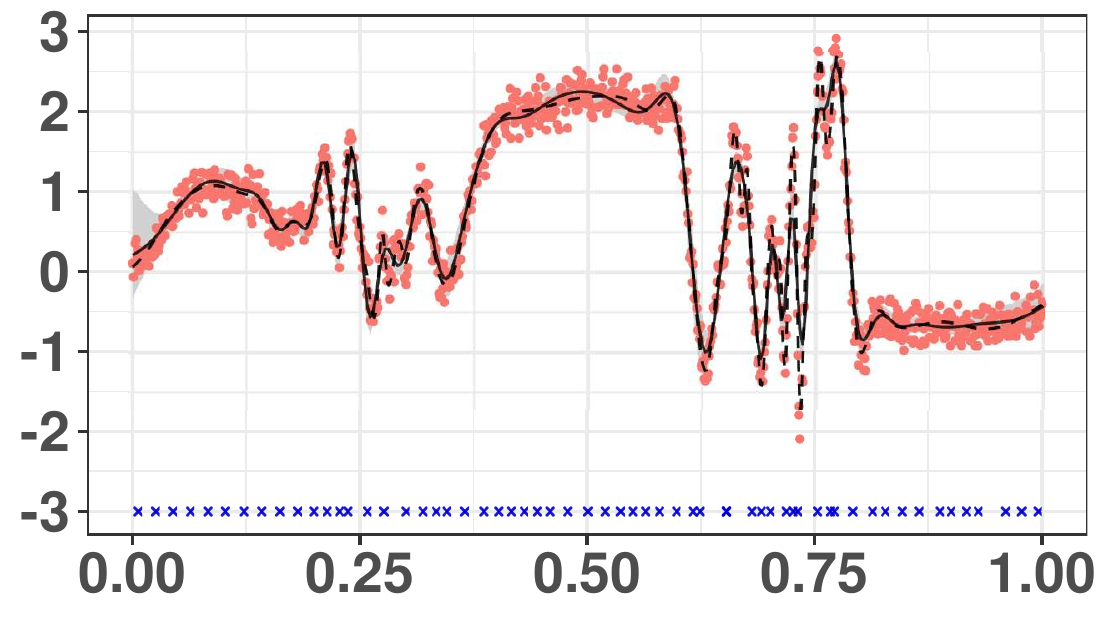}}}
\\
	\subcaptionbox{\scriptsize{M=30, J=5}}{{\includegraphics[scale=.444]{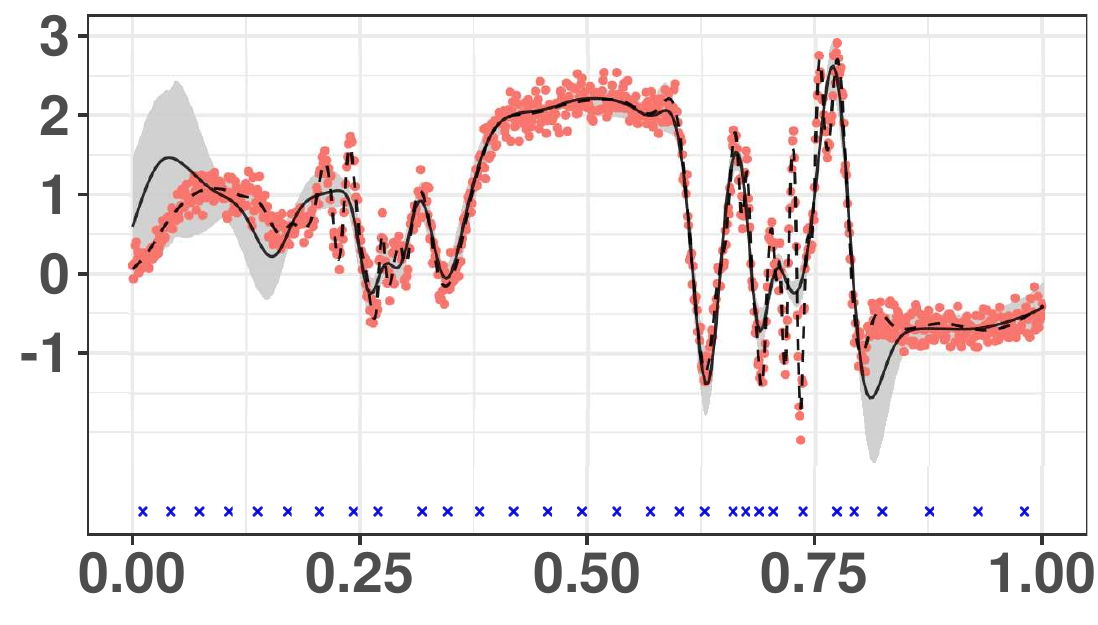}}}
	\subcaptionbox{ \scriptsize{M=45, J=5}}{{\includegraphics[scale=.444]{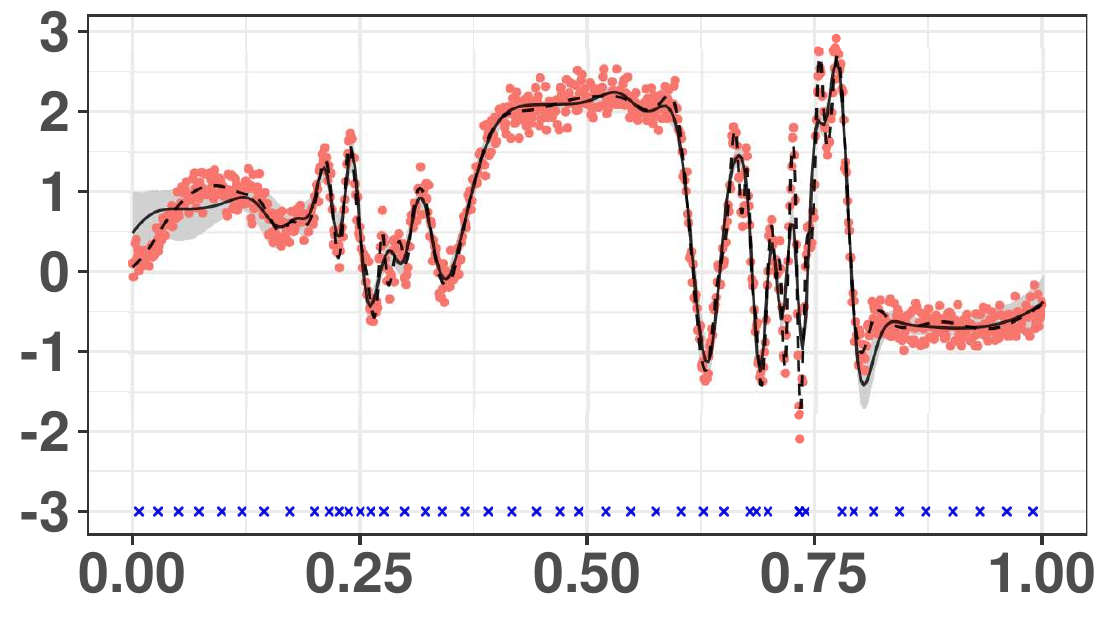}}}
	\subcaptionbox{ \scriptsize{M=60, J=5}}{{\includegraphics[scale=.444]{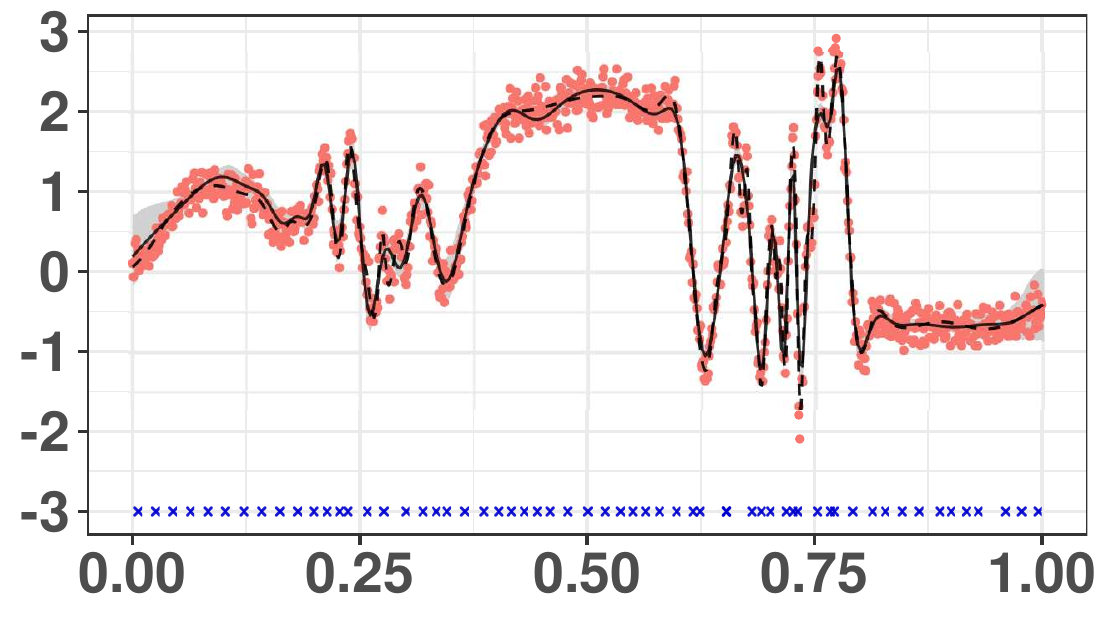}}}
	\\
	\subcaptionbox{ \scriptsize{M=30, J=8}}{{\includegraphics[scale=.444]{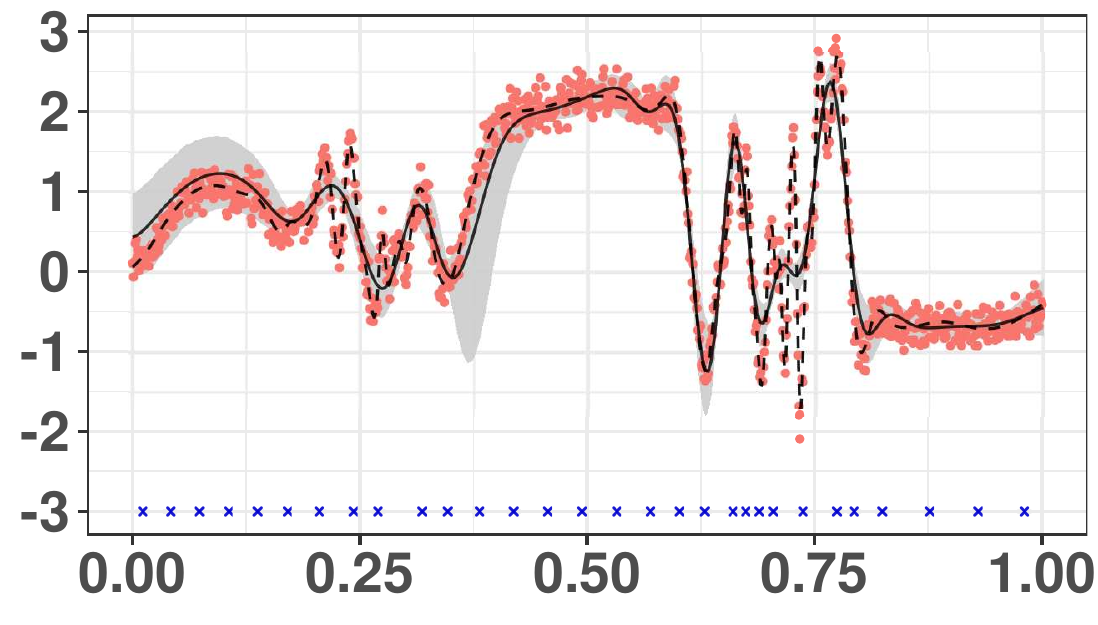}}}
	\subcaptionbox{ \scriptsize{M=45, J=8}}{{\includegraphics[scale=.444]{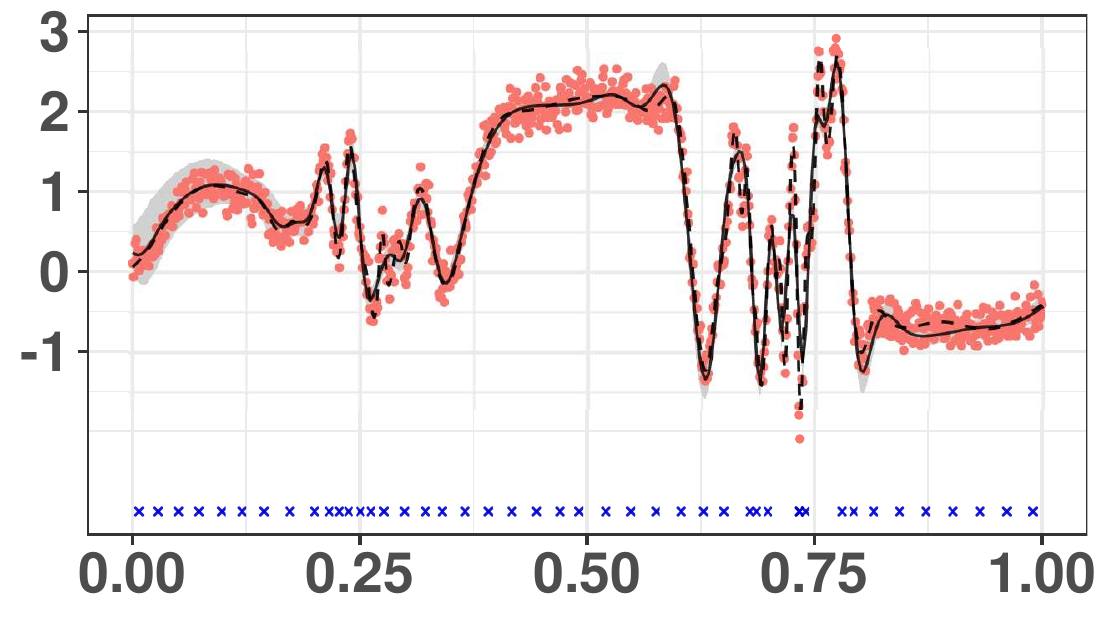}}}
	\subcaptionbox{ \scriptsize{M=60, J=8}}{{\includegraphics[scale=.444]{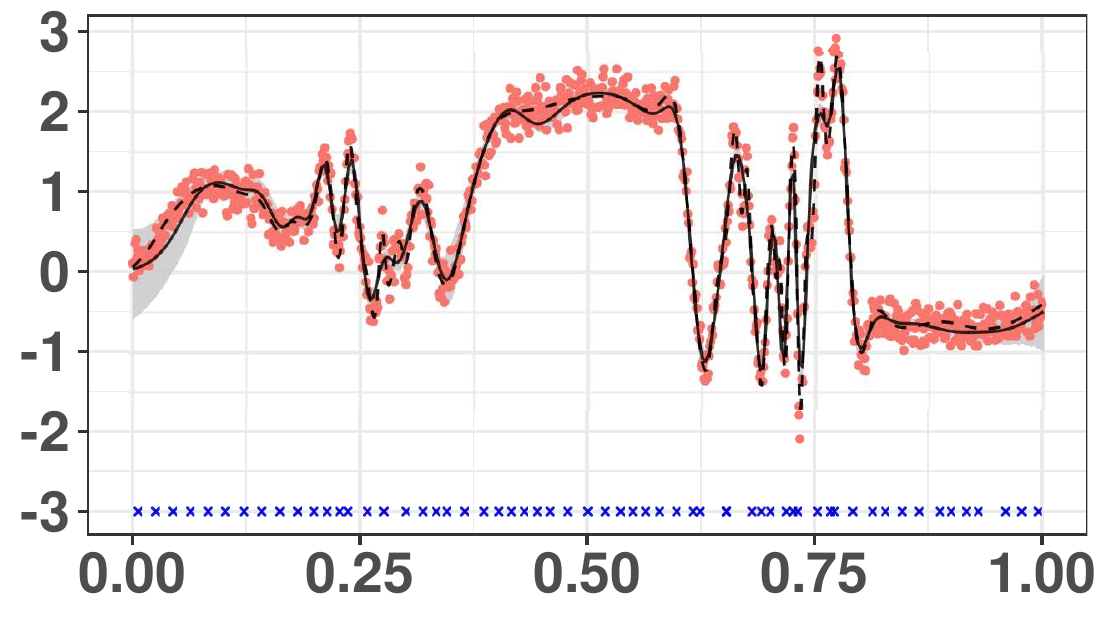}}}
	\\
	\subcaptionbox{\scriptsize{M=30, J=10}}{{\includegraphics[scale=.444]{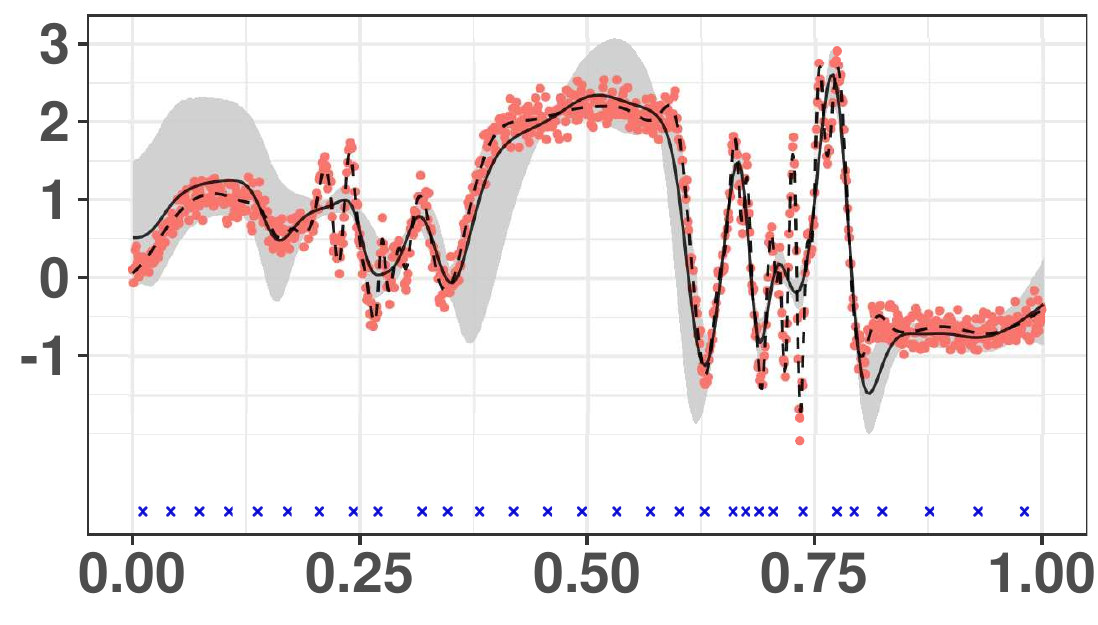}}}
	\subcaptionbox{\scriptsize{M=45, J=10}}{{\includegraphics[scale=.444]{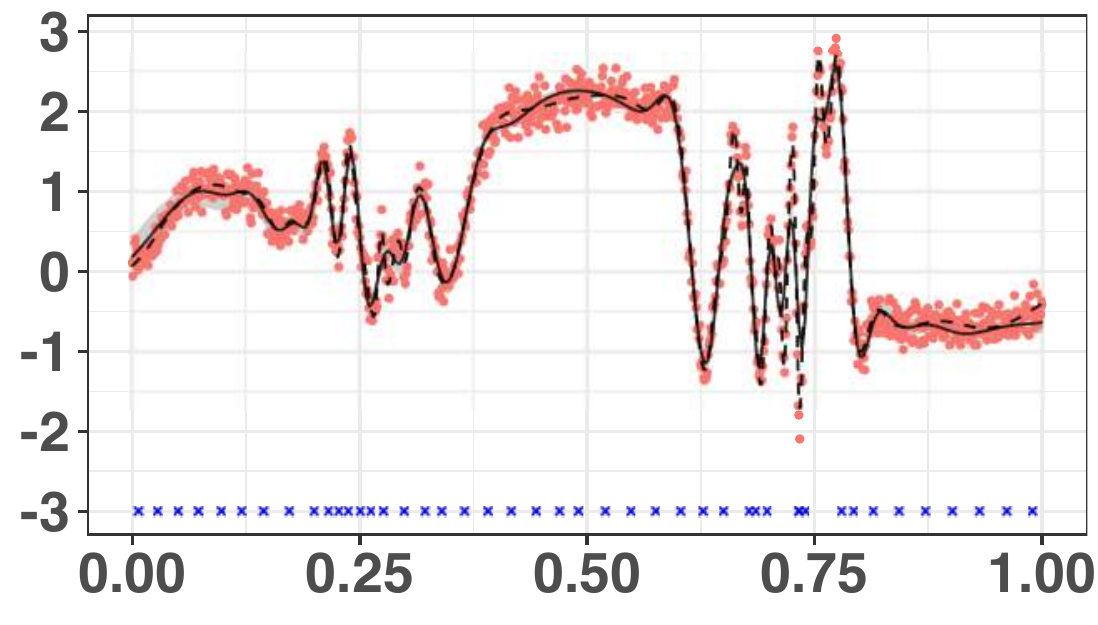}}}
	\subcaptionbox{ \scriptsize{M=60, J=10}}{{\includegraphics[scale=.444]{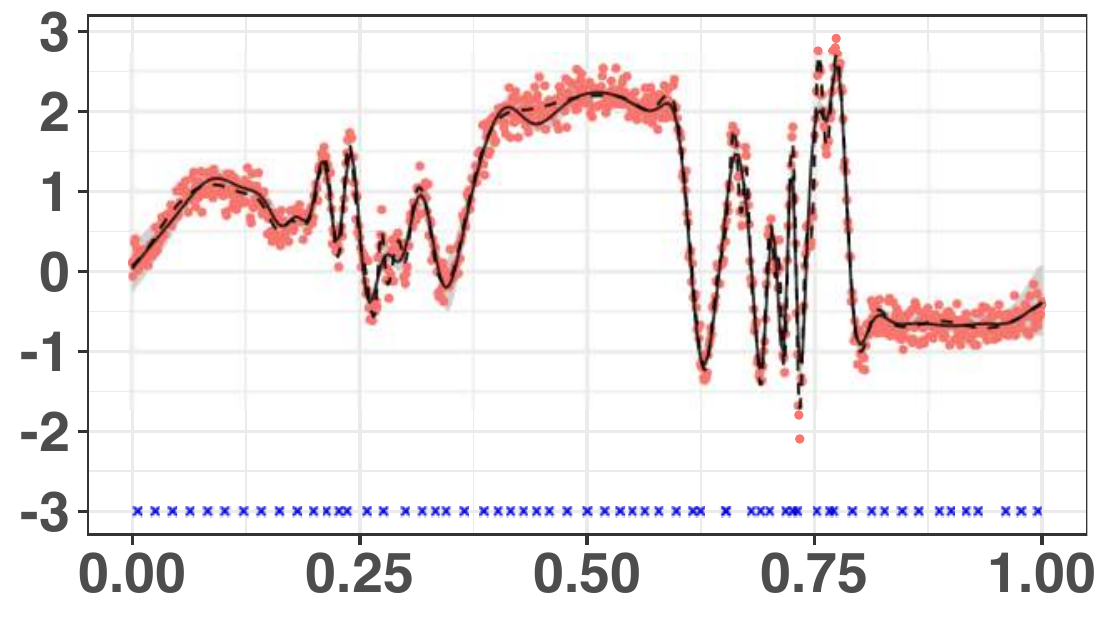}}}
	\\
	\subcaptionbox{\scriptsize{M=30, J=15}}{{\includegraphics[scale=.444]{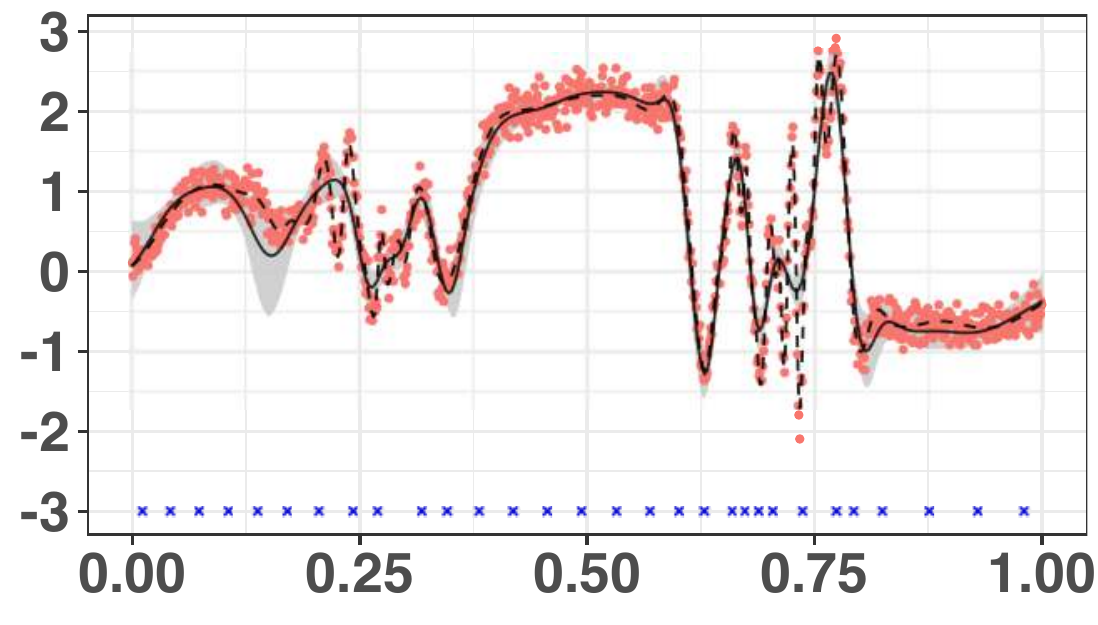}}}
	\subcaptionbox{\scriptsize{M=45, J=15}}{{\includegraphics[scale=.444]{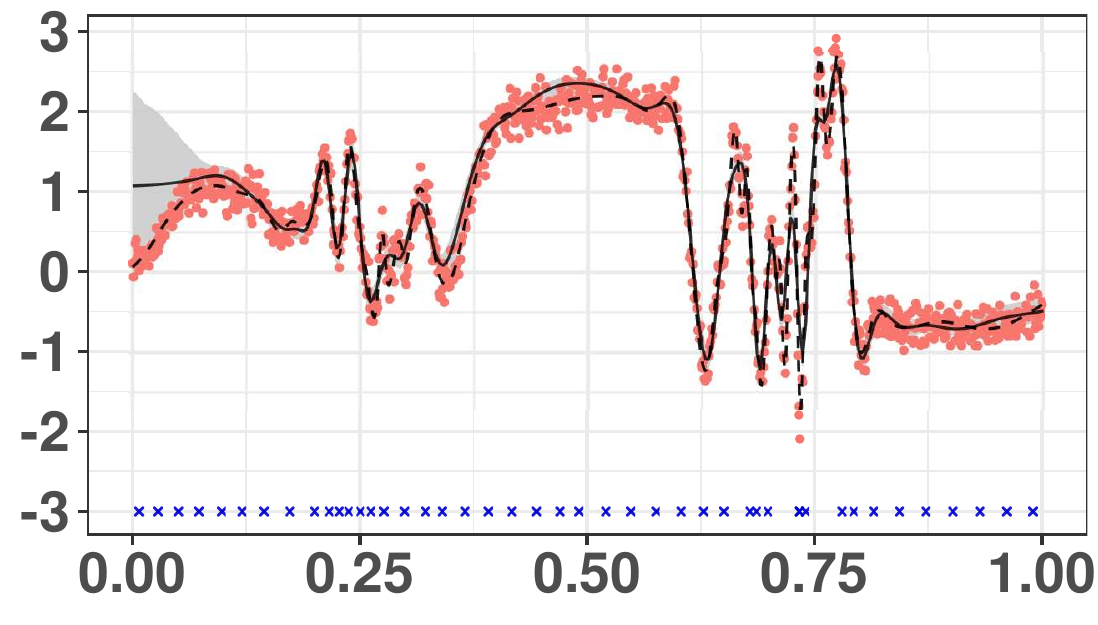}}}
		\subcaptionbox{\scriptsize{M=60, J=15}}{{\includegraphics[scale=.444]{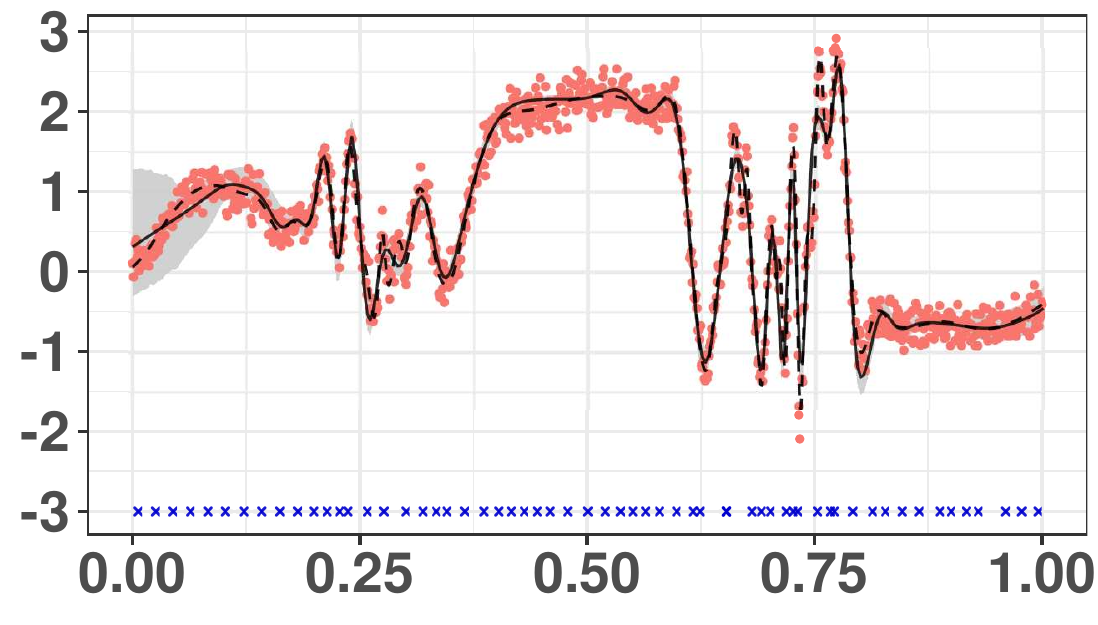}}}
	\\
	\subcaptionbox{\scriptsize{M=30, S-BP-PM}}{{\includegraphics[scale=.444]{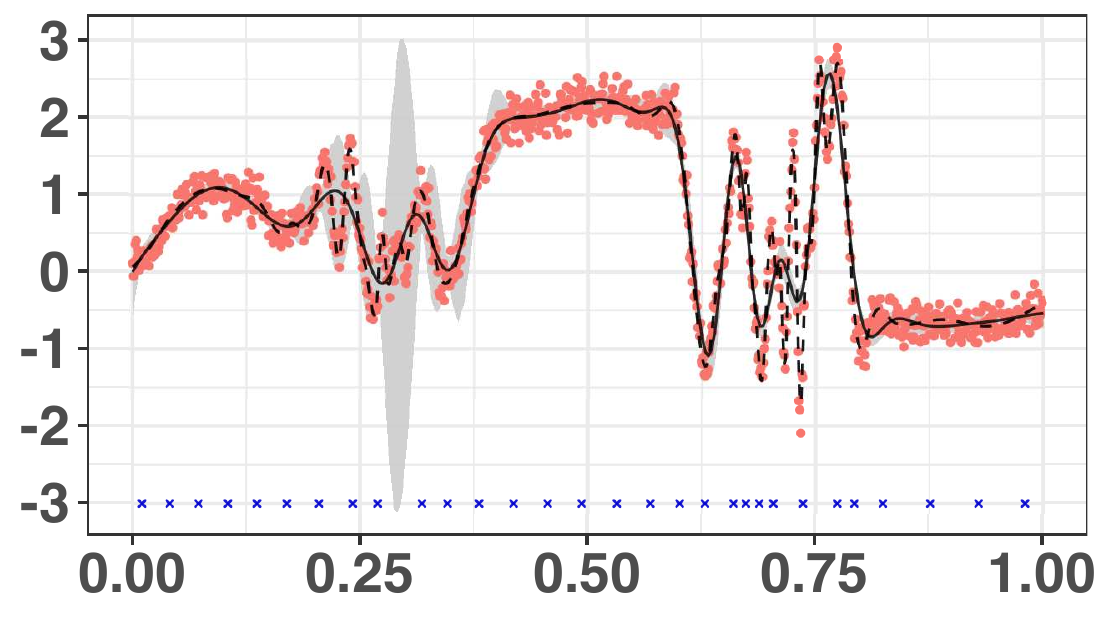}}}
	\subcaptionbox{\scriptsize{M=45, S-BP-PM}}{{\includegraphics[scale=.444]{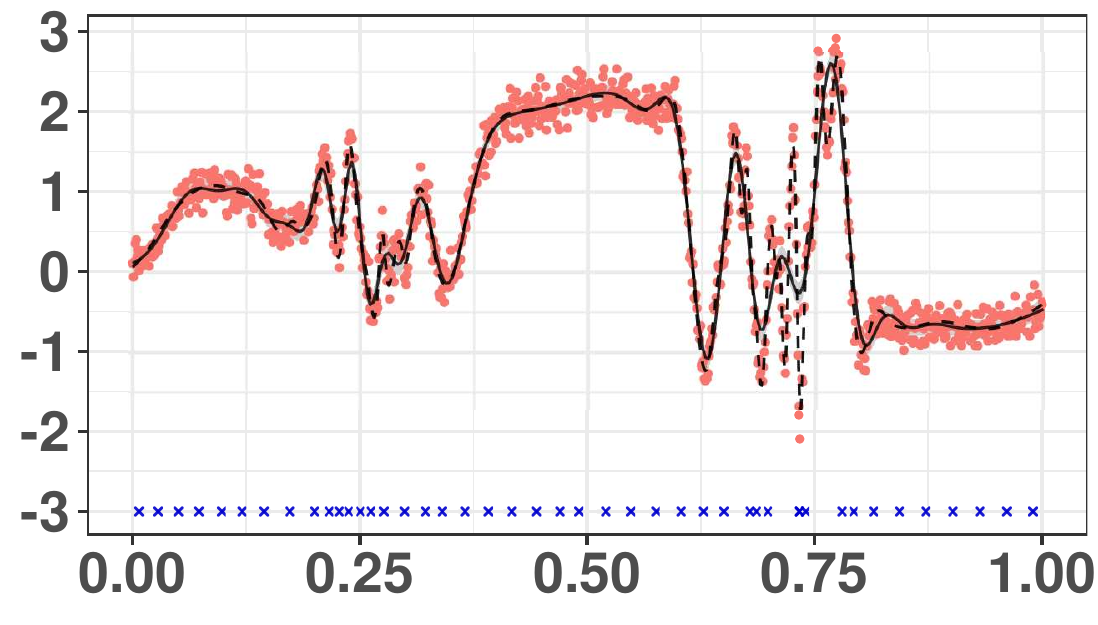}}}
	\subcaptionbox{\scriptsize{M=60, S-BP-PM}}{{\includegraphics[scale=.444]{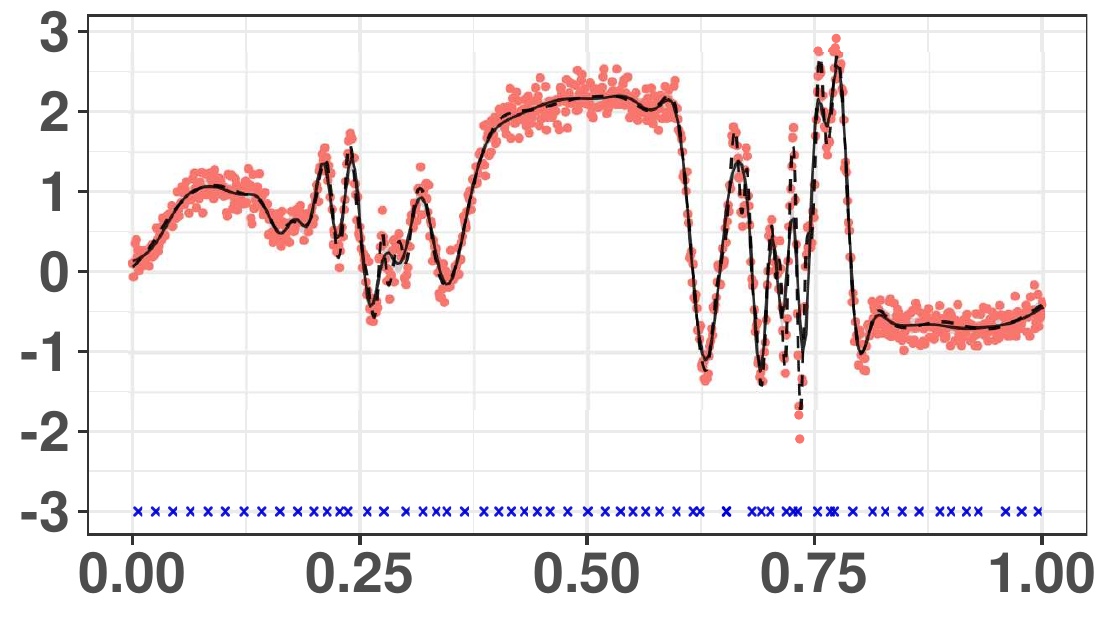}}}
	\\
	\caption{ Predictions for different numbers of inducing points. The solid line denotes the predictive mean and the grey area depicts 95\% HPD point-wise credible intervals. The dashed line denotes the true process.}% 
	\label{fig:z_pred}
\end{figure}

\begin{table}[!htbp]
  \centering
  	\begin{threeparttable}
  \scriptsize
    \begin{tabular}{llcc}
    	\toprule
          &       & \multicolumn{1}{l}{Avg. time (min)} & \multicolumn{1}{l}{Avg. evaluations} \\
    \midrule
    \multicolumn{2}{c}{Full MCMC} & 15.13 & 9.71 \\
    	\cmidrule(lr){1-4} 
    \multirow{5}[0]{*}{$M=30$} & $J=4$   & 1.41  & 9.53 \\
          & $J=5$   & 1.69  & 9.53 \\
          & $J=8$   & 2.53  & 9.53 \\
          & $J=10$  & 3.06  & 9.53 \\
          & $J=15$  & 4.34  & 9.53 \\
          	\cmidrule(lr){2-4} 
          & S-BP-PM & 0.48 & 11.32 \\ 
          	\cmidrule(lr){1-4} 
    \multirow{5}[0]{*}{$M=45$} & $J=4$   & 1.90  & 8.84 \\
          & $J=5$   & 2.42  & 9.35 \\
          & $J=8$   & 3.77  & 9.34 \\
          & $J=10$  & 4.82  & 9.91 \\
          & $J=15$  & 6.21  & 8.73 \\
          	\cmidrule(lr){2-4} 
           & S-BP-PM & 0.73 & 11.98\\
          	\cmidrule(lr){1-4} 
    \multirow{5}[0]{*}{$M=60$} & $J=4$   & 2.68  & 9.42 \\
          & $J=5$   & 3.62  & 9.79 \\
          & $J=8$   & 5.25  & 9.02 \\
          & $J=10$  & 6.56  & 9.18 \\
          & $J=15$  & 9.44  & 9.13 \\
          	\cmidrule(lr){2-4} 
           & S-BP-PM & 1.17 & 12.08\\
          	\bottomrule\\
    \end{tabular}
    \end{threeparttable}
        \caption[Average time (in minutes) and likelihood evaluations required in the MCMC scheme.]{Average time (in minutes) and likelihood evaluations required in the MCMC scheme. The average time required for 100 iterations is reported in minutes. The average number of likelihood evaluations in the elliptical slice sampler per iteration is reported.}
        	\label{tab:comptime}%
\end{table}%
\end{document}